\documentclass[aps,rmp,twocolumn,groupedaddress]{revtex4-1}

\usepackage[dvips]{graphicx}
  \setkeys{Gin}{draft=false}
\usepackage{color}
\usepackage{amsmath,amssymb}
\usepackage{mathtools}
\DeclarePairedDelimiter\abs{\lvert}{\rvert}
\newcommand{\suchthat}{\mid}

\newcommand{\optcite}[1]{\ignorespaces}
\newcommand{\optcitep}[1]{\ignorespaces}

\usepackage{mathrsfs}
\renewcommand{\vec}[1]{\mathbf{#1}}
\newcommand{\fourier}[1]{\hat{#1}}
\newcommand{\ensavg}[1]{\langle {#1} \rangle}
\newcommand{\domain}{\mathcal{D}}

\newcommand{\lapgreen}{G}
\newcommand{\partfun}{\mathcal{Z}}
\newcommand{\meanfield}[1]{\ensuremath{\mathscr{#1}}}  
\newcommand{\thermo}[1]{\ensuremath{#1}}		       

\usepackage{mathtools}
\usepackage{graphicx}
\usepackage{mathrsfs}
\usepackage{amsmath}
\usepackage{amsfonts}
\usepackage{amssymb}

\begin{document}

%
%

\title{Mathematical and Physical Ideas for Climate Science}

\author{Valerio Lucarini}
\email[]{valerio.lucarini@uni-hamburg.de}
\altaffiliation{Also at: Department of Mathematics and Statistics, University of Reading, Reading, RG6 6AX, UK.}
\author{Richard Blender, Salvatore Pascale}
\author{Francesco Ragone}
\altaffiliation{Now at:  Klimacampus, Institut f\"{u}r Meereskunde, University of Hamburg, 20146 Bundestrasse 53, Hamburg, Germany.}
\author{Jeroen Wouters}
\altaffiliation{Now at: Laboratoire de Physique, \'{E}cole Normale Sup\'{e}rieure de Lyon, 69364 46, all\'{e}e d'Italie, Lyon, France.}
\affiliation{Klimacampus, Meteorologisches Institut, University of Hamburg, 20144 Grindelberg 5, Hamburg, Germany.}
\author{Corentin Herbert}
\affiliation{National Center for Atmospheric Research, P.O. Box 3000, Boulder, CO, 80307, USA.}

\begin{abstract}
The climate is a forced and dissipative nonlinear system featuring non-trivial dynamics of a vast range of spatial and temporal scales. The understanding of the climate's structural and multiscale properties is crucial for the provision of a unifying picture of its dynamics and for the implementation of accurate and efficient numerical models. We present some recent developments at the intersection between climate science, mathematics, and  physics, which may prove fruitful in the direction of constructing a more comprehensive account of climate dynamics. We describe the Nambu formulation of fluid dynamics, and the potential of such a theory for constructing sophisticated numerical models of geophysical fluids. Then, we focus on the statistical mechanics of quasi-equilibrium flows in a rotating environment, which seems crucial for constructing a robust theory of geophysical turbulence. We then discuss ideas and methods suited for approaching directly the non-equilibrium nature of the climate system. First, we describe some recent findings on the thermodynamics of climate and characterize its energy and entropy budgets, and discuss related methods for intercomparing climate models and for studying tipping points. These ideas can also create a common ground between geophysics and astrophysics by suggesting general tools for studying exoplanetary atmospheres. We conclude by focusing on non-equilibrium statistical mechanics, which allows for a unified framing of problems as different as the climate response to forcings, the effect of altering the boundary conditions or the coupling between geophysical flows, and the derivation of parametrizations for numerical models.\end{abstract}

%
%

%
\maketitle

%
%


%
%

%


%
%


\section{Introduction}

The Earth's Climate provides an outstanding example of a high-dimensional forced and dissipative complex system. The dynamics of such system is chaotic, so that there is only a limited time-horizon for skillful prediction, and is non-trivial on a vast range of spatial and temporal scales, as a result of the different physical and chemical properties of the various components of the climate system and of their coupling mechanisms \citep{Peixoto:1992}. 

Thus, it is extremely challenging to construct satisfactory theories of climate dynamics and it is virtually impossible to develop numerical models able to describe accurately climatic processes over all scales. Typically, different classes of models and different phenomenological theories have been and are still being developed by focusing on specific scales of motion \citep{Holton,vallis_atmospheric_2006}, and simplified parametrizations are developed for taking into account at least approximately what cannot be directly represented \citep{palmer_stochastic_2009}. 

As a result of our limited understanding of and ability to represent the dynamics of the climate system, it is  hard to predict accurately its response to perturbations, were they changes in the opacity of the atmosphere, in the solar irradiance, in the position of continents, in the orbital parameters, which have been present for our planet during all epochs \citep{saltzman_dynamical}. The full understanding of slow- and fast-onset climatic extremes, such as drought and flood events, respectively, and the assessment of the processes behind tipping points responsible for the multi stability of the climate system are also far from being accomplished. 

Such limitations are extremely relevant for problems of paleoclimatological relevance such as the onset and decay of ice ages or of snowball-conditions, for contingent issues like anthropogenic global warming, as well as in the perspective of developing a comprehensive knowledge on the dynamics and thermodynamics of general planetary atmospheres, which seems a major scientific challenge of the coming years, given the extraordinary development of our abilities to observe exoplanets \citep{Dvorak}. 

Climate science \textit{at large} has always been extremely active in taking advantage of advances in basic mathematical and physical sciences, and, in turn, in providing stimulations for addressing new fundamental problems. The most prominent cases of such interaction are related to the development of stochastic and chaotic dynamical systems, time series analysis, extreme value theory, radiative transfer, and fluid dynamics, among others. At this regard, one must note that the year 2013 has seen a multitude of initiatives all around the world dedicated to the theme \textit{Mathematics of Planet Earth} (see \texttt{http://mpe2013.org}), and, in this context, climate-related activities  have been of great relevance.

In this review we wish to present some interdisciplinary  research lines at the intersection between climate science, physics and mathematics, which are  extremely promising for advancing, on one side, our ability to understand and model climate dynamics, and represent correctly climate variability and climate response to forcings. On the other side, the topics presented here provide examples of how problems of climatic relevance may pave the way for new, wide-ranging investigations of more general nature. 

The  literature related to the scientific interface mentioned above is enormous, and the selection of the material we present here is partial and non-exhaustive. We leave almost entirely out of this review very important topics such as extreme value theory \citep{ghil2011}, multiscale techniques \citep{klein2010}, adjoint methods and data assimilation \citep{wunsch2012},  partial different equations \citep{cullen2006},  linear and nonlinear stability analysis \citep{vallis_atmospheric_2006}, general circulation of the atmosphere \citep{schneider2006}, macroturbulence \citep{shaun2013}, networks theory \citep{donges2009}, and many relevant applications of dynamical systems theory to geophysical fluid dynamical problems \citep{kalnay2003,dijkstra2013}. 

Let us now mention what we are going to cover in this review and give a motivation to the specific perspective we have chosen. We are  motivated by the desire of bridging the gap between some extremely relevant results in mathematical physics, statistical mechanics, and theoretical physics, and open problems and issues of climate science, hoping to stimulate further investigations and interdisciplinary activities. Our selection of topics will focus on the concepts of energy, entropy, symmetry, coupling, fluctuations, and response. 

We will first concentrate on the properties of inviscid and unforced flows relevant for geophysical fluid dynamics (GFD). In  section \ref{nambu}, we provide an overview of a very powerful  formulation of hydrodynamics based on the formalism introduced by \citet{Nambu:1973} and present its applications in a geophysical context, suggesting how these ideas help clarifying somewhat \textit{hidden} properties of fluid flows, and how the Nambu formulation of GFD  could lead to a new generation of numerical models, to be used in a variety of weather and climate applications  In  section \ref{onsager}, starting from the classical investigation by \citet{Onsager1949} of the dynamics of point vortices, we will show how to develop an equilibrium statistical mechanical theory of turbulence for GFD flows and will discuss its relevance for interpreting observed climatic phenomena.

Equilibrium methods allow investigating  many properties of GFD flows. Nonetheless, at this point we cannot ignore anymore the \textit{elephant in the room}, \textit{i.e.}, the fact that the dynamics of the climate system cannot be assimilated to an inviscid and unforced GFD flow, because forcing and dissipative processes are of extreme relevance. Thus, we move towards  the paradigm of non-equilibrium systems. In  section \ref{prigogine}, taking inspiration from the points of view of \citet{Prigogine61} and of \citet{Lor67}, we explore how through classical non-equilibrium thermodynamics one can construct tools for assessing the energy budget and transport  of the climate system, define and estimate the efficiency of the \textit{climate machine}, and study the irreversible processes by evaluation the climatic material entropy production. This allows for characterizing the large scale properties of climate, for developing tools for   auditing climate models,  for gathering information on tipping points, and for exploring the properties of general planetary atmospheres. In  section \ref{ruelle}, we address the  non-equilibrium statistical mechanics formulation of climate dynamics, and explore how the formalism of response theory allows for addressing in a rigorous framework the climatic response to perturbations, taking inspiration from the work of  \citet{ruelle_differentiation_1997}. We will show how it is possible to construct operators useful for the prediction - in an ensemble sense - of climate change. A last aspect of GFD we want to discuss in a statistical mechanical setting is the derivation of parametrizations providing a surrogate description of the effect of fast, small scale variables, which are hard to represent explicitly in numerical models, on the larger scale, slow variables of more direct climatic relevance. Thus, in  section \ref{morizwanzig}, we present averaging and homogenization techniques, describe how projector operator methods due to \citet{mori_transport_1965} and \citet{zwanzig_memory_1961} provide powerful tools for  deriving parametrizations and firm ground to the inclusion of  stochastic terms and memory effects, and discuss how response theory can be used to derive similar results. 

Finally, in  section \ref{conclusions} we  draw our conclusions and present some perspectives of future research.

 
 \section{Beyond the Hamiltonian paradigm: Nambu representation of Geophysical Fluid Dynamics}
\label{nambu}

%
%
%
%


Hamiltonian formalism constitutes the backbone of most physical theories. 
In the case of a discrete autonomous  system, the basic idea is to provide a full description of the degrees of freedom by defining a set of canonical variables $q$ and of the related  momenta $p$ ($q$,  $p \in R^N$, \textit{i.e.}, they are $N$-dimensional vectors), and by identifying the time evolution to a  flow in phase space such that the canonical Hamiltonian function $\mathcal{H}$ acts as
a streamfunction,
$\dot{q}=\nabla_p \mathcal{H}, \dot{p}=-\nabla_q \mathcal{H}$, 
where $H(q,p)$ corresponds to the energy of the system, whose value is constant in time. The flow is inherently divergence-free (\textit{solenoidal}), so that the phase space does not contract nor expands, as implied by the Liouville Theorem \citep{LandauMechanics}. The time evolution of any function $X(q,p)$ can be expressed as:
\begin{equation}
\frac{d}{d t }X =\dot{X}=\{X,\mathcal{H}\}_P=\nabla_q X \cdot \nabla_p \mathcal{H} - \nabla_p X \cdot \nabla_q \mathcal{H},
\end{equation}
where $\{,\}_P$ are the so-called Poisson brackets and $\cdot$ indicates the usual scalar product. As suggested by Noether's theorem, the presence of symmetries in the system implies the existence of so-called physically conserved quantities $X_i$, such that $\dot{X}_i=0=\{X_i,\mathcal{H}\}_P$. An autonomous system possesses  time invariance and its   energy  is constant, while in a system possessing translational invariance, the total momentum $M$ is also constant. A system can possess many constants of motions, called \textit{Casimirs}, apart from energy, but the Hamiltonian plays a special role as it is the only function of phase space appearing \textit{explicitly} in the definition of the evolution of the system  \citep{LandauMechanics}. 

 \citet{Nambu:1973} 
presented a generalization  of  canonical Hamiltonian theory for discrete systems.  
The dynamical equations are constructed in 
order to satisfy Liouville's Theorem and are 
written in  terms  of two or more conserved quantities. The Nambu approach has been extremely influential in various fields of mathematics and physics and is viable to extension to the case of continuum, so that it can be translated into a field theory. The construction of a Nambu field theory for geophysical fluid dynamics went through
two decisive steps.  The first was the discovery of a Nambu representation
of 2D and 3D incompressible hydrodynamics
 \citep{Nevir:1993}. The second important step
was the finding that the Nambu representation can be
used to design conservative numerical algorithms 
in geophysical models, and that classical heuristic methods devised by Arakawa for constructing accurate numerical models actually reflected deep symmetries coming from the Nambu structure of the underlying dynamics of the flow \citep{Salmon:2005}.

The physical basis for the relevance of the Nambu theory for describing and simulating conservative geophysical fluid dynamics comes from the existence of relevant   \textcolor{black}{conserved quantities apart energy when forcing and dissipative terms are disregarded from the evolution equations}. 
Such a property is found  in several models relevant for studying  geophysical flows, and  are valid for 2D and 3D hydrodynamics, 
Rayleigh-B\'{e}nard convection,
quasi-geostrophy, shallow water model, and extends to the fully
baroclinic 3D atmosphere. In other terms, the Nambu representation provides the natural description of geophysical fluid dynamics and is superior to the more traditional approaches based essentially on Euler equations, just like the action-angle representation of the dynamics of a spring is superior to the simple description provided by the second Newton's law of motion. 



\subsection{Hydrodynamics in 2D and 3D}

In incompressible hydrodynamics enstrophy (in 2D) and helicity (3D) are
known as integral  conserved  quantities  besides energy
\citep{Kuroda:1991}.   
%
%
\citet{Nevir:1993}  adapted  Nambu's formalism  to incompressible
nonviscous hydrodynamics by using enstrophy and helicity 
in the dynamical equations. 
 
\subsubsection{Two-dimensional hydrodynamics}\label{2dhdsection}

The evolution of two-dimensional incompressible inviscid and unforced flows described by the velocity field $\mbox{\boldmath{$u$}}$ is governed by the
vorticity equation 
	\begin{equation}  \label{vorteq2D}
   		\frac{\partial \omega}{\partial t} =\partial_t \omega
   		= - \mbox{\boldmath{$u$}} \cdot \nabla
   		\omega
	\end{equation}
where customary symbols are used for indicating partial derivatives, the vorticity $\omega$ can be expressed, in Cartesian coordinates $(x,y)$, as $\omega=v_x-u_y$, and incompressibility is described by $\nabla \cdot \mbox{\boldmath{$u$}}=0$, where 
$\nabla \cdot\mbox{\boldmath{$U$}} =\partial_x U_x +\partial_y U_y$  is the divergence of the vector field $\mbox{\boldmath{$U$}}$. As a result, we can write  $\mbox{\boldmath{$u$}}=S\mbox{\boldmath{$\nabla$}}\psi=(-\partial_y \psi,\partial_x \psi)$ , where $S$ is the symplectic matrix $[0, -1; 1, 0]$, $\psi$ is the streamfunction, and $\nabla \phi=(\partial_x \phi,\partial_y \phi)$ is the gradient of the function $\phi$. Note that $\omega=\nabla^2 \psi$. In this section,  we consider a compact domain (\textit{e.g.}, a square of side $L$) with  periodic boundary conditions.

The Hamiltonian $\cal H$ is the kinetic energy
	\begin{equation}  \label{H2D}
   		{\cal H }= \frac{1}{2}	\int  
   		\mbox{\boldmath{$u$}}^2 \, \mbox{d} A
   		= -\frac{1}{2}  \int   \omega \psi \, \mbox{d} A,
	\end{equation}
%
and is a functional of velocity.
In general, a functional  $\cal F[\phi]$ maps a function $\phi$ of the phase space into a number. The functional derivative 
$\delta \cal F/\delta \phi$  the change of the functional $\cal F$ 
with respect to a change in the  function $\phi$.  
%
The functional derivative can be defined by considering the first term in the  expansion
	\begin{equation}  \label{functderiv}
   		{\cal F} [\phi + \delta \phi]
		-{\cal F}[\phi] =\delta{\cal F} [\phi]= \int  
		\frac{\delta {\cal F}}{\delta \phi(x)} 
		\delta \phi(x) \mbox{d} x
		+ \dots
	\end{equation}

 \textcolor{black}{The functional derivative $\delta \cal H/\delta \omega$ for
(\ref{H2D}) is explicitly calculated by 
	\begin{equation}  \label{dHdzeta}
   		\delta {\cal H} = \int \nabla \psi \cdot \delta \nabla \psi   
		\, \mbox{d} A  =  \nonumber
		 \int \nabla \cdot (\psi \delta \nabla \psi)   
		\, \mbox{d} A 
		- 
		\int \psi  \delta \omega    
		\, \mbox{d} A 
	\end{equation}
Since the first integral vanishes due to the boundary conditions, and since $\omega=\nabla^2 \psi$, 
we obtain  $\delta \cal H/\delta \omega= -\psi$.}

Equation (\ref{vorteq2D}) says that vorticity is transported across the domain by a non-divergent flow. One can prove easily that any functional of the vorticity is conserved  
	\begin{equation}  \label{Cf}
   		{\cal C} = \int  s(\omega) \, \mbox{d} A.
	\end{equation}
%
%
%
where the integration is performed over the whole domain of the system. The most familiar of such functional is the total enstrophy of the flow:
	\begin{equation}  \label{Enstrophy}
   		{\cal E} = \frac{1}{2} \int   \omega^2 \, \mbox{d}  A
	\end{equation}
The functional derivative of the enstrophy is
simply   $\delta \cal E/\delta \omega = \omega$.  
	
Since $\mbox{\boldmath{$u$}}=S\mbox{\boldmath{$\nabla$}}\psi=(-\partial_y \psi,\partial_x \psi)$, the 2D vorticity equation can be expressed as
	\begin{equation}  \label{Z2Dnew}
   		\frac{\partial \omega}{\partial t}
   		=  -{\cal J} 
   		\left(\psi,\omega\right)= - {\cal J} 
   		\left(
   		\frac{\delta {\cal E }}{\delta \omega},
   		\frac{\delta {\cal H}}{\delta \omega}
   		\right),
	\end{equation}
with the antisymmetric Jacobi operator 
\begin{equation}  \label{JacOp}
     {\cal J}(a,b)=\partial_x a \, \partial_y b -\partial_y a \, \partial_x b=-{\cal J}(b,a).
	\end{equation}
Relating $\psi$ and $\omega$ to the functional derivatives of two conserved quantities amounts to expressing the evolution equation in a Nambu form
using the enstrophy ${\cal E}$.

The time-evolution of an arbitrary functional of vorticity
${\cal F}={\cal F}[\omega]$ 
is determined by 
	\begin{equation} 
   	\frac{d {\cal F}}{d t}   	= -\int   \frac{\delta {\cal F}}{\partial \omega}   	{\cal J} \left(	\frac{\delta {\cal E}}{\delta \omega},   	\frac{\delta {\cal H}}{\delta \omega}   	\right) \mbox{d} A =  \left\{ {\cal F}, {\cal E}, {\cal H} \right \}   \label{Nambra2D}
	\end{equation}
which defines a Nambu bracket for the three functionals
involved.
The bracket is anti-symmetric in all arguments, 
$\{{\cal E}, {\cal H}, {\cal F} \}	=- \{{\cal H}, {\cal E}, {\cal F} \}$, etc.
Using rearrangements  of these functionals and 
partial integration it can be shown that the Nambu bracket is cyclic
	\begin{equation} \label{cycbra2D}
	\{{\cal F}, {\cal E}, {\cal H} \}
	=\{{\cal E}, {\cal H}, {\cal F} \}
	=\{{\cal H}, {\cal F}, {\cal E} \}
	\end{equation} 
The cyclicity of this bracket is 
a main ingredient in Salmon's application
of Nambu mechanics  \citep{Salmon:2005} 
to construct conservative  numerical codes
(see Section \ref{shawat}).

In the following the relationship between  
Nambu mechanics and Hamiltonian theory of two-dimensional flows is briefly summarized. 
%
%
As mentioned above, a Hamiltonian description of the dynamics is obtained when we can write
	\begin{equation}  \label{FFHHam}
   	\frac{d {\cal F}}{d t}
	=\{{\cal F}, {\cal H} \}_P
	\end{equation} 
with an antisymmetric Poisson bracket, to be seen in general as an antisymmetric map 
in the space of functionals, such that $\{{\cal A}, {\cal B} \}_P=-\{{\cal B}, {\cal A} \}_P$. Deriving such a bracket amounts to defining the dynamics of the system.

The Poisson bracket for 2D hydrodynamics
\citep{Salmon:1988,Shepherd:1990}
is easily obtained from the Nambu bracket 
if the dependency $\delta {\cal E}/ \delta  \omega=\omega$
is evaluated 
	\begin{equation} \label{PoiNam}
	\{{\cal F}, {\cal H} \}_P
	=
	\{{\cal F}, {\cal E}, {\cal H} \}
	= 
	\int   \omega
   	{\cal J} 
   	\left(
   	{\cal F}_\omega,
   	{\cal H}_\omega
   	\right)
	\,
	\mbox{d} A
	\end{equation} 
where we indicate ${\cal H}_\omega=\delta H/\delta \omega$; here cyclicity is used, see Eq. \ref{cycbra2D}.

The Poisson bracket used in Eulerian hydrodynamics 
is degenerate because of the presence of an infinite number of so-called \textit{Casimirs}, \textit{i.e.}, the functionals defined in Eq. \ref{Cf}, which are automatically conserved
so that  $\{{\cal C}, H \}_P = 0$. In this case, we talk about noncanonical Hamiltonian mechanics.

The relationship (\ref{PoiNam}) demonstrates 
that noncanonical Hamiltonian mechanics 
is embedded in Nambu mechanics.
The main extension is that in
Nambu mechanics  two functionals acting as an Hamiltonian, 
the enstrophy and the energy, are used
(\ref{Z2Dnew}), and that the Nambu bracket 
(\ref{Nambra2D}) is nondegenerate and
void of Casimir functionals. 

\subsubsection{Three-dimensional incompressible hydrodynamics}

The dynamics of incompressible unforced and inviscid fluid flows in three dimension is determined by the 
vorticity $\mbox{\boldmath{$\omega$}}= \nabla \times \mbox{\boldmath{$u$}}$ evolution equation: 
	\begin{equation}\label{domega}
		\frac{\partial \mbox{\boldmath{$\omega$}}}{\partial t}	
		= \mbox{\boldmath{$\omega$}} \cdot \nabla \mbox{\boldmath{$u$}}
		 - \mbox{\boldmath{$u$}} \cdot \nabla \mbox{\boldmath{$\omega$}}
	\end{equation}
where $\mbox{\boldmath{$u$}}$ is the velocity field and $\nabla \cdot \mbox{\boldmath{$u$}} = 0$. Note that in cartesian coordinates we have that the curl of $\mbox{\boldmath{$U$}}$ ($\nabla \times U$) can be expressed as $(\nabla \times U)_i = \epsilon_{ijk}\partial_jU_k$ where $\epsilon_{ijk}$ is the standard totally antisymmetric Levi-Civita symbol and $\nabla \cdot U=\partial_x U_x+\partial_y U +\partial_z U_z$ is the divergence in three dimensions.
Similarly to the two-dimensional case,  the total energy
	\begin{equation}	\label{H}
		{\mathcal{H}} = \frac{1}{2}	 
		\int  \mbox{\boldmath{$u$}}^2 
		\, \mbox{d}V
		= - \frac{1}{2} \int  
		\mbox{\boldmath{$\omega$}} \cdot \mbox{\boldmath{$A$}}
		\, \mbox{d} V
	\end{equation}
is conserved, where we have introduced $\mbox{\boldmath{$A$}}$ as the vector potential such that 
$\mbox{\boldmath{$u$}}
= -\nabla \times \mbox{\boldmath{$A$}}$. Note that in deriving the second identity we use integration by parts and consider periodic boundary conditions. It is important to note that the total helicity
	\begin{equation} \label{helic}
		h = \frac{1}{2} 
		\int   
		\mbox{\boldmath{$\omega$}} \cdot \mbox{\boldmath{$u$}}
		\,
		\mbox{d} V
	\end{equation}
is also conserved, while \textit{e.g.}, the enstrophy is not.
Following the procedure detailed in Eq. \ref{functderiv}, we derive that the functional derivative of the energy 
with respect to the vorticity is given by
$\delta {\mathcal{H}} / \delta \mbox{\boldmath{$\omega$}} 
= - \mbox{\boldmath{$A$}}$ and
for helicity  
$\delta  h/\delta\mbox{\boldmath{$\omega$}} 
= \mbox{\boldmath{$u$}}$ 
(compare the 2D version (\ref{dHdzeta})).

The Nambu form of the vorticity equation	is
	\begin{equation} \label{xiK}
		\frac{\partial \mbox{\boldmath{$\omega$}}}{\partial t}	
		= K \left(\frac{\delta h}{\delta \mbox{\boldmath{$\omega$}}},
			\frac{\delta \mathcal{H}}{\delta \mbox{\boldmath{$\omega$}}}
			\right)=-K \left(\mbox{\boldmath{$u$}},\mbox{\boldmath{$A$}}\right)
	\end{equation}
with 
	\begin{equation}	  \label{K12}
		K(\mbox{\boldmath{$U_1$}},\mbox{\boldmath{$U_2$}}) 
		=-\nabla \times\left[ (\nabla \times \mbox{\boldmath{$U_1$}})
		\times (\nabla \times \mbox{\boldmath{$U_2$}}) \right]
	\end{equation}
Considering that $\mbox{\boldmath{$\omega$}}=\nabla \times \mbox{\boldmath{$u$}}$ and using some  standard vector calculus identities, we obtain that Eq. \ref{xiK} agrees with Eq. \ref{domega}. We can derive the evolution equations for functional $\mathcal{F}=\mathcal{F}[\mbox{\boldmath{$\omega$}}]$ as follows:
	\begin{align} 
	\frac{d \mathcal{F}}{d t}	
		&=-\int   \left(\nabla \times \frac{\delta \mathcal{F}}{\delta
		\mbox{\boldmath{$\omega$}}} \right) 
		\times
		\left(\nabla \times \frac{\delta h}{\delta
		\mbox{\boldmath{$\omega$}}} \right)  
		\nonumber 
		 \cdot 
		\left(\nabla 
		\times \frac{\delta {\mathcal{H}}}{\delta
		\mbox{\boldmath{$\omega$}}} \right)
		\, \mbox{d} V	\nonumber \\
		&= \{\mathcal{F},h,\mathcal{H}\}     \label{FhH}
	\end{align}
where the  last equation defines the Nambu bracket 
for 3D incompressible hydrodynamics based on the vorticity equation.
Helicity is no longer  a  hidden  conserved  quantity but enters the
dynamics  on  the  same  level  as  the   Hamiltonian.  Therefore, the Nambu mechanics is able to account explicitly for conservation laws of the system, and, correspondingly, to its symmetries.

\subsection{Geophysical fluid dynamics}

%
A Nambu representation can be constructed also for  some of the
 most important mathematical models relevant for geophysical fluid dynamics on large scales:
the quasi-geostrophic potential vorticity equation \citep{Nevir:2009},
the shallow water model \citep{Salmon:2005,Sommer:2009}, 
%
and the baroclinic stratified atmosphere \citep{Nevir:2009}.
%
%
Other models of geophysical relevance can also be treated in this way, as, most notably, the Rayleigh-B\'{e}nard equations for two-dimensional convection, which have been studied in detail in 
\citet{Bihlo:2008} and
\citet{Salazar:2010}. We will not treat this latter case in this review.


\subsubsection{Quasi-geostrophic approximation}\label{qgsection}

Quasi-geostrophic (QG) theory is one of the most important and most studied pieces of geophysical fluid dynamics and is of crucial relevance for studying the large-scale dynamics of the Earth's atmosphere and ocean, and, more recently, of planetary atmospheres \cite{Holton,pedlosky,klein2010}. QG dynamics is relevant when, within a good approximation,  the fluid motions are 1) hydrostatic and 2) the Coriolis acceleration balances the horizontal pressure gradients. This is typically realized, \textit{e.g.}, in  the atmospheric midlatitudes. In absence of dissipative processes and of forcings, QG dynamics is described by the  material conservation
of the QG potential vorticity. We consider customary Cartesian coordinates plus time $(x,y,z,t)$, where $x$ indicates the zonal direction, $y$ the meridional direction, and $z$ the vertical direction as defined by gravity as in \citet{Holton}. The evolution equation reads as follows:
\begin{equation}
\frac{\partial Q}{\partial t}
+ \frac{1}{f_0} \mathcal{J}(\Phi,Q)
=0
\end{equation}
where $\mathcal{J}$ is the Jacobian (\ref{JacOp}).
$Q$ is the QG
approximation of Ertel's potential vorticity 
\begin{equation}
Q = \omega_g + \frac{f_0}{N^2}
\frac{\partial^2 \Phi}{\partial z^2 }
+f
\end{equation}
with the geostrophic vorticity $\omega_g = 1/f_0 \nabla_h^2 \Phi$, 
geopotential $\Phi$, $ \nabla_h^2$ is the Laplacian operator limited to the $x$ and $y$ directions, 
Brunt-V\"{a}is\"{a}l\"{a} frequency   $N$,
and Coriolis parameter $f=f_0+\beta y$. The geostrophic velocity $\mbox{\boldmath{$u_g$}}$ has nonzero components only along the $x$ and $y$ directions, so  that we can write $\mbox{\boldmath{$u_g$}}=(\mbox{\boldmath{$u^h_g$}},0)$, where   $\mbox{\boldmath{$u^h_g$}}=1/f_0S\mbox{\boldmath{$\nabla_h$}}\Phi=1/f_0(-\partial_y \Phi,\partial_x \Phi)$, where $\nabla_h$ is the gradient operator limited to the $x$ and $y$ directions.

The first conserved integral 
is the total energy of the system
\begin{equation}
{\mathcal{H}} = \frac{1}{2}
\int 
\left[
\left(
\frac{\nabla_h \Phi}{f_0}
\right)^2
+ 
\left(
\frac{1}{N}
\frac{\partial \Phi}{\partial z}
\right)^2
\right]
\mbox{d}V
\end{equation}
where the first term is the density of kinetic energy and the second term is the density of potential energy. At each level $z$ the geopotential acts as a stream function in defining the geostrophic velocity field, while the vertical derivative of the geopotential is proportional to the temperature fluctuations of the system \citep{Holton}. The second conserved integral is the potential enstrophy
\begin{equation}\label{PEnstrophy}
\mathcal{E} = \frac{1}{2}
\int  Q^2 \mbox{d}V
\end{equation}
which is defined similarly to the enstrophy in Eq. \ref{Enstrophy}.  One can prove that QG dynamics can be written in a Nambu form as follows:
	\begin{equation}  \label{Q2Dnew}
   		\frac{\partial Q}{\partial t}
   		= - \mathcal{J} 
   		\left(
   		\frac{\delta \mathcal{E }}{\delta Q},
   		\frac{\delta {\mathcal{H}}}{\delta Q}
   		\right)
	\end{equation}
Thus, the mathematical structure   
is analogous to the two-dimensional 
vorticity equation (\ref{Nambra2D}).
Moreover, we can construct the evolution of any functional ${\cal F}[Q]$ 
by defining the Nambu bracket as follows:
\begin{equation}
\frac{d \mathcal{F}}{d t}
=
-\int 
\frac{\delta \mathcal{F}}{\delta Q}
J\left(
\frac{\delta \mathcal{E}}{\delta Q},
\frac{\delta {\mathcal{H}}}{\delta Q}
\right)
\mbox{d} V
=
\{\mathcal{F, E, H}\}
\end{equation}
with 
$\delta \mathcal{E}/\delta Q=Q$
and
$\delta {\mathcal{H}}/\delta Q= -\Phi/f_0$.

\subsubsection{Shallow water model}  \label{shawat}

Roughly speaking, shallow water equations are useful two-dimensional approximations of Navier-Stokes equations often used for describing some fluid motions where the horizontal scale of motion is much larger than its vertical extent,  such as in the case of tidal waves or tsunami in the ocean, or Rossby and Kelvin waves in the atmosphere.
%
%
Here the single layer model is summarized \citep{Sommer:2009}.
The dynamics 
is given by the evolution of the vorticity $\omega$
and the divergence $\mu= \nabla \cdot \mbox{\boldmath{$u$}}$
of the horizontal velocity $\mbox{\boldmath{$u$}}$
\begin{equation}
\partial_t \omega
=-\nabla \cdot (\omega \mbox{\boldmath{$u$}})
\end{equation}
\begin{equation}
\partial_t \mu
= k \cdot \nabla \times (\omega_a \mbox{\boldmath{$u$}})
+ \nabla^2 (\mbox{\boldmath{$u$}}^2/2+gh_T)
\end{equation}
\begin{equation}
\partial_t h_T
=  - \nabla \cdot (h \mbox{\boldmath{$u$}})
\end{equation}
where $h_T$ is  the total height of the fluid. The shallow water model possesses two conserved integrals,
the total energy, given by the sum of kinetic and potential energy
\begin{equation}
{\mathcal{H}}
= \frac{1}{2}
\int 
\rho\left(
h_T
\mbox{\boldmath{$u$}}^2 + g h_T^2
\right)
\mbox{d}A
\end{equation}
and potential enstrophy
\begin{equation}
\mathcal{E}
= \frac{1}{2}
\int 
\rho q^2
\mbox{d}A
\end{equation}
with the absolute potential vorticity
$ q = \omega_a/h_T$, $\omega_a=\omega+f $.
The functional derivatives of the conserved integrals are
$\delta \mathcal{H}/\delta \omega =  -\rho \psi$,
$\delta \mathcal{H}/\delta \mu =  -\rho \gamma$,
$\delta \mathcal{H}/\delta h_T=  \rho \Psi$,
$\delta \mathcal{E}/\delta \omega =  \rho q$,
$\delta \mathcal{E}/\delta \mu = 0$,
$\delta \mathcal{E}/\delta h_T =  -(1/2)\rho q^2$,
where $\rho$ is the density of the fluid, $\psi$ is the streamfunction, $\gamma$ the
velocity potential for
$h \mbox{\boldmath{$v$}}= S \nabla\psi+\nabla \gamma$,
$\Psi=(1/2) \mbox{\boldmath{$u$}}^2 +gh_T$ is the Bernoulli function,

%
%

The Nambu representation of the shallow water model was derived
by \citet{Salmon:2005} and is a bit more cumbersome than in, \textit{e.g.}, QG case.
\citet{Sommer:2009} present a numerical simulation 
of these equations
on a spherical grid, and \citet{Nevir:2009} published the multilayer
shallow water equations.
%
%
In the case of a single layer shallow water equations, the dynamics of any functional $\mathcal{F}$ 
is determined by the sum of three Nambu brackets
\begin{equation}
\frac{d}{d t} \mathcal{F}
=
\{\mathcal{F},{\mathcal{H}},\mathcal{E} \}_{\omega,\omega,\omega}
+
\{\mathcal{F},{\mathcal{H}},\mathcal{E} \}_{\mu,\mu,\omega}
+
\{\mathcal{F},{\mathcal{H}},\mathcal{E} \}_{\omega,\mu,h_T}
\end{equation}
The first bracket is 
  \begin{equation}  \label{FHEzzz}
   \{\mathcal{F},,{\mathcal{H}},\mathcal{E} \}_{\omega,\omega,\omega}
    = \int J(F_\omega,,{\mathcal{H}}_\omega) \mathcal{E}_\omega \mbox{d}A
  \end{equation}
where  $\mathcal{X}_\omega =  \delta \mathcal{X} / \delta \omega$. Such first bracket is 
 analogous 
to the 2D Nambu bracket (\ref{Nambra2D}) (apart from the sign).
For the other brackets we refer to 
\citep{Salmon:2005,Sommer:2009}.
\citet{Salmon:2007} calculated the Nambu brackets
based on the velocities instead of vorticity.
%
\subsubsection{Baroclinic atmosphere}

\citet{Nevir:2009} published the equations determining
the dynamics of a baroclinic dry atmosphere in Nambu form
(denoted as Energy-vorticity theory of ideal fluid mechanics).
The Nambu representation encompasses the 
Eulerian equation of motion in a rotating frame, 
the continuity
equation, and the first law of thermodynamics.
The Nambu dynamics uses three brackets for energy, helicity,
energy-mass, and energy-entropy.
Due to its special role in all three brackets the integral of Ertel's
potential enstrophy 
is coined as a super-Casimir.

The Nambu form shows an elegant
structure where fundamental processes are combined by additive terms.
Incompressible, 
barotropic or baroclinic atmospheres 
are associated to additive contributions.
Thus approximations are simply attained by the neglect
of terms.

In absence of forcings and of dissipative processes, the momentum equation, the continuity equation 
and the first law  of thermodynamics equation are \cite{Peixoto:1992}
\begin{equation} \label{vbcl0}
\partial_t \mbox{\boldmath{$u$}}
=
-\mbox{\boldmath{$u$}} \cdot
\nabla \mbox{\boldmath{$u$}}
-2 \mbox{\boldmath{$\Omega$}} \times
\mbox{\boldmath{$u$}} 
-\frac{1}{\rho} \nabla p - \nabla \Phi
\end{equation}
\begin{equation} \label{rhobcl0}
\partial_t \rho
=
- \nabla \cdot (\rho \mbox{\boldmath{$u$}})
\end{equation}
\begin{equation} \label{sbcl0}
\partial_t s
=
- \mbox{\boldmath{$u$}}  \cdot \nabla s
\end{equation}
where  $\mbox{\boldmath{$u$}}$ is velocity, 
$\mbox{\boldmath{$\Omega$}}$
 the angular  velocity of the earth,
$\Phi$ is the sum of the gravitational and centrifugal
potential of the earth, 
$\rho$ is density and
$s$ is the specific entropy per unit mass, determined by the equation of state of the gas.

These equations possess four conservation laws.
The first is the total energy
\begin{equation}
{\mathcal{H}} = \int 
 \rho e 
\mbox{d}V; \quad e=\frac{1}{2}\mbox{\boldmath{$u$}}^2
+  i + \Phi
\end{equation}
where $e$ is the specific total enery and $i$ is its  internal energy component. 
The absolute helicity is
\begin{equation}
h_a
= \int 
\mbox{\boldmath{$u$}}_a
 \cdot
\mbox{\boldmath{$\omega$}}_a
\mbox{d}V 
\end{equation}
where the absolute velocity is
$\mbox{\boldmath{$u$}}_a=\mbox{\boldmath{$u$}}
+\mbox{\boldmath{$\Omega$}} \times \mbox{\boldmath{$r$}}$
and
$\mbox{\boldmath{$\omega$}}_a = \nabla \times 
\mbox{\boldmath{$v$}}
+2 
\mbox{\boldmath{$\Omega$}}$.
with the angular  velocity of the earth 
$\mbox{\boldmath{$\Omega$}}$, and 
$\mbox{\boldmath{$r$}}$ is the position vector.
%
The total mass and entropy are given by
\begin{equation}
\mathcal{M} = \int  \rho \mbox{d}V, 
\quad
\mathcal{S}= \int  \rho s \mbox{d}V
\end{equation}
and the total 
potential enstrophy is defined starting from Ertel's  potential vorticity $\Pi$ 
\begin{equation}
\mathcal{E}_\rho
= \int  \rho  
\Pi^2 \mbox{d}V, 
\quad
\Pi
= \frac{\mbox{\boldmath{$\omega$}}_a \cdot \nabla \theta}{\rho},
\end{equation}
analogously to the definition in the QG context given in Eq. \ref{PEnstrophy}.
The functional derivatives of the conservation laws
are
$\delta \mathcal{H}/\delta \mbox{\boldmath{$u$}} 
= \rho \mbox{\boldmath{$u$}} $,
$\delta \mathcal{H}/\delta \rho = (1/2) \mbox{\boldmath{$u$}}^2 + i+p/\rho - Ts + \phi$,
$\delta \mathcal{H}/\delta \sigma = T$,
$\delta \mathcal{M}/\delta \mbox{\boldmath{$u$}}= 0$,
$\delta \mathcal{M}/\delta \rho = 1$,
$\delta \mathcal{M}/\delta \sigma = 0$,
$\delta \mathcal{S}/\delta  \mbox{\boldmath{$u$}}= 0$,
$\delta \mathcal{S}/\delta \rho = s$,
$\delta \mathcal{S}/\delta \sigma = 1$,
$\delta h_a/\delta \mbox{\boldmath{$u$}} = \mbox{\boldmath{$\omega$}}_a$,
$\delta h_a/\delta \rho = 0$,
and
$\delta h_a/\delta \sigma = 0$,
where $T$ is temperature, and 
$\sigma = \rho s$. 

An arbitrary functional $\mathcal{F}$
of $\mbox{\boldmath{$u$}}, \rho$ and $\sigma$
evolves according to the sum of three brackets which are defined below
\begin{equation} \label{bracks}
\frac{d}{d t} \mathcal{F}
=
\{\mathcal{F},h_a,\mathcal{H}\}_h
+
\{\mathcal{F},M,\mathcal{H}\}_m
+
\{\mathcal{F},S,\mathcal{H}\}_s
\end{equation}
%
%

%
The three brackets are defined below. The first one is the so-called helicity bracket:
\begin{equation}  \label{FhHbra}
\{
\mathcal{F}, h_a, {\mathcal{H}}
\}_h
=
-\int 
\left[
\frac{1}{\rho}
\frac{\delta \mathcal{F}}{\delta  \mbox{\boldmath{$u$}}}
\cdot
\left(
\frac{\delta h_a}{\delta  \mbox{\boldmath{$u$}}}
\times 
\frac{\delta \cal{H}}{\delta  \mbox{\boldmath{$u$}}}
\right)
\right]
\mbox{d}V;
\end{equation}
the second is the so-called mass bracket: \begin{align}
\{
\mathcal{F}, & \mathcal{M}, {\mathcal{H}}
\}_m
=
-\int 
\left[
\frac{\delta \mathcal{M}}{\delta \rho}
\frac{\delta \mathcal{F}}{\delta  \mbox{\boldmath{$u$}}}
\cdot
\nabla
\frac{\delta {\mathcal{H}}}{\delta \rho}
\right.  \nonumber \\
& + \left.
\frac{\delta \mathcal{F}}{\delta \rho}
\nabla
\cdot
\left(
\frac{\delta \mathcal{M}}{\delta \rho}
\frac{\delta {\mathcal{H}}}{\delta  \mbox{\boldmath{$u$}}}
\right)
\right]
\mbox{d}V 
 +cyc(\mathcal{F},\mathcal{M},{\mathcal{H}});
\end{align}
where $cyc$ indicates permutations in cyclic order of the arguments. The third one is the so-called entropy bracket:
\begin{align}  \label{FSHsbar}
\{
\mathcal{F}, & \mathcal{S}, {\mathcal{H}}
\}_s
=
-\int 
\left[
\frac{\delta \mathcal{S}}{\delta \rho}
\frac{\delta \mathcal{F}}{\delta  \mbox{\boldmath{$u$}}}
\cdot
\nabla
\frac{\delta {\mathcal{H}}}{\delta \sigma}
\right.  \nonumber \\
& + \left.
\frac{\delta \mathcal{F}}{\delta \sigma}
\nabla
\cdot
\left(
\frac{\delta \mathcal{S}}{\delta \rho}
\frac{\delta {\mathcal{H}}}{\delta  \mbox{\boldmath{$u$}}}
\right)
\right]
\mbox{d}V 
+cyc(\mathcal{F},\mathcal{M},{\mathcal{H}})
\end{align}
For a barotropic flow the first law of 
thermodynamics  is physically not relevant
and the entropy bracket
is discarded in  (\ref{bracks}) because the functional derivatives with respect to $\sigma$ vanish.
The continuity equation remains unapproximated 
and the pressure gradient term 
is replaced by the gradient of enthalpy.
Note the different brackets for helicity
(\ref{FhHbra}) and vorticity (\ref{FhH})
in 3D hydrodynamics.

\subsection{Conservative algorithms and numerical models}

\citet{Salmon:2005,Salmon:2007} recognized that the 
existence of  a Nambu bracket with two conserved integrals
allows the design of  high-precision numerical algorithms for studying geophysical flows. The idea is in fact simple: just like in the usual case we aim at writing numerical codes able to conserve energy when dissipation and forcing are neglected, Nambu mechanism provides encouragement and conceptual support for expanding this point of view by encompassing other important physical quantities. 
%
%
The approach is useful in GFD 
turbulence simulations because these flows are
characterized by the existence of conservation laws
besides total energy.
In particular,  the conservation of enstrophy 
inhibits spurious accumulation
of energy at small scales.

For the numerical design of conservative codes
based on a Nambu structure
the following remarks are noted:
\begin{itemize}
\item
A Nambu form of the continuous
physical system is required.
\item
The quantities used in the Nambu bracket
are conserved.
\item
The discrete form of the Jacobian
needs to preserve antisymmetry  (\ref{FFHHam}).
\item
The approach is applicable to 
any kind of discretization, 
e.g. for finite differences, finite volumes,
or spectral models.
\item
Arbitrary approximations
of the conservation laws are possible;
these approximations are conserved exactly.
\item For the barotropic vorticity equation
the classic Arakawa Jacobian could be retrieved by
equally weighting the cyclic permutations of the Nambu bracket. In other terms, Arakawa found heuristically a discrete Nambu representation of barotropic dynamics \citep{dubi2007}.

\end{itemize}
In recent years, various authors have provided promising examples of  actual implementations of GFD codes which take into explicit consideration the underlying Nambu dynamics of the unforced and inviscid case.  \citet{Salmon:2007} presents the first 
numerical simulation of a shallow water model derived from the Nambu brackets formalism.
The simulation is on a square rectangular grid
and the design on an unstructured triangular mesh is outlined.

\begin{figure}[t]
\includegraphics[scale=0.18,angle=0]{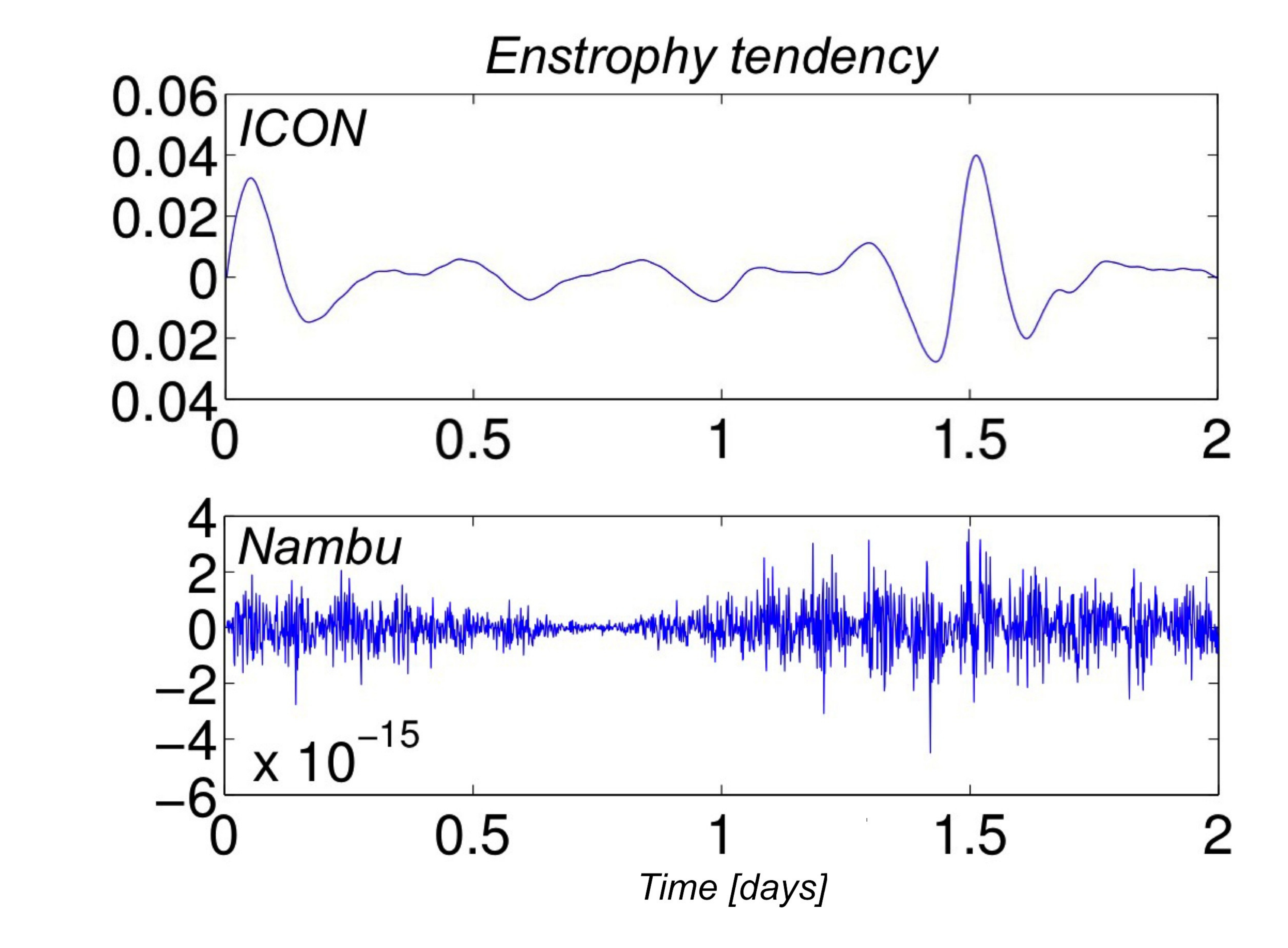} 
\caption{\label{fig-EE}
Enstrophy tendencies in the enstrophy conserving ICON shallow water model and the Nambu model of \citet{Sommer:2009} (courtesy of Matthias Sommer, Ludwig-Maximilians-Universit\"{a}t M\"{u}nchen). Note that the tendency in the Nambu model is of the order of the numerical accuracy.}
\end{figure}

\citet{Sommer:2009} report the first
simulation of a shallow water atmosphere
using Nambu brackets.
The authors use an isosahedric grid 
(as in the ICON model, ICOsahedric Non-hydrostatic model,
of the German Weather Service and the 
Max Planck Institute for Meteorology, Hamburg).
The construction of the algorithm is as follows
\citep{Sommer:2009}:
\begin{enumerate}
\item First the continuous versions of the Nambu-brackets and
conservation laws need to be obtained.
\item On the grid, the following expressions need to be
calculated:
functional derivatives, 
discrete operators (div and curl),
discretization of the Jacobian and the Nambu brackets.
\item Finally, the prognostic equations are obtained  
by inserting the variables in the brackets.
Various options are available for the time stepping is arbitrary; \citet{Sommer:2009}
use a leap-frog with Robert-Asselin filter.
\end{enumerate}
The authors find quasi-constant enstrophy 
and energy compared to
a standard numerical design
(Fig. \ref{fig-EE}). 

Along these lines, \citet{Gassmann:2008} suggest a radically new concept
for a global numerical simulation 
of the non-hydrostatic atmosphere
using the Nambu representation for
the energy-helicity bracket
$\{ \mathcal{F},h_a,{\mathcal{H}} \} $
 \citep{Nevir:1998}. 
Their suggestion incorporates  a careful 
description of Reynolds averaged 
subscale processes and budgets.
\citet{Gassmann:2013} describes a global non-hydrostatic 
dynamical core based on an icosahedral nonhydrostatic model
on a hexagonal C-grid. The model conserves mass and energy
in a noncanonical Hamiltonian framework,even if some 
still unsolved 
numerical problems  occur when 
the non-hydrostatic compressible equations 
are in a Nambu bracket form. The use of dynamical cores constructed according to the sophisticated version of fluid dynamics  discussed here might provide crucial for improving the ability of atmospheric models in representing correctly the global budgets of physically relevant quantities also in the case when forcing and dissipative processes are taken into account. As discussed by 
\citet{LucariniRagone} for the case of energy, this is far from being a trivial task.




\subsection{Perspectives}

Like Hamiltonian mechanics, the Nambu approach is
a versatile tool for the analysis and simulation of
dynamical systems. 
Here some possible research directions are outlined.

{\bf Modular modeling and approximations: }   
In several applications a Nambu representation 
can be found by adding brackets which 
conserve a particular Casimir,
this is already mentioned by \citet{Nambu:1973}; see the baroclinic atmosphere \citep{Nevir:2009}
and the classification by \citet{Salazar:2010}.
The dynamics is determined by these 'constitutive' 
Casimirs (a notion coined in \citep{Nevir:2009}) 
which are not conserved in the complete system.
Decomposition leads to subsystems where the
constitutive Casimirs are conserved.
An example is helicity which is constitutive 
in the baroclinic atmosphere and only conserved 
in the 3D incompressible flow.
The decomposition is directly associated 
with approximations \citep{Nevir:2009}.
Composition allows a process-oriented model design.




{\bf Statistical Mechanics:}
The statistical mechanics of fluids is characterized by
the existence of conservation laws besides total energy
 \citep{Bouchet2012}, see also section \ref{onsager} in this review. 
Thus these conservation laws have a two-fold impact:
They determine the dynamics in a Nambu bracket
and the canonical probability distribution
in equilibrium.

{\bf Dynamics of Casimirs:}
Casimir-functions of a conservative system
are ideal observables to characterize
the dynamics in the presence of forcing and dissipation. This might prove especially interesting when studying the response of a system to perturbations in the context of the Response theory proposed by \cite{ruelle_differentiation_1997,Ruelle:1998,ruelle_nonequilibrium_1998,ruelle_review_2009} and recently used in a geophysical context by various authors with promising results \citep{eyink2004,majda07,lucarini09,Lucarini:2011_NPG}; see also section \ref{ruelle} in this review.

As illuminating example, we mention the recent work of  \citet{Pelino:2007} and  \citet{Gianfelice:2012}, who have used 
recurrence maps of extremes of energy and
a Casimir in
a Lorenz-like map to assess predictability of the system and study
the properties of the invariant measure. 

\section{Equilibrium Statistical Mechanics for geophysical flows}
\label{onsager}
We have seen in the previous section that different models of geophysical flows have a specific mathematical structure: they are Hamiltonian systems, and have an infinite number of conserved quantities - the \textit{Casimirs}. The previous section has shown how one could take advantage of these features and construct theoretically rich representation of the dynamics and provide proposals for constructing new numerical codes of GFD flows.  This section goes in the direction of constructing a probabilistic description of GFD flows, basically taking the point of view that  due to the large amount of degrees of freedom involved, one can consider the state of the atmosphere and the ocean as random variables. Here we shall review the progress that has been made by using the simplest class of possible probability distributions: the equilibrium distributions which depend only on the conserved quantities.
 However, most of the standard applications of equilibrium statistical mechanics deal with dynamics on a finite dimensional phase space (e.g., a gas with a finite number of molecules), with a finite number of dynamical invariants (often just the energy). The equations describing the dynamics of geophysical flows violate both these constraints. 
Several solutions have thus been proposed: they are reviewed briefly in the next sections, going from the main fundamental ideas to selected geophysical applications. 

\subsection{Finite-dimensional models: Point vortices}\label{pointvorticessection}
\vspace{10pt}

\subsubsection{Negative temperature states and clustering of vortices}

\begin{figure}[tbhp]
\centering
\includegraphics[width=0.95\linewidth]{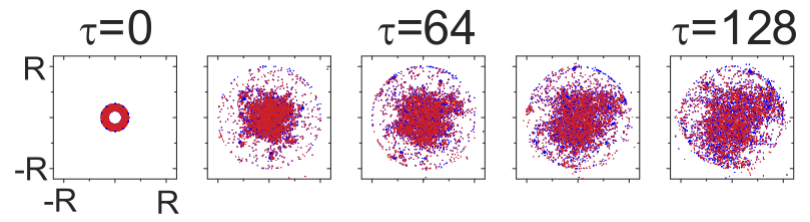}

\includegraphics[width=0.95\linewidth]{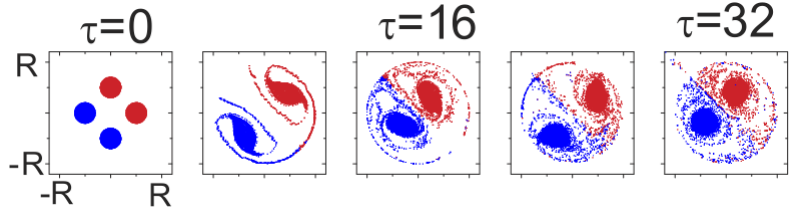}
\caption{Time evolution for a numerical simulation of two-sign point vortices (shown in red and blue), for positive temperatures (upper panel) and negative temperatures (lower panel)~\citep{Yatsuyanagi2005}. For negative temperatures, we observe the clustering of same-sign vortices, while for positive temperatures, positive and negative vortices are distributed homogeneously in the domain.}\label{pointvortfig}
\end{figure}

\citet{Onsager1949} was the first to understand that the coherent structures and persistent circulations that appear ubiquitously in planetary atmospheres and in the Earth's oceans could be explained on statistical grounds. His work focused on 2D incompressible, inviscid fluids given in equation \ref{vorteq2D}. 
To make the system tractable, he introduced an approximation of the vorticity field in terms of $N$ point vortices with \emph{circulation} $\gamma_i$ and position $\vec{r}_i(t)$: $\omega(\vec{r},t)=\sum_{i=1}^N \gamma_i \delta(\vec{r}_i(t)-\vec{r})$, where $\delta(x)$ is the usual Dirac's delta distribution. Introducing the Hamiltonian $\mathcal{H}=- \sum_{i<j} \gamma_i \gamma_j \lapgreen(\vec{r}_i,\vec{r}_j)$, where $\lapgreen$ is the Green function of the Laplacian (the response to an impulse source: $\Delta G(\vec{r}_i,\vec{r}_j)=\delta(\vec{r}_i-\vec{r}_j)$), the dynamics reads simply
\begin{equation}
\gamma_i \frac{d x_i}{dt} = \frac{\partial \mathcal{H}}{\partial y_i}, \quad \gamma_i \frac{d y_i}{dt} = - \frac{\partial \mathcal{H}}{\partial x_i}.
\end{equation}
This is a canonical Hamiltonian system with a finite number of degrees of freedom, for which the standard methods of statistical mechanics apply directly. In particular, the \emph{microcanonical} probability measure, acting as invariant - \textit{e.g.} unaltered by the dynamics - measure of the system, assigning a uniform probability to all the configurations with the same energy, is given by
\begin{equation}
\rho\{\vec{r}_i\}_{1 \leq i \leq N}) = \frac{\delta(\mathcal{H}(\{\vec{r}_i\}_{1 \leq i \leq N})-E)}{\Omega(E)},
\end{equation}
where $\Omega(E)$ is the \emph{structure function}, which measures the volume in phase space occupied by configurations with energy $E$. It is easily proved that, for a bounded domain, and hence a finite volume phase space, this function reaches a maximum for a given value of the energy. Hence, the thermodynamic entropy $S(E)=k_B \ln \Omega(E)$ decreases for a range of energies, and the statistical temperature $1/T=\partial S/\partial E$ becomes negative. Negative temperatures, although counter-intuitive, have since been commonly encountered in the study of other systems with long-range interactions~\citep{DauxoisLRIBook}, and correspond to self-organized states. Here, the energy increases when two same-sign vortices move closer, while it decreases for opposite signs. When the temperature is negative, configurations with maximum energy are favored. Hence, negative temperature equilibrium states exhibiting clusters of same-sign vortices are expected. This behavior has been confirmed by numerical simulations with up to $N=6724$ point vortices: see Fig. \ref{pointvortfig}.

\subsubsection{Mean-field equation}

The above argument is qualitative; to characterize the coherent structures which are expected to emerge from the clustering of same-sign vortices, we introduce the 
probability density $\rho_i(\vec{r},t)$ for a vortex with strength $\gamma_i$ to be found at point $\vec{r}$ at time $t$.
It satisfies the normalization $\int \rho_i(\vec{r},t)d\vec{r}=1$. We define a \emph{coarse grained} vorticity field $\overline{\omega}(\vec{r},t)=\sum_i \gamma_i \rho_i(\vec{r},t)$.
This probability density is expected to converge towards its statistical equilibrium: the equilibrium distribution maximizes the statistical entropy $\mathcal{S}=-\sum_i \int \rho_i(\vec{r})\ln \rho_i(\vec{r})d\vec{r}$. The solution of this variational problem is given by $\rho_i(\vec{r})=e^{\beta (\gamma_i \overline{\psi}(\vec{r})+\mu_i)}/\partfun$
where $\beta$ and $\beta \mu_i$ are the Lagrange parameters associated with conservation of global energy and normalization of each $\rho_i$, respectively, and $\overline{\psi}=\Delta^{-1}\overline{\omega}$ is the coarse-grained stream function, while the normalization factor $\partfun$ is called the \emph{partition function}. Averaging over this equilibrium distribution gives the coarse-grained vorticity field, which satisfies the \emph{mean-field equation}:
\begin{equation}
\overline{\omega}(\vec{r}) = \frac 1 \partfun \sum_i \gamma_i e^{\beta (\gamma_i \overline{\psi}(\vec{r})+\mu_i)}.
\end{equation}
This is an equation of the form $\omega=F(\psi)$, characteristic of the steady-states of the 2D Euler equations. A well-known particular case is that of $N$ vortices with circulation $1/N$ and $N$ vortices with circulation $-1/N$. In that case, the mean-field equation can be recast as $\omega=A\sinh(\beta\Psi)$, with $\Psi=\psi-(\mu_+-\mu_-)/2$~\citep{Montgomery1974}.


The theory can be generalized in a straightforward manner to quasi-geostrophic (QG) flows \citep{Miyazaki2011}. \citet{DiBattista2001a} have given solutions of the mean-field equation for a two-layer model - \textit{i.e.}, a QG model where the stream function is defined only at two discrete value of the vertical coordinate and the temperature is defined at the interface between such level \citep{Holton}  -  where the point vortices stand for \emph{hetons}, introduced by \citet{Hogg1985} as a model of individual convective towers in the ocean. They have shown that a background barotropic current (the \emph{barotropic governor}) confines potential vorticity and temperature anomalies, thereby suppressing the baroclinic instability, in agreement with numerical simulations~\citep{Legg1993}.

The point vortex model suffers from a number of limitations inherent to the approach. First of all, when we let the number of vortices tend to infinity (the \emph{thermodynamic limit}), we have to introduce an \emph{ad-hoc} scaling of the Lagrange parameters to retain the organized, negative temperature states. Besides, there is no unique way to approximate a vortex patch by a finite number of vortices. A consequence is also that the area of vorticity patches cannot be conserved in this singular formulation. We shall see in section \ref{mrstheory} that dealing directly with the vorticity field will solve these issues, while predicting a relation between vorticity and stream function very similar to the one obtained above.

\subsection{Finite-dimensional models: truncated Fourier modes}\label{galerkinsection}
\vspace{10pt}
\subsubsection{2D Turbulence}\label{kraichnan2dsection}

Rather than a discretization in physical space, one may consider a finite number of modes in Fourier space, as proposed by \citet{Lee1952} and \citet{Kraichnan1967} in the context of the Euler equations. For 2D flows --- for simplicity, we consider here a rectangular geometry with periodic boundary conditions; the case of a spherical geometry can be found in \citet{Frederiksen1980} --- writing the vorticity field as a truncated Fourier series $\omega(\vec{x})=\sum_{\vec{k}} \fourier{\omega}(\vec{k}) e^{i \vec{k} \cdot \vec{x}}$, the evolution in time of the Fourier coefficients follows an equation of the form $\partial_t \fourier{\omega}(\vec{k})=\sum_{\vec{p},\vec{q}} A_{\vec{k}\vec{p}\vec{q}} \fourier{\omega}(\vec{p})\fourier{\omega}(\vec{q})$, where the summation is restricted to a finite set of wave vectors $\mathcal{B}= \{ \vec{k} \in 2\pi/L \mathbb{Z}^3, k_{\text{min}} \leq k \leq k_{\text{max}} \}$ and $A_{\vec{k}\vec{p}\vec{q}} $ takes care of the quadratic nonlinearity terms. This dynamics preserves two quadratic quantities: the energy $E=\sum_{\vec{k}} \abs{\fourier{\omega}(\vec{k})}^2/(2k^2)$ and the enstrophy $\Gamma_2=\sum_{\vec{k}} \abs{\fourier{\omega}(\vec{k})}^2$. \citet{Kraichnan1967} suggested to consider the canonical probability distribution:
\begin{equation}
\rho(\{\fourier{\omega}(\vec{k})\}_{\vec{k} \in \mathcal{B}}) = \frac{e^{-\tilde{\beta} E-\alpha \Gamma_2}}{\partfun},
\end{equation}
In particular, the average energy at \emph{absolute equilibrium} is given by
\begin{equation}
\ensavg{E} = -\frac{\partial \ln \partfun}{\partial \tilde{\beta}} = \frac 1 2 \sum_{\vec{k} \in \mathcal{B}} \frac 1 {\tilde{\beta}+2\alpha k^2},
\end{equation}
which corresponds to an equipartition spectrum for the general invariant $\tilde{\beta} E + \alpha \Gamma_2$: $E(k)=\pi k/(\tilde{\beta}+2\alpha k^2)$. Inviscid numerical runs indeed relax to this spectrum~\citep{Fox1973,Basdevant1975}.
Note that the Lagrange parameters $\alpha$ and $\tilde{\beta}$ cannot take arbitrary values; they are constrained by the \emph{realizability} condition --- for the Gaussian integral defining $\partfun$ to converge. Here, this condition reads: $\tilde{\beta} + 2 \alpha k_{\text{min}}^2>0$ and $\tilde{\beta} + 2\alpha k_{\text{max}}^2>0$. In particular, when $\alpha>0$, negative temperatures can be attained. In this regime, which corresponds to $\ensavg{\Gamma_2}/(2\ensavg{E})$ small enough~\citep{Kraichnan1980}, the energy spectrum is a decreasing function of $k$. When $\tilde{\beta} \to -2\alpha k_{\text{min}}^2$, a singularity appears at $k \to k_{\text{min}}$, which means that the energy is expected to concentrate in the largest scales. Hence, statistical mechanics for the truncated system predicts that when the enstrophy is small enough compared to the energy, we expect the energy to be transferred to the large scales. 
\citet{Kraichnan1967} gives other arguments to support and refine this view; in particular he shows the existence of two inertial ranges, with a constant flux of energy and enstrophy, respectively, with the energy spectrum scaling as $E(k)\sim C \varepsilon^{2/3}k^{-5/3}$ and $E(k)\sim C' \eta^{2/3}k^{-3}$ respectively, where $\varepsilon$ and $\eta$ are the energy and enstrophy fluxes. In particular, the equilibrium energy spectrum at large scales is shallower than the energy inertial range spectrum. Assuming a tendency for the system to relax to equilibrium --- although the equilibrium is never attained in the presence of forcing and dissipation --- we thus expect the flux of energy to be towards the large scales; a process referred to as the \emph{inverse cascade} of 2D turbulence. 
Similarly, the transfer of enstrophy in the corresponding inertial range should be towards the small scales. 
The dual cascade scenario has been confirmed both by numerical simulations~\citep{Boffetta2007} and laboratory experiments~\citep{Paret1997}.

\subsubsection{Quasi-Geostrophic Turbulence}\label{kraichnanstratsection}

The dynamical equations of QG flow are very similar to the Euler equations, replacing vorticity by potential vorticity (see section \ref{qgsection}). In particular, they conserve similar quadratic invariants, and the theory can be extended in a straightforward manner \citep{Holloway1986,SalmonBook}. 
We will discuss in this section the effect of stratification and $\beta$ effect.

\begin{figure}[tbhp]
\centering
\includegraphics[width=1\linewidth]{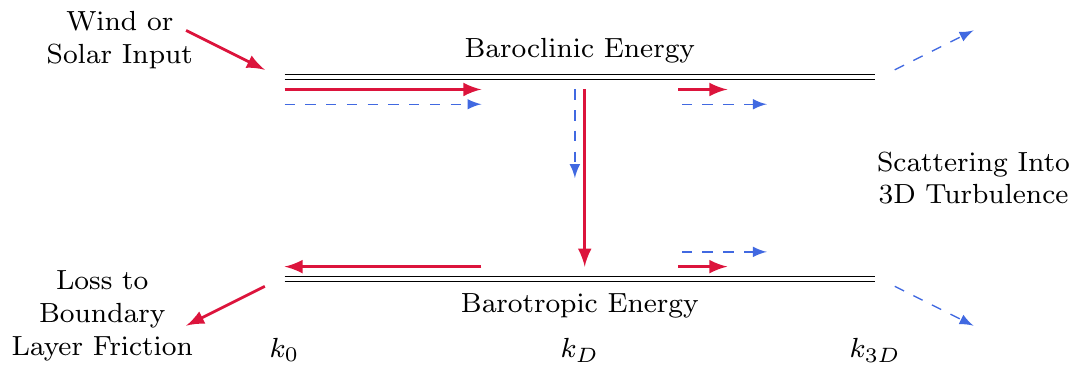}
\caption{Energy (solid arrows) and potential enstrophy (dashed arrows) flux diagram for two-layer quasi-geostrophic turbulence, taking inspiration from ~\citet{Salmon1978}. The energy injected in the baroclinic mode at large scales is cascaded downscale until the deformation scale is reached, then it is transferred to the barotropic mode and cascaded upscale like in 2D turbulence, in agreement with the predictions of equilibrium statistical mechanics.}\label{salmonfig}
\end{figure}

Perhaps the simplest framework to consider the role of stratification is the two-layer QG case. As in section \ref{kraichnan2dsection}, a canonical probability distribution can be constructed, taking into account the three invariants: the total energy $E$ and the potential enstrophies of each layer, $Z_1$ and $Z_2$. The corresponding partition function can be computed, and the spectrum studied in the various regimes, with similar results. In particular, negative temperature states are accessible, which correspond to condensation of the energy on the largest horizontal scales and the \citet{Fofonoff1954} solutions mentioned below. Maybe more interestingly, although the various forms of energy (kinetic energy $K_1,K_2$ in each layer and potential energy $P$) are not individually conserved, we can compute their average value at equilibrium, as \citet{Salmon1976} did. Alternatively, the standard decomposition in terms of the barotropic and baroclinic modes (constructed by taking the average and the difference of the streamfunctions in the two layers), with their kinetic energies $K_T$ and $K_B$, can be used. 
As \citet{Salmon1976} highlighted, the Rossby deformation scale $k_D=2\pi/R_D$ plays an important role. $R_D=NH/f_0$, where $H$ is the vertical extent of the domain and $f_0$ is the reference Coriolis parameter, defines the typical horizontal scale of perturbations of vertical extent equal to $H$, with $N/f_0\gg1$ defining the typical geometric aspect ratio.
At scales smaller than the deformation scale ($k \gg k_D$), the two layers behave essentially as two independent copies of 2D turbulence; the energy spectrum in each layer is the same as in the 2D case, the correlation at statistical equilibrium is low, and there is about as much energy in the barotropic mode and the baroclinic mode: $\ensavg{K_T(k)}/\ensavg{K_B(k)} \sim 1$. Besides, the potential energy is small compared to the kinetic energy: $\ensavg{P(k)}/\ensavg{K_T(k)}=O(k_D/k)$. 
At scales larger than the deformation radius ($k \ll k_D$), the system rather behaves as a unique barotropic layer: the amount of energy in the two layers is about the same, but the energy is essentially in the barotropic mode, with negligible energy in the baroclinic mode, and a statistical correlation between the two layers of order 1.
This theoretical analysis goes in strong support of the standard picture of two-layer QG turbulence, developed on phenomenological grounds~(\citet{Salmon1978,Rhines1979}, see also \citet[chap. 9]{vallis_atmospheric_2006} and Fig. \ref{salmonfig}), and is in agreement with numerical simulations~\citep{Rhines1976}.
These results have been extended to an arbitrary number of layers and to continuously stratified flows by \citet{Merryfield1998}. Although the equilibrium mean, vertically integrated stream function remains similar to the two-layer case, the distribution of the statistics on the vertical differs as higher-order moments are considered. The ratio of potential to kinetic energy for instance, can become significantly underestimated, especially in the limit of strong stratification ($k_D \to 0$) where the two-layer model does not capture well the possibility that an important fraction of the energy may be trapped near the bottom.

The second dominant effect in geophysical flows, in addition to stratification, is rotation. 
The Coriolis force introduces a linear term in the equations, which does not affect directly the previous analysis of the nonlinear energy transfers: the conserved quantities remain the same and the statistical theory is easily extended by replacing relative vorticity with absolute vorticity. However, the variation of the Coriolis force with latitude is responsible for the appearance of Rossby waves, which modify the physical interpretation of the predicted cascade of energy. As anticipated by \citet{Rhines1975} and verified numerically \citep[e.g.,][]{Vallis1993}, the Rossby waves deflect the inverse energy cascade: they dominate over nonlinear effects in a part of Fourier space and prevent access to low wavenumbers along one direction in Fourier space. This leads to the preferential formation of zonal flows.

\subsubsection{Beyond balanced motion}\label{kraichnanunbalancedsection}

Although the large-scale motions of the atmosphere and oceans of the Earth are very close to geostrophic and hydrostatic balance, these relations break up when moving down to the mesoscale, and the transfers of energy due to turbulence, or the non-linear interaction of inertia-gravity waves, might not follow the inverse cascade scenario described in sections \ref{kraichnan2dsection}-\ref{kraichnanstratsection}. As a matter of fact, a downscale transfer of energy is needed in the ocean to feed enhanced vertical mixing~\citep[e.g.,][]{Ledwell2000} or small-scale dissipation in the ocean interior~\citep{Nikurashin2013}. Such processes are necessary to close the energy budget of the ocean~\citep{Wunsch2004}.
It is therefore natural to ask how equilibrium statistical mechanics can help understanding how energy is exchanged by nonlinear interactions between the slow, balanced motions and the fast, wave motions.

\citet{Errico1984} first observed a tendency for unforced inviscid flows described by hydrostatic primitive equations to reach an energy equipartition state, in which the energy in the fast wave modes is comparable to that in the slow balanced modes.
The study by \citet{Warn1986}, in the context of the shallow water equations, essentially confirms that QG flows are not equilibrium states, and that a substantial part of the energy may end up in the fast (surface) wave modes at statistical equilibrium, implying a direct cascade of energy to the small scales. 
\citet{Bartello1995} has obtained analytically the equilibrium energy spectrum for the Boussinesq equations (neglecting the nonlinear part of potential vorticity), in the presence of rotation
, confirming the direct cascade of energy. In particular, there is no negative temperature states in this case, due to the presence of the inertia-gravity waves. In fact, numerical simulations~\citep{Pouquet2013} indicate that turbulence with rotation and stratification might have at the same time an inverse and a direct cascade of energy. A natural interpretation would be that vortical modes are responsible for the inverse cascade while waves cascade energy downscale simultaneously. \citet{Bartello1995} had discussed the possibility of a wave-vortical mode decoupling on the basis of resonant triadic interactions. Without any assumptions on the dynamics, another interpretation in the statistical mechanics framework uses an analogy with metastable states: restricting the equilibrium probability distribution to the slow manifold yields an inverse cascade, while taking into account the whole phase space including the waves results in a direct cascade~\citep{Herbert2014c}.

\subsection{The mean-field theory for the continuous vorticity field}\label{mrstheory}
\vspace{10pt}


\subsubsection{Mean-field theory}\label{meanfieldtheorysection}

Above, we have considered finite-dimensional models conserving at most two quadratic quantities, the energy and the enstrophy. In fact, the majority of the flows considered above --- and in particular 2D and QG flows --- conserve an infinite family of invariants, called Casimir invariants: for any function $s$, $\int s(\omega) d\vec{r}$ is conserved (see Eq. \ref{Cf}). The specific case $s_n(x)=x^n$ corresponds to the moments of the vorticity distribution. 
Instead, the conservation of $s_\sigma(x)=\delta(x-\sigma)$ implies that the area $\gamma(\sigma)$ where the vorticity takes value $\sigma$ is conserved. This is due to the absence of a vortex stretching term, in contrast with full 3D flows; here the vorticity (or potential vorticity, in the QG case) patches are stirred in such a way that their area remains conserved. The theory developed by \citet{Miller1990} and \citet{Robert1991a} (see also \citet{Bouchet2012} for a review) introduces a coarse-grained vorticity field $\overline{\omega}$, which corresponds to the macroscopic state of the flow. This coarse-grained vorticity field can be predicted based on the invariants using statistical mechanics. 
To do so, we introduce $\rho(\sigma,\vec{r})$, the probability density for the vorticity field to take value $\sigma$ at point $\vec{r}$. The coarse-grained vorticity field is given by $\overline{\omega}(\vec{r})= \int_{-\infty}^{\infty} \sigma \rho(\sigma,\vec{r})d\sigma$. The invariants of the system are the energy
\begin{align}
\meanfield{E}[\rho] &=\int_{\domain^2} d\vec{r} d\vec{r'} \int_{\mathbb{R}^2} d\sigma d\sigma' \sigma\sigma' \lapgreen(\vec{r},\vec{r'})\rho(\sigma,\vec{r})\rho(\sigma',\vec{r'}),\label{mfenergyeq}
\shortintertext{with $G$ the Green function of the Laplacian, and the Casimir invariants}
\meanfield{G}_n[\rho]&= \int_\domain d\vec{r} \int_\mathbb{R} d\sigma \sigma^n \rho(\sigma,\vec{r}),
\shortintertext{or equivalently, the vorticity levels}
\meanfield{D}_\sigma[\rho] &= \int_\domain d\vec{r} \rho(\sigma,\vec{r}).
\end{align}
The idea of the theory is to select the probability distribution $\rho$ which maximizes a \emph{mixing entropy} $\meanfield{S}[\rho]=-\int_\domain \int_\mathbb{R} d\vec{r}d\sigma \rho(\sigma,\vec{r}) \ln \rho(\sigma,\vec{r})$, under the constraints of conservation of the invariants, and pointwise normalization $\meanfield{N}[\rho](\vec{r})=\int_\mathbb{R}d\sigma \rho(\sigma,\vec{r})=1$. Hence, we are interested in the variational problem:
\begin{align}
&\max_{\rho, \meanfield{N}[\rho](\vec{r})=1} \{ \meanfield{S}[\rho] \suchthat \meanfield{E}[\rho] = \thermo{E}, \forall n \in \mathbb{N}, \meanfield{G}_n[\rho]=\thermo{\Gamma}_n \},
\shortintertext{or equivalently,}
&\max_{\rho, \meanfield{N}[\rho](\vec{r})=1} \{ \meanfield{S}[\rho] \suchthat \meanfield{E}[\rho] = \thermo{E},  \forall \sigma \in \mathbb{R}, \meanfield{D}_\sigma[\rho]=\thermo{\gamma}(\sigma) \}. \label{rsmvpeq}
\end{align}
The solutions of this variational problem correspond to the most probable states for a given set of conserved quantities.

The critical points of the variational problem \eqref{rsmvpeq} are simply given by $\delta \meanfield{S} - \int d\vec{r} \zeta(\vec{r}) \delta \meanfield{N}(\vec{r}) - \tilde\tilde\beta \delta \meanfield{E} - \int d\sigma \alpha(\sigma) \delta \meanfield{D}_\sigma=0$, where $\tilde\beta$ and $\alpha(\sigma)$ are the Lagrange multiplier associated with the conservation constraints. Easy computations yield the solution
\begin{align}
\rho(\sigma,\vec{r}) &= \frac 1 {\partfun} e^{\tilde\beta \sigma \overline{\psi}(\vec{r})-\alpha(\sigma) },
\shortintertext{so that the coarse-grained vorticity is given by}
\overline{\omega} &= F(\overline{\psi}),\quad \text{with } F(\overline{\psi})= \frac 1 {\tilde\beta} \frac{\delta \ln \partfun}{\delta \overline{\psi}},\label{mfeq}
\end{align}
and $\partfun(\overline{\psi}) = \int_\mathbb{R} d\sigma e^{\tilde\beta \sigma \overline{\psi}-\alpha(\sigma)}$. To compute the equilibrium states of the system, one should solve the partial differential equation \eqref{mfeq}, referred to as the \emph{mean-field equation}, and check afterwards that the obtained critical points are indeed maxima of the constrained variational problem by considering the second derivatives. This will automatically ensure that the equilibrium states are nonlinearly stable steady states~\citep{Chavanis2009}. 


\subsubsection{Equilibrium states for 2D and barotropic flows}

\begin{figure}
\centering
\includegraphics[width=0.95\linewidth]{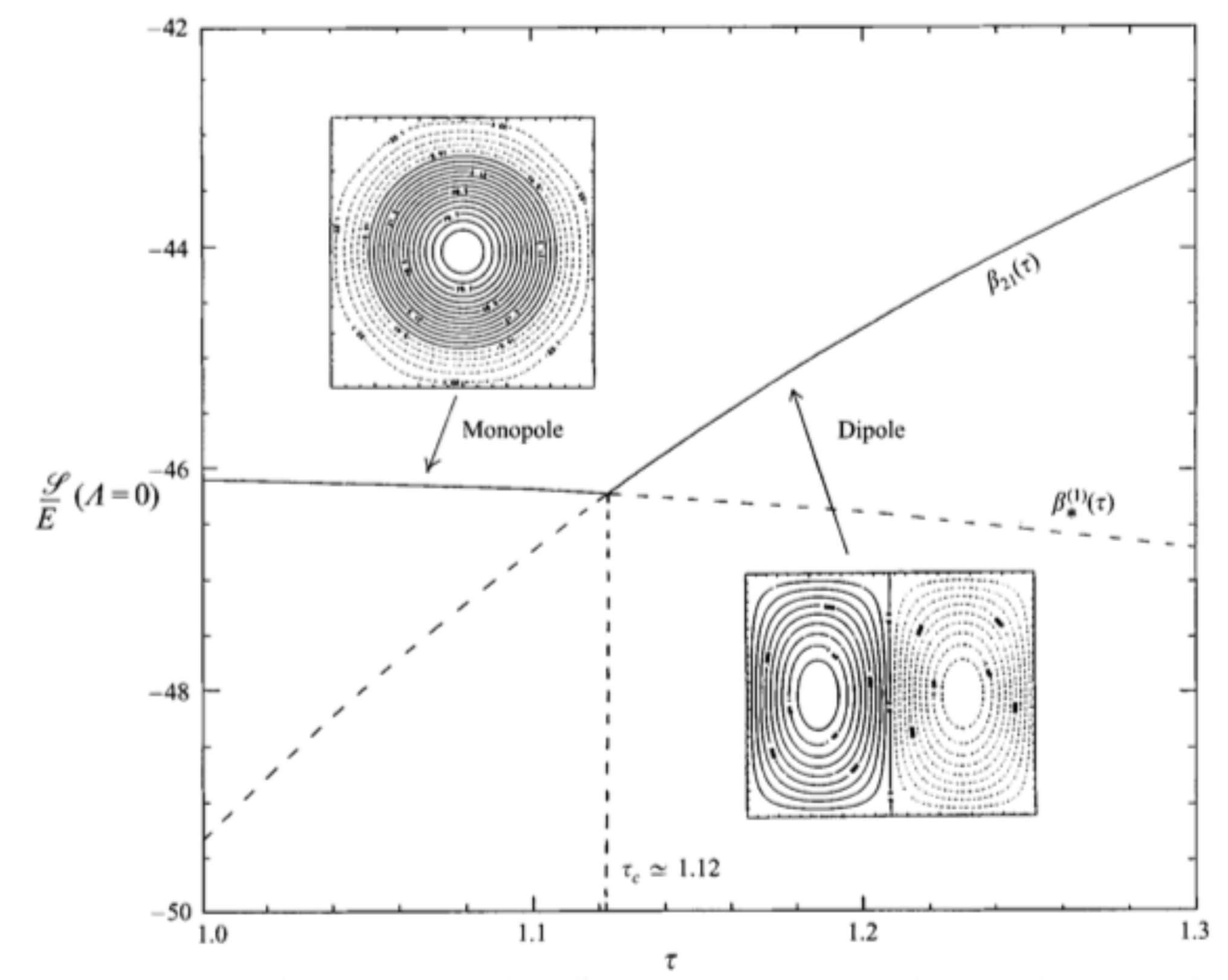}
\caption{Maximum entropy states as a function of the aspect ratio for a rectangular domain, in the linear (\emph{strong mixing}) $\omega$-$\psi$ limit~\citep{Chavanis1996a}. For $\tau < \tau_c$, the equilibrium is a monopole, while for $\tau>\tau_c$, it is a dipole.}\label{chavanisfig}
\end{figure}

The mean-field equation \eqref{mfeq} is in general difficult to solve; one issue is that the $\overline{\omega}$ - $\overline{\psi}$ relation is in general nonlinear. Most of the analytical solutions have been obtained in the linear case, by decomposing the fields on a basis of eigenfunctions of the Laplacian on the domain $\domain$. This technique was first introduced in a rectangular domain by \citet{Chavanis1996a}, who showed that the statistical equilibrium is either a monopole or a dipole, depending on the aspect ratio (Fig. \ref{chavanisfig}). 
\begin{figure*}[tbhp]
\centering
\includegraphics[width=0.65\textwidth]{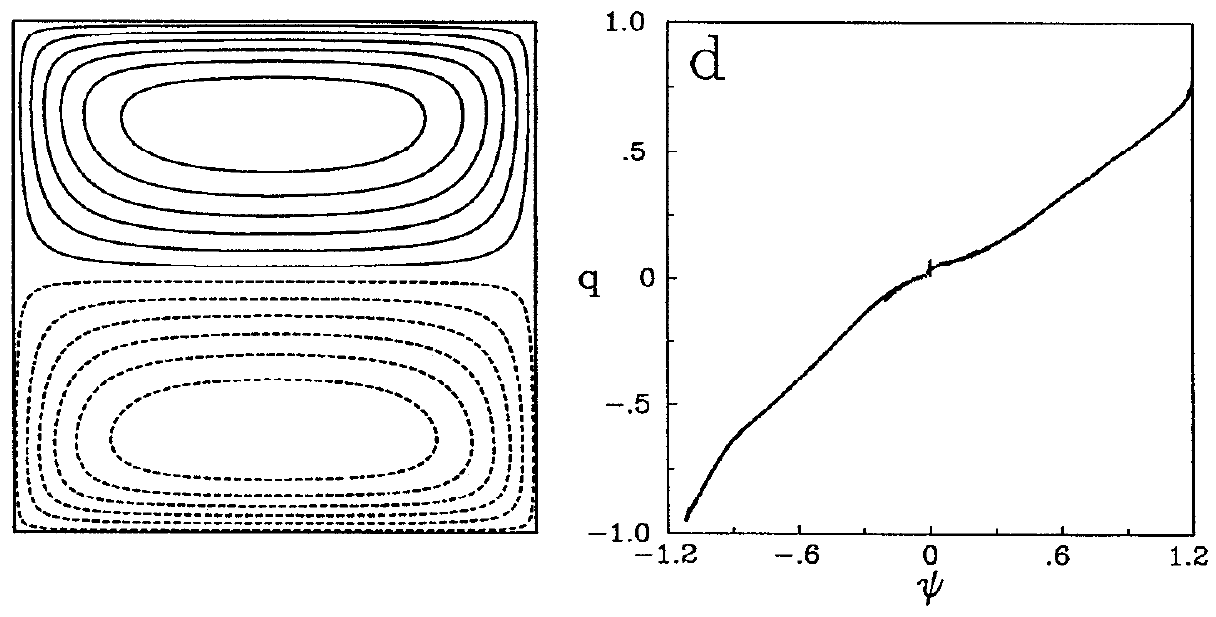}
\caption{Convergence towards the statistical equilibrium in inviscid truncated barotropic flow on a $\beta$-plane~\citep{Wang1994}. Left: Stream function. Right: scatterplot of the $q$-$\psi$ relation.}\label{fofonofffig}
\end{figure*}
The same method was extended to the case of barotropic flows, replacing vorticity by potential vorticity. Taking into account the $\beta$ effect, \citet{Fofonoff1954}  flows are obtained as statistical equilibria in a rectangular basin \cite{Naso2011,Venaille2011a}. Such solutions correspond to flows with two gyres (anticyclonic in the northern basin, cyclonic in the southern basin) in a rectangular basin, see Fig. \ref{fofonofffig}. The relative vorticity is confined to a boundary layer, whose with decreases with the total energy or when the $\beta$ effect (\textit{i.e.}, the relative strength of the gradient of the planetary vorticity) increases. The flow is westward in the interior of the basin, with an eastward compensating flow near the boundaries.

Different geometries can be studied: in a rotating sphere, the equilibria, in the linear limit, can be either solid-body rotations, dipole flows~\citep{Herbert2012b} or quadrupoles, taking into account conservation of angular momentum~\citep{Herbert2013b}. In the latter case, a perturbative treatment of the nonlinearity in the $\overline{\omega}$ - $\overline{\psi}$ relationship leads to the same flow topology, but sharper vortex cores~\citep{Qi2014}.
\citet{Bouchet2009} have also considered the role of a small nonlinearity in the $\overline{\omega}$ - $\overline{\psi}$ relationship for a rectangular domain of aspect ratio close to 1, with periodic boundary conditions, thereby obtaining two topologies for the equilibrium states: dipole and unidirectional flows. Adding a small stochastic forcing generates transitions from one to the other equilibrium.

\subsubsection{Stratified flows}\label{stratRSMsection}

In addition to the 2D and quasi-2D cases mentioned above, the theory has also been applied to stratified fluids (essentially in the quasi-geostrophic regime). \citet{Herbert2014b} has obtained and classified the statistical equilibria of the two-layer QG model in the framework of the Robert-Miller-Sommeria theory, and updated the discussion of the vertical distribution of energy at statistical equilibrium (see section \ref{kraichnanstratsection}): in particular, it is shown that even at statistical equilibrium, there will remain some residual energy in the baroclinic mode, unless the initial vertical profile of fine-grained enstrophy is uniform. 
In the context of continuously stratified flows, \citet{Venaille2012b} has taken up the thread initiated by \citet{Merryfield1998} (see section \ref{kraichnanstratsection}) and shown that bottom-trapped currents are indeed statistical equilibria of the Robert-Miller-Sommeria theory. Still in the continuous case, \citet{Venaille2012a} have also studied the vertical distribution of energy at statistical equilibrium, focusing on the tendency to reach barotropic equilibrium states; as also observed in the two-layer model, the constraint of conservation of fine-grained enstrophy prevents complete elimination of energy in the baroclinic mode. As the $\beta$ effect increases, barotropization is facilitated, until we enter a regime dominated by waves. It is well known that  baroclinic dynamics is hindered by strong values $\beta$ \citep{Holton}.

\subsection{Subgrid scale parameterization}

Results from  equilibrium statistical mechanics  have found  practical applications in the development of parameterization methods. 
 \citet{Holloway1992} suggested to replace the  usual sub-grid scale parameterizations in ocean models, where, \textit{e.g.}, viscous forces are represented with terms of the form $\nu_* \Delta \vec{u}$), where  $\nu_*$ is the eddy viscosity. He proposed to replace such formula with  $\nu_* \Delta (\vec{u}-\vec{u^*})$, so that viscosity relaxes the system  towards the statistical equilibrium state $\vec{u}^*$. Such a parameterization has been implemented, tested and commented in a number of studies~\citep[e.g.]{Cummins1994}. For more perspective on this type of subgrid-scale parameterizations, the reader is referred to \cite{Holloway2004} and \cite{Frederiksen2008}.

Along similar lines, \citet{Kazantsev1998} have proposed more generally to treat the subgrid scales so as to maximize the entropy production, inspired by the relaxation equations formulated in the Robert-Miller-Sommeria theory as an algorithm to construct equilibrium states~\citep{Chavanis1997}.
Note also that it has been shown in direct numerical simulations of ideal 3D turbulence that the small scales thermalize progressively, and act as a sort of effective viscosity in the ideal system, leading to the appearance of transient Kolmogorov scaling laws~\citep{Cichowlas2005}. This seems to be consistent with the above suggestions for subgrid scale parameterizations.


%
%

\section{Climate as a forced-dissipative thermodynamic system}
\label{prigogine}
In the previous sections the focus has been on identifying symmetry properties and conservation laws of GFD flows and relate these to  dynamical features and statistical mechanical properties. Neglecting forcing and dissipation has led us to study reversible equations whose statistical properties can be described using equilibrium statistical mechanics. 

Indeed, this provides the backbone of the properties of GFD flows and are of great relevance for studying more realistic physical conditions. Nonetheless, at this stage, a reality-check  is  necessary. The atmosphere and the oceans  are  out-of-equilibrium systems, which exchange irreversibly matter and energy  from their surrounding environment   and re-export it in a more degraded  form at higher entropy. For example, Earth absorbs short-wave radiation (low-entropy solar photons emitted at a temperature of  $\approx 6000$ K) which is then   re-emitted  to space as infrared radiation (high entropy thermal photons emitted at at a temperature $\approx 255$ K). In addition  to that, spatial gradients in chemical concentrations and temperature as well as their  associated  internal matter and energy fluxes   can be  established  and maintained for long time within non-equilibrium  systems (e.g. the  temperature contrast   between the polar and equatorial regions and the associated large-scale,  atmospheric and oceanic circulation). In this and in the next sections we will take such a point of view. 

The basis of the physical theory of climate was  established  in a seminal paper  by   \cite{Lor55}, who elucidated how the mechanisms of energy forcing, conversion and dissipation  are related to  the general circulation of the atmosphere. Oceanic and atmospheric large  scale flows results from the conversion of available potential energy - coming from the differential heating due to the inhomogeneity of the absorption of solar radiation- into kinetic energy through different mechanisms of instability \textcolor{black}{due to the presence of large temperature gradients \citep{char48,Eady49}. Such instabilities create a negative feedback, as they tend to reduce the temperature gradients they feed upon by favoring the mixing between masses of fluids at different temperatures.}
\textcolor{black}{Furthermore, in a forced and dissipative system like the Earth's climate, entropy is continuously produced by irreversible processes \citep{Prigogine61, deGroot84}.  Contributions to the total \emph{material entropy production}, which is related to the non-radiative irreversible processes  \citep{Goody00, Kleidon09}, come from: dissipation of kinetic energy due to viscous processes,  turbulent diffusion of heat and chemical species, irreversible phase transitions  associated to various processes relevant for the hydrological cycle, and  chemical reactions relevant for the biogeochemistry of the planet.}

It is important to note that the study of the climate \textit{entropics}  has been revitalized after  \citet{Paltridge, Paltridge2} proposed a principle of maximum entropy production (MEPP) as a constraint on the climate system. While the scientific community disagrees on the validity of such a point of view - see, \textit{e.g.}, \citet{Goody07} -  the discussion revolving around MEPP has led the scientific community  to refocus on the importance  of a thermodynamical approach -- as complementary to the dynamical one -- in providing  physical insights for student the climate system. In this paper we will not discuss MEPP  and  other non-equilibrium variational principles \citep[for an updated review see][]{DewarBeyond}.

\subsection{Climatic energy budget and energy flows}
\label{energy}
\vspace{10pt}

\subsubsection{Energy Budget}
\label{conserv}

%
 %
We first focus on developing equations describing the energy budget of the climate system. The total specific (per unit mass) energy of a geophysical fluid  is given by the sum of internal, potential, kinetic and latent energy. This can be expressed as  $e=\mathbf{u^2}/2+i+\phi+Lq$ for the atmosphere, where $\mathbf{u}$ is the velocity vector, $i=c_v T$ is the nertanl energy, with $c_v$ is the specific heat at constant volume for the gaseous atmospheric mixture and $T$ is its temperature,  $\Phi$ is the gravitational (plus centrifugal) potential,  $L$ is the latent heat of evaporation, and $q$ is the specific humidity. In this formula, we neglect the heat content of the liquid and solid water and the heat associated to the phase transition between solid and liquid water. The approximate expression for the specific energy of the ocean reads  $e=\mathbf{u^2}/2+i+\Phi$, where $i=c_W T$ is the specific heat at constant volume of water (we neglect the effects of salinity and of pressure), while we can consider $e=c_S T+\phi$ as the specific energy for solid earth or ice. The conservation of energy and the conservation of mass imply that \citep{Peixoto:1992}:
\begin{equation}
\frac{\partial \rho e}{\partial t}=-\nabla\cdot (\mathbf{J_h}  + \mathbf{F_R} +\mathbf{F_S} + \mathbf{F_L})-\nabla(\tau\cdot\mathbf u)
\label{energy1}
\end{equation}
where $\rho$ is the density; $p$ is the pressure; $\mathbf{J_h} = (\rho e + p)\mathbf u$  is the total enthalpy transport; $\mathbf{F_R}$, $\mathbf{F_S}$, and $ \mathbf{F_L}$ are the vectors of the radiative, turbulent sensible, and turbulent latent heat fluxes, respectively; and $\mathbf{\tau}$ is the stress tensor. By expressing Eq. (\ref{energy1})  in spherical coordinates $(r, \lambda, \varphi)$, and assuming the usual thin shell approximation $r=R+z$, $z/R\ll1$, where $R$ is the Earth's radius and $z$ is the vertical coordinate of the fluid, we have \citep{Peixoto:1992}:
\begin{equation}
[\dot{E}]=-\frac{1}{R\cos\varphi}\frac{\partial T_T}{\partial \varphi} + [F_{R}^{TOA}]
\label{energy3}
\end{equation}
where $[X](\varphi,t)\equiv\int X(\lambda,\varphi,t) d\lambda$,  $F_{R}^{TOA}$ is the net radiation at the top of the atmosphere (with the convention  that the value is positive when there is an excess of incoming over outgoing radiation) and  the meridional enthalpy transport has been defined as:
\begin{equation}
T_T(\varphi,t)\equiv\iint      J_{h\varphi}(\varphi,\lambda,z,t) R\cos\varphi     dz d\lambda.
\label{energy4}
\end{equation}
Equation (\ref{energy3}) relates the rate of change of the vertically and zonally integrated total energy  to the divergence of the  meridional transport by the atmosphere and oceans  and the zonally integrated radiative budget at the top-of-the-atmosphere.
Integrating  along $\varphi$ ($\{ X \}=\int  X d\varphi$), the   expression for the time derivative of the net global energy balance is straightforwardly derived: 
\begin{equation}
\{ [F_{R}^{TOA}] \}=\{ [\dot{E}] \}.
\label{energy4bis}
\end{equation}
\begin{figure}[thbl]
\centering
\includegraphics[width=20pc,trim=0pc 0pc 0pc 1pc, clip]{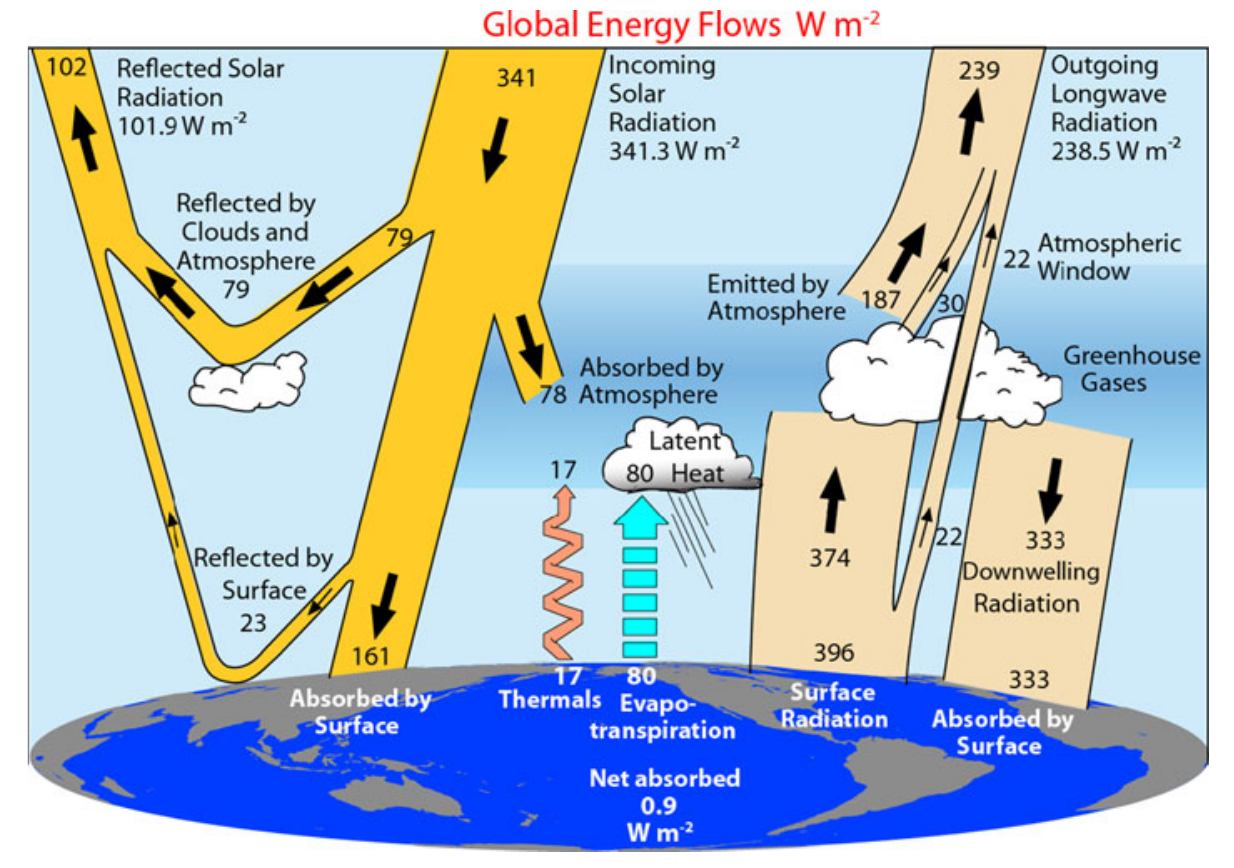}
\caption{Global annual mean earthÕs energy budget for 2000-2005 ($W m^{-2}$).  From \citet{TF2012}.  
 \label{energydiagr}
}
\end{figure}
Similar relationships can be written for the atmosphere, ocean and land provided that  energy fluxes of sensible, latent heat as well as radiative fluxes are taken into account at the surface \citep[][]{Peixoto:1992}. \textcolor{black}{A schematic view of the surface and TOA energy fluxes for present day Earth \citep{TF2012} can be see in Fig. \ref{energydiagr}.} Under steady state conditions, the long term average $\overline{ \dot{E}}=0$. Therefore from equation (\ref{energy4bis}) the stationarity condition  implies that
\begin{equation}
\overline{ \{ [F_{R}]_{toa} \}}=0.
\label{energy4tris}
\end{equation}
Equation (\ref{energy4tris}) describes the basic fact that he climate system, at steady state, does not on the average receives nor emits energy.

 \begin{figure}[thbl]
\centering
a) \includegraphics[width=11pc,trim=0pc 0pc 0pc 1.5pc, clip]{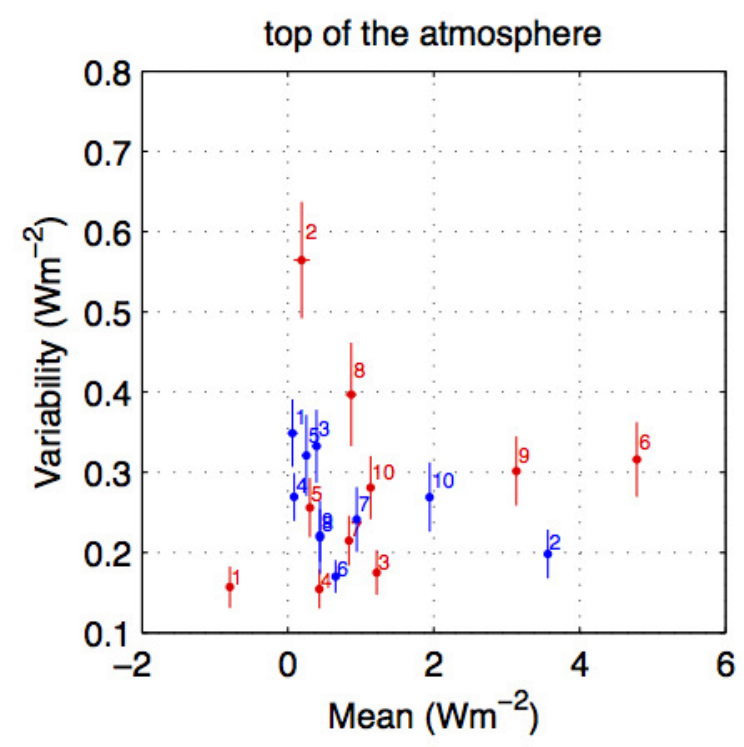}\\
 b) \includegraphics[width=11pc,trim=0pc 0pc 0pc 1.5pc, clip]{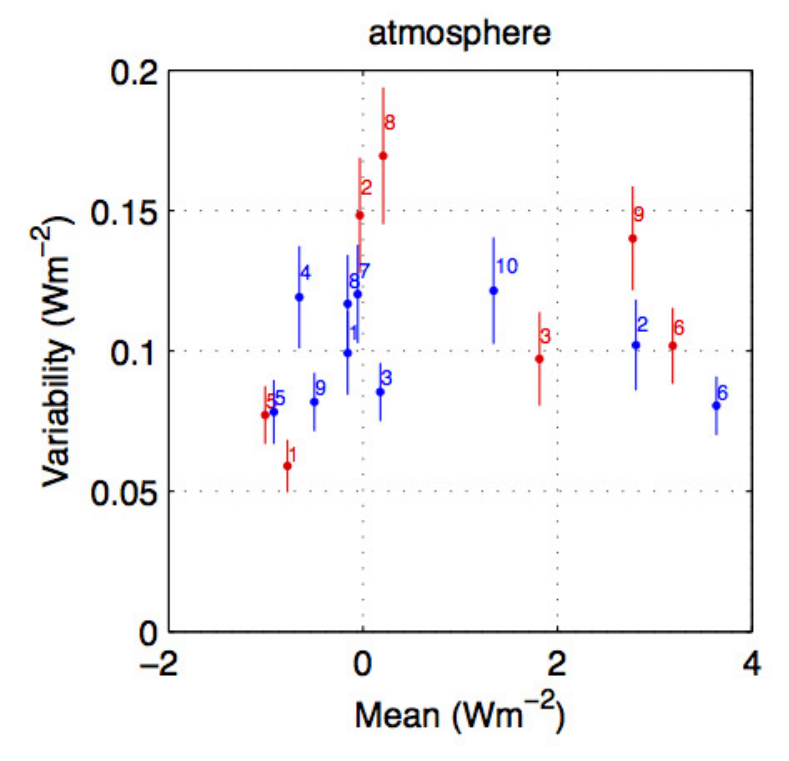}\\
 c)  \includegraphics[width=11pc,trim=0pc 0pc 0pc 1.5pc, clip]{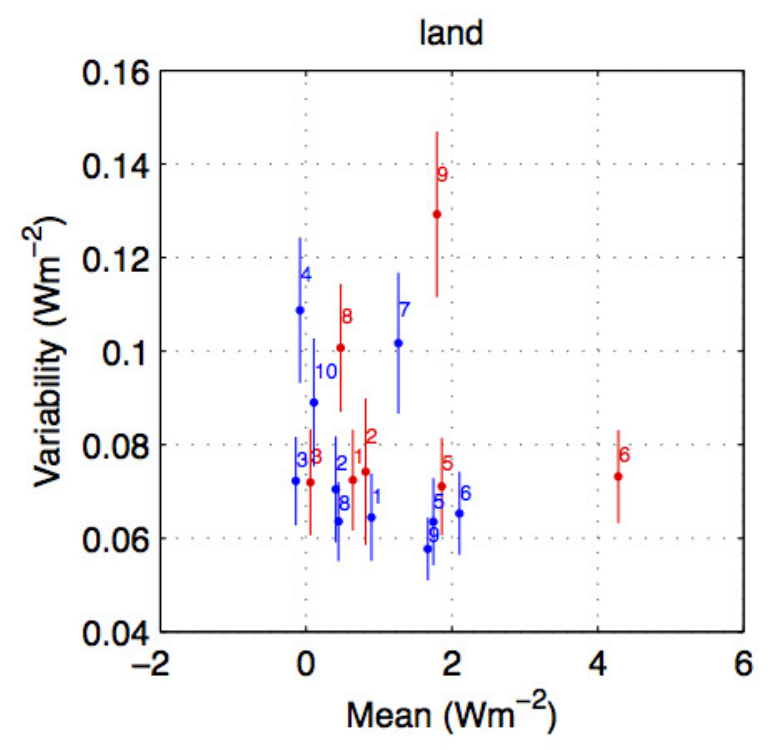}\\
 d) \includegraphics[width=11pc,trim=0pc 0pc 0pc 1.5pc, clip]{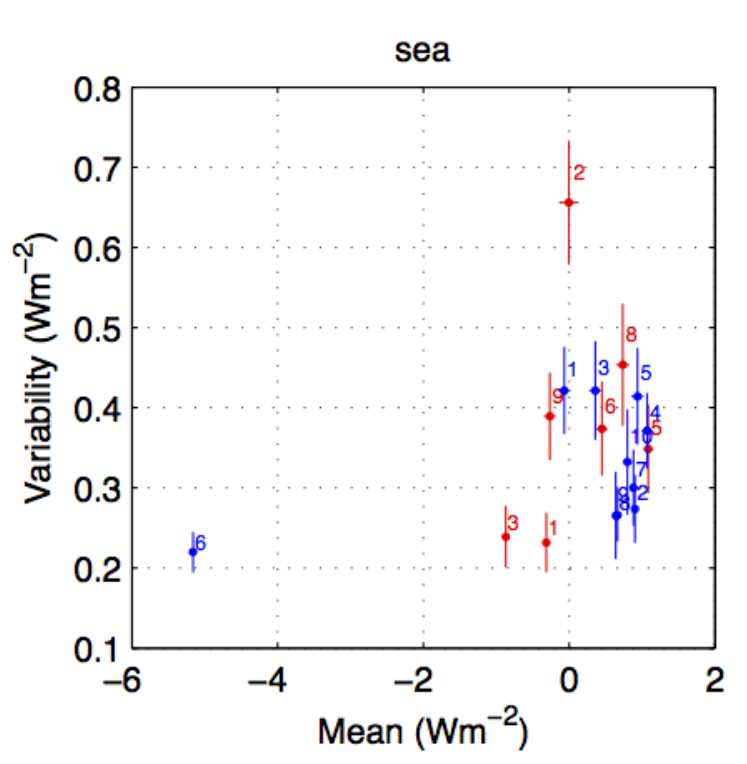}
 \caption{ Mean and standard deviation of globally averaged  top-of-the-atmosphere radiative budget (a),    atmosphere energy budget (b), ocean (c) and   land energy budget (d)   for inter-comparable   CMIP3 (red) and CMIP5 (blue) climate models control  simulations.  Updated from \citet{LucariniRagone}.  
 \label{toa}
}
\end{figure}
\textcolor{black}{These constraints can be used for auditing climate models. At observational level non-zero energy balances are found at TOA and at the surface \citep{TF2012,Wild2013}, due to the fact the the actual Earth is not at a stationary state, most notably because of the ongoing greenhouse gas forcing. However,} a physically consistent  climate model should feature a vanishing  net energy balance when its parameters are held fixed and statistical stationarity is eventually obtained.  \citet{LucariniRagone} analyzed the behavior of more than  twenty atmosphere-ocean coupled climate models (PCMDI/CMIP3 intercomparison project, http://www-pcmdi.llnl.gov/)   under steady state conditions  (preindustrial scenario) and found that   models'  energy balances are wildly different with global balances spanning between $-0.2$ and $2$ W\,m$^{-2}$, with a few ones featuring imbalances larger than $3$ W\,m$^{-2}$. The analysis of similar budgets for  the last generation of climate models (CMIP5 intercomparison project, \citet{Taylor}) does not show a significant improvement (Fig.~\ref{toa}).  Spurious energy biases may be   associated with non-conservation of water in the atmospheric branch of the hydrological cycle \citep{Liepert,Liepert2} and in the water surface fluxes \citep{Lucarini08D,Hasson13}, with  the fact that dissipated kinetic energy  is not  re-injected in the system as thermal energy \citep{Becker,Lucarini:2009_PRE_KF}, as well as with nonconservative numerical schemes \citep{Gassman}.

\subsubsection{Meridional enthalpy transport}
\label{trasp}
The next step in constructing the energetics of the climate system is the study of the large scale transport of various forms of energy. The meridional distribution of the radiative fields at the top-of-the-atmosphere poses a strong constraint on the meridional general circulation \citep{Stone78}.  As clear from  equation  (\ref{energy3}), the stationarity condition (\ref{energy4tris}) leads to the  following indirect relationship for  $T_T$:
\begin{equation}
 \overline{T_T }(\varphi)= -2\pi\int_{\varphi}^{\pi/2}\! R^2  \cos\varphi^\prime \overline{(F_{R})_{toa}(\varphi^\prime)}.
\label{energy5}
\end{equation}
In other terms, the flux $T_T$ transports enthalpy from the low-latitudes, which feature a positive imbalance between the net input of solar radiation -- determined by planetary  albedo, determined mostly by \textit{i.e.}, clouds \citep{Donohe2} and by surface properties -- and the output of longwave radiation, to the high-latitudes, where a corresponding negative imbalance is present.  Atmospheric and oceanic circulations act as responses needed to equilibrate such an imbalance  \citep[][]{Peixoto:1992}.

The  climatic meridional enthalpy  transport  $T_T(\varphi)$  reduces the temperature difference between the low and high latitude regions with respect to what imposed by the radiative-convective equilibrium picture.   \citet{Stone78} showed that $T_T$  depends essentially on the mean planetary albedo and on  the equator-to-pole contrast of the  incoming solar forcing, while being  mostly independent  from dynamical  details of atmospheric and oceanic circulations.  As emphasized by \citet{Enderton}, if one assumes drastic changes in the meridional distributions of planetary albedo differences emerge with respect to Stone's theory.  A  comprehensive  thermodynamic theory of the climate system able   to   predict  the   peak location and strength of the meridional transport, the partition between atmosphere and ocean \citep{Rose}  and to accommodate the variety of processes contributing to it, is still missing.

 Besides theoretical difficulties,  observational estimations  of $T_T$, $T_A$ and $T_O$ also poses non-trivial challenges.  For simplicity, we here refer to $T_T$. There is still  not an accurate estimate of such a fundamental quantity for testing the output of climate models, despite  the efforts of   several authors  \citep{TreCar, Wun,Fas2,Fas,  Mayer}. The precision of the estimates relies on  the knowledge of the \textit{boundary} fluxes  $\mathbf{F_R}$, $\mathbf{F_S}$, and $ \mathbf{F_L}$  and on the reanalysis datasets. \citet{Wun}, by using measurements of the radiative fluxes at the top of the atmosphere  and previous estimates of the oceanic enthalpy transport, gave a range of values of $3.0-5.2$ PW ($1$ PW $=10^{15}$ W) for the maximum  of the total poleward transport in the  Northern Hemisphere (NH) and $4.0-6.7$ PW for the maximum of the total poleward transport in the  Southern Hemisphere (SH).   \citet{Fas},  by combining measurements of  top-of-the-atmosphere  radiative fields   with different reanalyses and ocean datasets, found the range   to be   $4.7-5.1$ PW for the SH maximum  and   $4.6-5.6$ PW for  NH. \citet{Mayer}, using two reanalysis datasets (ERA-40 and the more recent ECMWF reanalysis ERA-Interim), constrained the two peaks in narrower confidence intervals:      $5.1-5.6$ PW  in the SH ($4.4-4.9$ PW  in the NH)     for the ERA-40 data and  $5.1-5.6$ PW in the SH ($4.4-4.9$ PW  in the NH)  for the  ERA-Interim data.  Unfortunately   reanalysis datasets are affected by mass and energy conservation (e.g.  $+1.2$ W\,m$^{-2}$ at the top-of-the-atmosphere and $+6.8$ W\,m$^{-2}$ over oceans  in ERA-
\begin{figure}[thbl]
\centering
\includegraphics[width=20pc,trim=0pc 0pc 0pc 0pc, clip]{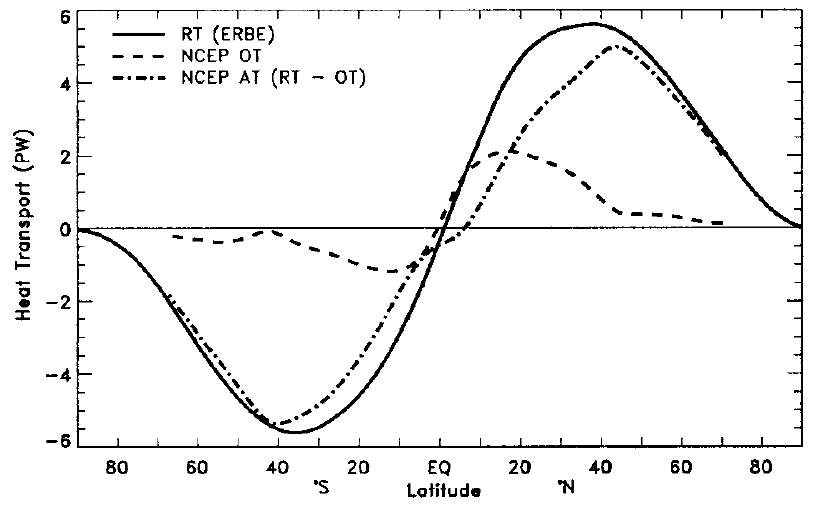}
\caption{ Annual meridional enthalpy transports of ocean (dashed), atmosphere (dash-dotted) and total (solid) estimated from satellite and reanalysis data ($PW$).  From \citet{TreCar}.  
 \label{TransportTreCar}
}
\end{figure} 
Interim, \cite{Mayer})  problems  that may potentially bias  the transport estimates. 
Furthermore, these estimates are dependent on  the analysis method and the  model used  -- \citet{TreCar}, using  other reanalysis dataset  (NCEP), found a value of the  maxima  $0.6$ PW larger in the NH  than those found with the  ECMWF  reanalysis. \textcolor{black}{Estimates from \citet{TreCar} are shown in Fig. \ref{TransportTreCar}.}

\begin{figure}
\centering
 a)\includegraphics[width=19pc,trim= 1.6pc 0pc 0pc 0pc,clip]{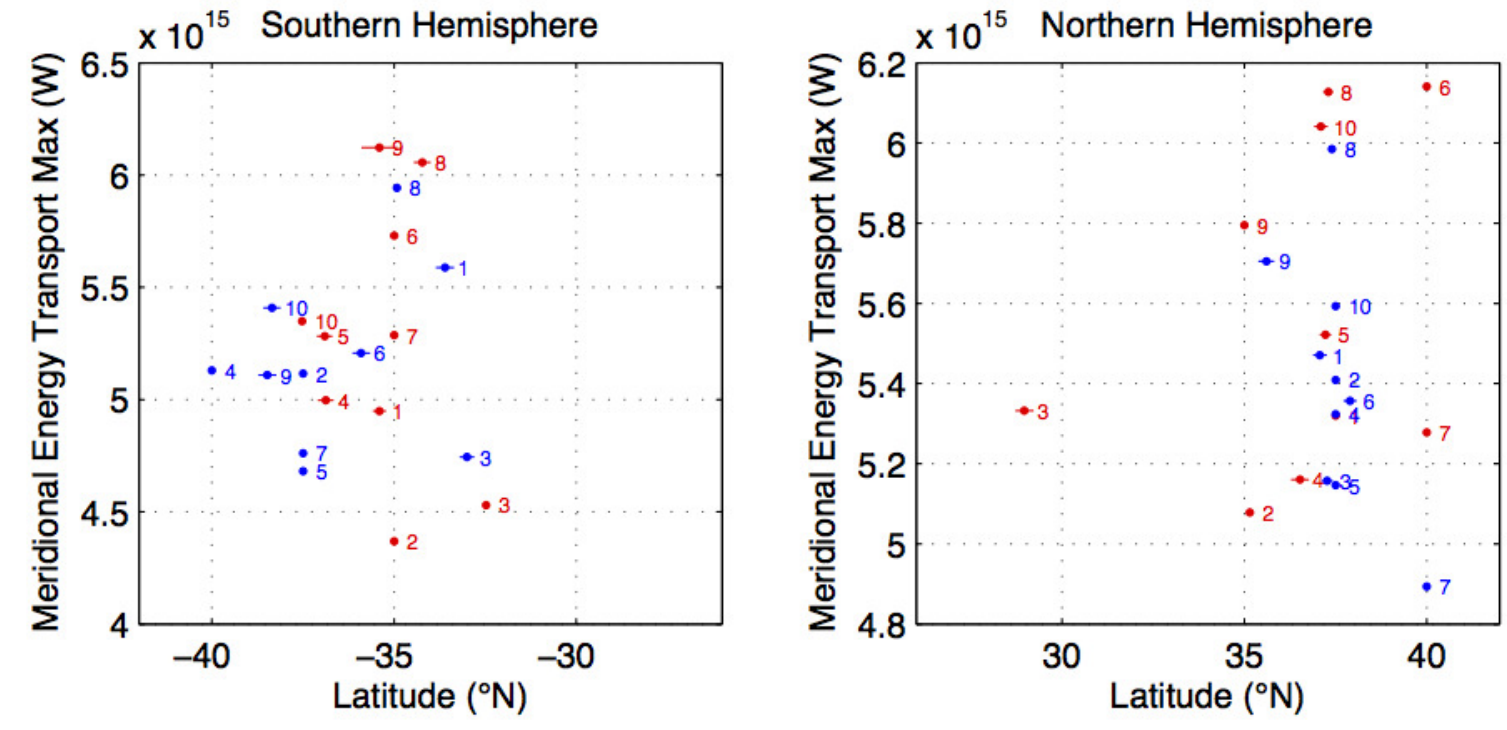}
 b)\includegraphics[width=19pc,trim= 1.5pc 0pc 0pc 0pc,clip]{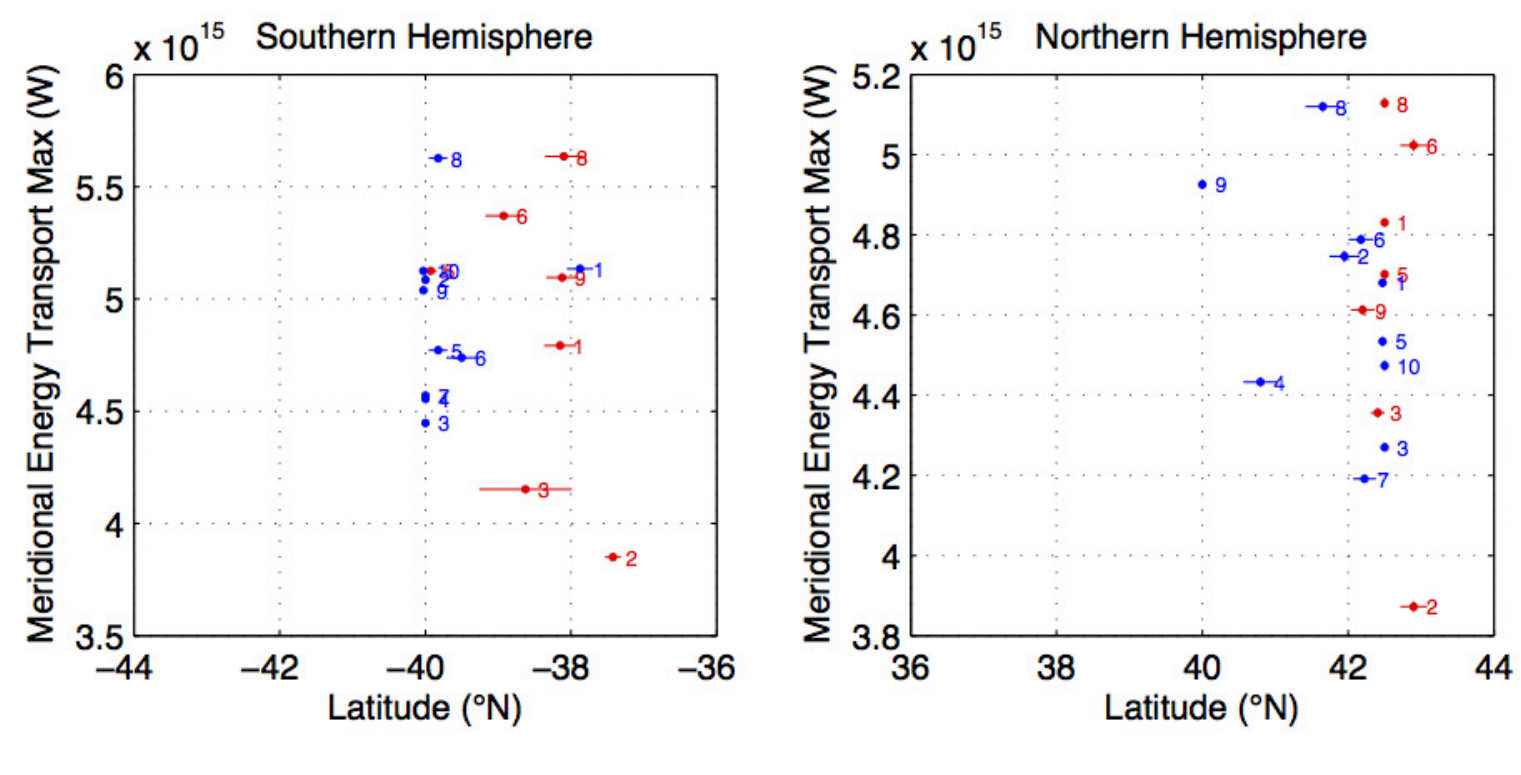}
 c)\includegraphics[width=19pc,trim= 0pc 6pc 0pc 0pc,clip]{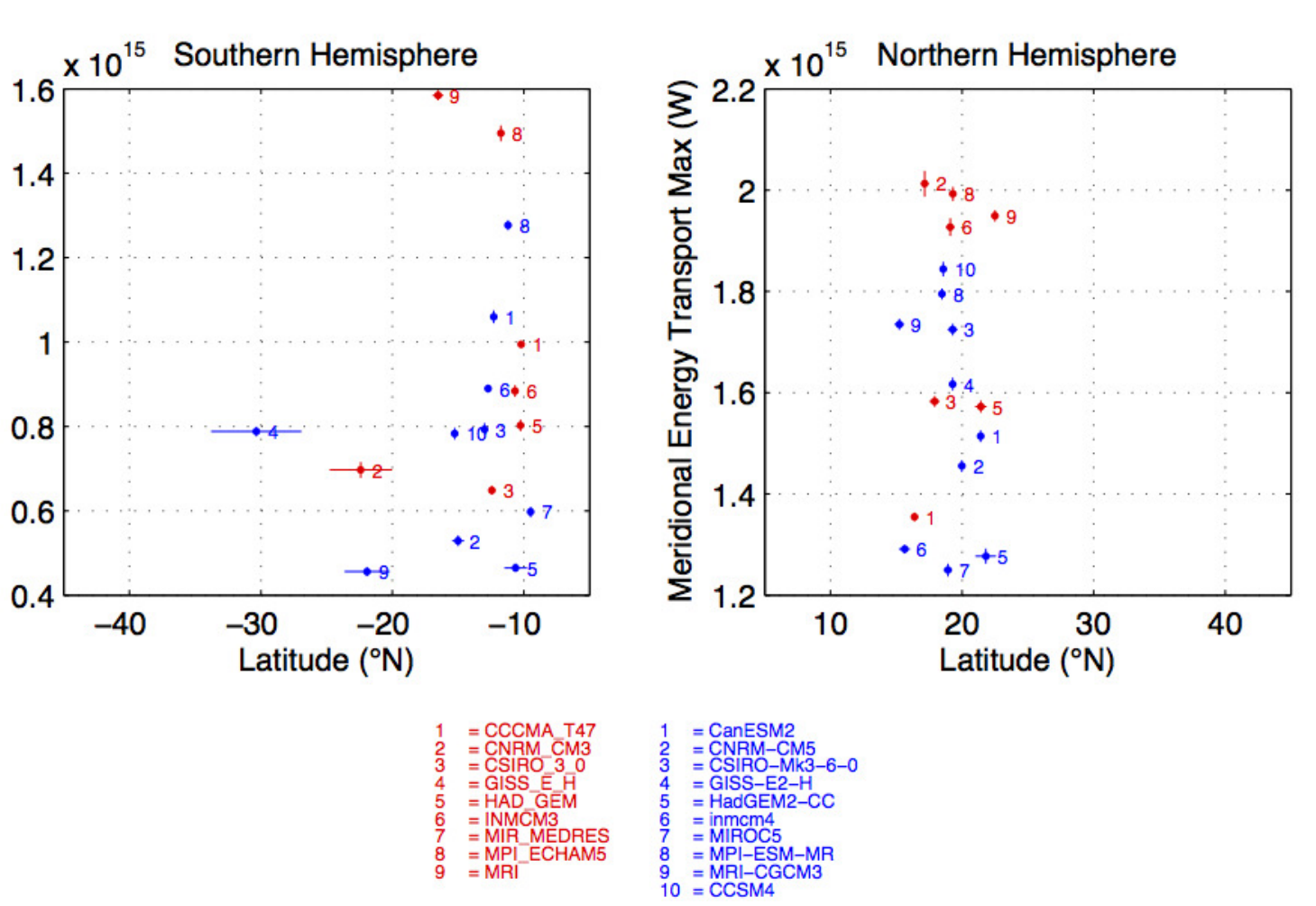}
  \includegraphics[width=28pc,trim= 10pc 0pc 6pc 18.5pc,clip]{max_sea-eps-converted-to.pdf}
 \caption{ Value and position of the peak of the poleward meridional enthalpy transport in the pre-industrial scenario for the whole climate (a), atmosphere (b) and ocean (c) for the some of the CMIP3 (red) and CMIP5 (blue) general circulation models. Updated from \citet{LucariniRagone}. \label{max}
}
\end{figure}

The use of numerical climate model does not help to reduce such uncertainties \citet{LucariniRagone} analyzed  a large  dataset of coupled climate models (PCMDI/CMIP3, http://www-pcmdi.llnl.gov/)  and   found a  large spread in the meridional enthalpy transports peaks  with discrepancies of the order of 15-20 $\%$ around a typical value of about $5.5$ PW. State-of-the-art climate models (CMIP5 intercomparison project,  \citet{Taylor}) show little improvement in terms of mutual agreement  (Fig.~\ref{max}).  \citet{ Donohe2} attributed such a large spread in $T_T$ to intermodel differences in the meridional  contrast of absorbed solar radiation, which, in turn, is mainly due to   the inter-model difference in the  shortwave optical properties of the atmosphere; for an intercomparison of the cloud distribution is different climate models see \citet{Probst2012}. Figure ~\ref{max} also shows that, while the disagreement among models for the peak of the atmospheric transport is comparable to that for the peak of the total transport,  enormous differences emerge when comparing oceanic transports. 

Interesting information emerge when looking at the position of the peaks of the transport. \citet{Stone78} predicted that the position of  the maximum of $T_T$ is well constrained by the geometry of  the system and weakly dependent of longitudinal homogeneities, and, accordingly in Fig. \ref{max} both CMIP3 and CMIP5 models feature small spread in the position of the peak of $T_T$, with minute differences between the two hemispheres, except one outlier. Similarly, the spread among models is small with respect to the position of the peak of $T_A$ in both hemispheres and of $T_O$ in the Northern Hemispheres, while a larger uncertainty exists in the position of the peak of $T_O$ in the Southern Hemisphere.   

\subsection{The maintenance of thermodynamical \\disequilibrium}
\label{secondlaw}

The basic understanding of the  maintenance of  the atmospheric  general circulation   was achieved nearly sixty years ago by  \citet{Lor55,Lor67} through    the concepts of available potential energy  and atmospheric energy cycle.  The concept of available potential energy, first introduced by \citet{Marg} to study  storms,
 is defined  as $A=\int c_p(T-T_r) dV$, where $T_r$ is the temperature field of the reference state, obtained by an isentropic redistribution of the atmospheric mass so that the isentropic surfaces become horizontal and the mass between the two isentropes remains the same. By its own definition, this state minimizes  the total potential energy   at constant entropy. Such  a  definition  is somewhat arbitrary and different definitions lead to  different formulations of atmospheric energetics \cite{Tailleux}.  For example, the choice  of a reference state  maximizing entropy at constant energy \citep{Dutton73}  leads in a natural way to the concept of exergy. {Exergy is the part of the internal energy measuring  the departure of the  system from  its  thermodynamic and mechanical equilibrium, \textit{i.e.},a state of maximum entropy at constant energy.}, and is a commonly used concept in  heat engines theory \citep{Rant}.
 
\citet{Lor67} proposed the following picture  of the transformation of energy  in the atmosphere. We define $E=P+K$, where $K=(1/2)\int  dV \rho \mathbf{u}^2  $ represents the total kinetic energy and $P = \int  dV \rho (c_V T+\Phi)$  the dry static energy  and $V$ is the atmospheric domain. {Under hydrostatic approximation one can show that $\int  dz \rho (c_v T+gz) = \int  dz \rho (c_v T+RT ) =  \int  dz \rho  c_p T$ \citep[see e.g.][]{Lor67}.}. In the Lorenz framework  one considers the hydrological cycle as a forcing to the atmospheric circulation. This amounts to separating the budget of the moist static energy and of the part related to the phase changes of water. See \citet{Peixoto:1992}, Chap. 13. We obtain: 
 \begin{align}
\dot{P} = & -W(P,K)+\dot{\Psi}+D, \label{prima}\\
\dot{K}  = & -D+W(P,K), \label{seconda}
\end{align}
where  $D =\int dV \rho\epsilon^2 > 0   $ is the dissipation of kinetic energy due to turbulent cascades to small scales and to the wind shear associated to falling hydrometeors, $W(P,K) = - \int dV  \rho \mathbf{u}\cdot\nabla p $ is the potential-to-kinetic energy conversion rate,  and $\dot{\Psi}=\int dV  \rho  \dot{q}_{nf} $ is the non-frictional  diabatic  heating due to the convergence of turbulent sensible heat fluxes, condensation/evaporation inside the atmosphere,  and convergence of radiative fluxes. The conversion term $W$ can be  interpreted as  the  instantaneous work performed by the system. In this respect, Eq. (\ref{prima}) represents the statement of  the first law of thermodynamics for the atmosphere.  Equations (\ref{prima})-(\ref{seconda}) imply that $\dot{E}=\dot{P}+\dot{K} = \dot{\Psi}$ and therefore the frictional heating $D$ does not increase the total energy since it is just an internal conversion between kinetic and potential energy. Stationarity implies that $\overline{\dot{P}}=\overline{\dot{K}}=0$ and therefore $\overline{D}=\overline{W}$, which is referred to as the intensity of the Lorenz energy cycle. One has to note that the latter can be expressed as the average rate of generation of available potential energy, $\overline{G}=\int dV \overline{\rho  \dot{q}_{nf}(1-T/T_r)}$, where  $T_r$ is the temperature field of the reference state \citep{Lor67}.  

The strength of the   Lorenz energy cycle  is a fundamental non-equilibrium  property of the atmosphere, which, just as  the meridional enthalpy transport (Sect.~\ref{trasp}), is   known with a certain degree of  uncertainty for the present climate. Reanalysis datasets (with all associated problems, see Sect.~\ref{trasp})   constrain $\overline{D}$   in the range $1.5-2.9$ W\,m$^{-2}$ \citep{Li}.  On the other hand,  general circulation models   feature values of $\overline{D}$ ranging from     $2$ to $3.5$ W\,m$^{-2}$  \citep{Marques11}.  Numerical simulations  show  that a CO$_2$ doubling causes a decrease of $\overline{G}$ of nearly $10\%$ \citep{ LucACP}.   Warming patterns can alter $G$ either by affecting the gross static stability (stronger stability implies a weaker energy cycle, as clear from the theory of baroclinic instability) or the meridional temperature/diabatic heating distribution.   \citet{Hernandez2}    show that the decrease in $G$ is mostly associated with changes in the gross static stability changes rather than with meridional temperature gradient changes. 
   
\textcolor{black}{Another aspect to be considered is that the intensity of the Lorenz energy cycle is formulated assuming hydrostatic conditions. Therefore, the Lorenz energy cycle in itself neglects any systematic transfer of potential into kinetic energy occurring through non-hydrostatic, small scale motions \citep{Stein}. Along these lines,  \citet{Pauluis_Dias} suggest that small scales processes such as precipitation  may significantly contribute to $\overline{D}$, which might therefore be considerably underestimated when computed for models that do not treat explicitly convection.}

\textcolor{black}{In the case of the ocean, available potential energy is generated through thermohaline forcings due to the correlation of density inhomogeneities and density forcings (e.g. through heat and freshwater fluxes) at surface. In addition to that, mechanical energy enters the ocean through direct transfer of kinetic energy by surface winds (and though tidal effects). Kinetic energy is dissipated through a variety of frictional processes, occurring mostly at the bottom of the ocean, and, similarly, available potential energy is lost through diffusion mostly due to small scale eddies \citep{Wunsch2014}. The understanding of the details of the oceanic Lorenz energy cycle is still at a relatively early stage. Estimates of dissipation and generation terms range within $(1-2)\times10^{-2}$ W\,m$^{-2}$  \citep{Oort94,  VonStorch,Tailleux}.}

\subsubsection{Atmospheric heat engine and efficiency}
\label{engine}
\citet{Johnson00} proposed an interesting construction for further elucidating the idea the the climate can be seen as a heat engine. We define the total  diabatic heating  $\dot{q}= \dot{q}_{nf}+\rho\epsilon^2$ and splitting the  atmospheric domain $V$ into the subdomain $V^+$  in which $\dot{q} =\dot{q}^+ >0,$ and $V^-$, where $\dot{q} =\dot{q}^- <0$, it can be  seen from equation (\ref{prima}) that:
\begin{equation}
\overline{W}= \int_{V^+} \overline{ \dot{q}^+ \rho} dV + \int_{-} \overline{ \dot{q}^- \rho } dV  \equiv \overline{\Phi^+} + \overline{ \Phi^-}, 
\label{piumeno}
\end{equation}
with $\overline{\Phi^+} > 0$ and $\overline{\Phi^-}< 0$ by definition.   Therefore the atmosphere can be interpreted as a heat engine, in which  $ \Phi^+ $ and $ \Phi^- $  are the net heat gain and loss, and  $W$ the mechanical work. The efficiency of the atmospheric heat engine,\textit{i.e.},the capability of generating mechanical work given a certain heat input, can therefore be defined as:
\begin{equation}
\label{eff1}
\eta = \frac{(\overline{\dot{\Phi}^+}+ \overline{\dot{\Phi}^-})}{(\overline{\dot{\Phi}^+})} = \frac{\overline{W}}{\overline{\dot{\Phi}^+}}.
\end{equation}
The analogy between the atmosphere and a (Carnot) heat engine can be pushed further if we introduce the total rate
of entropy change of the system, $\dot{S}=\int  dV \rho  \dot{q}/T = \dot{S}^+ + \dot{S}^-$. In a steady state the following expression holds:
 \begin{equation}
 \overline{ \dot{S}} = \frac{\overline{ \dot{\Phi}^+} }{T^+}+\frac{\overline{ \dot{\Phi}^-} }{T^-}=0,
 \end{equation}
 where $T^\pm\equiv \overline{\dot{\Phi}^\pm}/ \int_{V^\pm} dV \overline{\rho \dot{q}^\pm /T } $    from which it follows  that  $\eta = 1 - T^-/T^+$.  Johnson's approach provides a self-consistent treatment of the heat engine of a geophysical fluid and extends closely related thermodynamical theories of hurricane dynamics \citep{Emanuel91}.
 
In  Emanuel's theory    a mature hurricane  is depicted as an ideal Carnot engine driven by the thermal disequilibrium between the sea-surface temperature $T_s$ and the cooling temperature $T_0$ with an efficiency    $1-T_0/T_s\approx 1/3$. A similar  approach  was extended  also to moist convection   \citep{Emanuel, Renno}   for determining the  the wind speed reached by the convective system for a certain rate of  heat input $F_{in}$ from the sea,  $W=F_{in}(1-T_0/T_s)$.  \textcolor{black}{Such an approach has been used to study large scale, open systems like the Hadley cell \citep[][]{Adams} and the monsoonal circulation \citep{JohnMons}.}
 
\subsubsection{Entropy production in the Climate System}
\label{entropy}
We wish now to emphasize a different aspect of the climate's thermodynamics, namely the study of its irreversibility by the investigation of its material entropy production, \textit{i.e.},the entropy produced by the geophysical fluid, neglecting the change in the properties of the radiative fields \citep{Goody00,Ozawa03}. The entropy budget of the fluid can be rewritten as:
\begin{equation} 
\dot{S}  =  -\int dV \frac{\nabla \cdot \mathbf{F_R}}{T}  +\dot{S}_{mat},
\label{claus1}    
\end{equation} 
so that we separate the contribution coming from the absorption of the radiation from other effects related to the other irreversible processes occurring in the fluid. Note that, in the previous formula, we refer to the entropy budget of the whole climate, not of the atmosphere, as done, instead, in the previous section. 
%
%
%
%

The material entropy production, $\dot{S}_{mat}$ can be expressed as $\dot{S}_{diff} + \dot{S}_{fric} + \dot{S}_{hyd}$, \textit{i.e.}, the sum of   contributions associated with heat diffusion, frictional heating and the hydrological cycle (due to diffusion of water and phase-changes) respectively. \textcolor{black}{ Detailed estimates of the entropy budget of the climate system and of the material entropy production ($\overline{\dot{S}_{mat}} \approx50$ mW\,m$^{-2}$\,K$^{-1}$) can be found in  \cite{Goody00, Pascale11}. Oceanic entropy production due to small-scale mixing in the interior gives a small contribution ($\approx 1$ mW\,m$^{-2}$\,K$^{-1}$) to $\overline{\dot{S}_{mat}}$ \citep{Pascale11}. Therefore we will limit the discussion to processes occurring in the interior and at the boundaries of the atmosphere.}

 Entropy production due to  heat diffusion $\overline{\dot{S}_{diff}}=-\int dV\overline{ \nabla\cdot {\mathbf{J_S}}/T }$ is generally small ($\approx 2$ mW\,m$^{-2}$\,K$^{-1}$, \citep{Kleidon09}) and associated mostly with dry atmospheric convection occurring nearby the surface and with vertical mixing in the mixed layer of the ocean. The entropy production due to frictional heating - $\overline{\dot{S}_{fric}}=\int dV\overline{\rho \epsilon^2/T  }\approx 10$ mW\,m$^{-2}$\,K$^{-1}$ \citep{Fraedrich08,Pascale11} - is associated with turbulent energy cascades bringing kinetic energy from large scales down to  scales (millimeters or less for geophysical flows) where  viscosity can efficiently operate. Finally,  $\dot{S}_{hyd}$ is  due  to irreversible processes associated with the hydrological cycle --  evaporation of liquid water  in unsaturated air, condensation of water vapor in supersaturated air  and molecular diffusion of water vapour \citep{Pauluis, Pauluis0} and requires the knowledge of relative humidity ${H}$ and the molecular fluxes of water vapor $\mathbf{J}_v=\mathbf{F_L}/L$:
\begin{eqnarray}
 \dot{S}_{hyd}&=\int dV (C-E)R(\ln {H} \nonumber \\ &+\mathbf{J}_{v}\cdot\nabla p_w)-\int_{z=surf} dA J_{v,z}R\ln{H}.
\label{hyd}
\end{eqnarray}
where $C$ and $E$ indicate condensation and evaporation, respectively, and $p_w$ is the partial pressure of the water vapor. \textcolor{black}{The importance of  $\dot{S}_{diff}$ and  $\dot{S}_{hyd}$ in the context of thermodynamic theories of moist convection is extensively discussed in \citet{Pauluis0} and \citet{Pauluis2011}. The impact of water vapor on the production of kinetic energy in deep convection can be described as a steam engine and it is to lower the maximum possible amount of work which can produced by an equivalent Carnot cycle \citep{Emanuel, Renno}  acting between the same temperature reservoirs.} An indirect  estimate of (\ref{hyd}) can be obtained from the entropy budget for  water $$\dot{S}_W=\int_{V_W} dV\rho_W\dot{s}^w=\int_{V_W} dV\rho_W\frac{\dot{q}_W}{T}+ \dot{S}_{hyd}$$  as discussed in \citet{Pauluis0}, where $\dot{S}_W$ is the rate of change of entropy of water  and $\dot{q}_W$ the neat heating amount of heat per time that the water substance receives from its environment (i.e. through evaporation and condensation). At steady state $\overline{\dot{S}_W}=0$ and so $\overline{\dot{S}_{hyd}}=\int_{V_W} dV\overline{\rho_W\dot{q}_W/T} \approx 37$ mW\,m$^{-2}$\,K$^{-1}$ \citep{Pascale11}.  Therefore, it is possible to compute the material entropy production by considering the exclusively heat exchanges and the temperature at which such exchanges take place, thus bypassing the need for looking into the complicated details of phase separation processes. 

\textcolor{black}{Furthermore, in climate models  aphysical  entropy sources  due to diffusive/dispersive numerical advection schemes  and parameterizations are also present \citep{Johnson97,Egger}. In particular, \citet{WoollingsThuburn} showed that dispersive dynamical cores can lead to negative numerical entropy production. More generally, it has been argued that parameterizations of sub-grid turbulent fluxes of heat, water vapor and momentum should conform to the second law of thermodynamics, and therefore should lead to locally positive definite entropy production, this being generally not the case \citep{Gassmann:2013}.}

\subsection{Applications and future perspectives}
\label{app}
\subsubsection{Auditing Climate Models}

\textcolor{black}{At steady state, we have that   $\overline{\dot{S}}=0$. Hence, from Eq. \ref{claus1}  we derive:
\begin{equation} 
\overline{\dot{S}_{mat}}  =  \int -dV \overline{\frac{\nabla \cdot \mathbf{F_R}}{T}}.
\label{claus2}    
\end{equation} 
Usually, this is referred to as  \textit{indirect formula} for computing the material entropy production \cite{Goody00}, because it provides an alternative way for estimating the material entropy production of the geophysical fluid  by only looking at the correlation between radiative heating rates and temperature fields).} Therefore, this formula allows for computing the material entropy production due to fluid motions bypassing all the complex fluid dynamical behavior of the system. See  \citet{lucapasca13} for an in-depth discussion of different ways for computing the material entropy production and of the effect of coarse graining the thermodynamic fields. Starting from Eq. (\ref{claus2}) it is possible to derive for Earth conditions an approximate formula for the long term average of the material entropy production, and to disentangle the contributions due to horizontal and vertical processes \citep{Lucarini11} as $\overline{\dot{S}_{mat}} \approx \overline{\dot{S}_{mat}^v} + \overline{\dot{S}_{mat}^h}$, where
\begin{equation}
\overline{\dot{S}_{mat}^h} = - \int_A dA {\frac{\boldsymbol{\nabla}_h \cdot \overline{\boldsymbol{\Upsilon}} }{T_E}} =  -\int_A dA \frac{\overline{F_{R}^{TOA}}}{T_E}
\label{eq:shor}
\end{equation}
where $\boldsymbol{\Upsilon}=\int dz \rho(z)\mathbf{J_h}$ is the vertically integrated atmospheric enthalpy flux introduced in Eq. (\ref{energy1}), 
$F_{R}^{TOA}=\overline{F_{R}^{TOA,SW}}-\overline{F_R^{TOA,LW}}$, where $SW$ and $LW$ refer to the short- and long-wave contributions, respectively, and $T_E=\left(\overline{F_R^{TOA,LW}}/\sigma\right)^{1/4}$ is the emission temperature at a given location. The contribution to the material entropy production coming from vertical processes can instead be written as: 
\begin{equation}
\overline{\dot{S}_{mat}^v} = \int_A dA \left( \overline{F_{R}^{surf}}\right) \left( \frac{1}{T_s}- \frac{1}{T_E}\right) 
\label{eq:sver}
\end{equation}
where $\overline{F_{R}^{surf}}=\overline{F_{R}^{surf,SW}} +  \overline{F_{R}^{surf,LW}}$ is the net radiation at surface (defined as positive when the there is a net incoming radiation into the atmosphere) , $SW$ and $LW$ refer to the short- and long wave components, and, $T_s$ is the surface skin temperature defined as $T_s=\left(\overline{F_R^{surf,LW}}/\sigma\right)^{1/4}\sim T_{surf}$ \citep{Lucarini11}. Equations (\ref{eq:shor})-(\ref{eq:sver}) allow one to compute the material entropy production due to internal irreversible processes making use only of 2D radiative fields at the boundaries of the relevant planetary fluid envelope (surface and top of the atmosphere). This makes Eqs. (\ref{eq:shor})-(\ref{eq:sver}) suitable for the post-processing of data hosted in publicly available archives of GCMs output, intercomparison studies, and studies of observational datasets of the Earth and other planets (where radiative data are the only available source of information). Instead, direct computations of  $\overline{\dot{S}_{mat}}$ require the knowledge of the full 3-D time-dependent heating and temperature fields, making their applicability nontrivial for numerical models and unfeasible for observations. 

\begin{figure}[tbhp]
\centering
a)\includegraphics[width=18pc,trim= 2pc 10pc 2pc 10pc,clip]{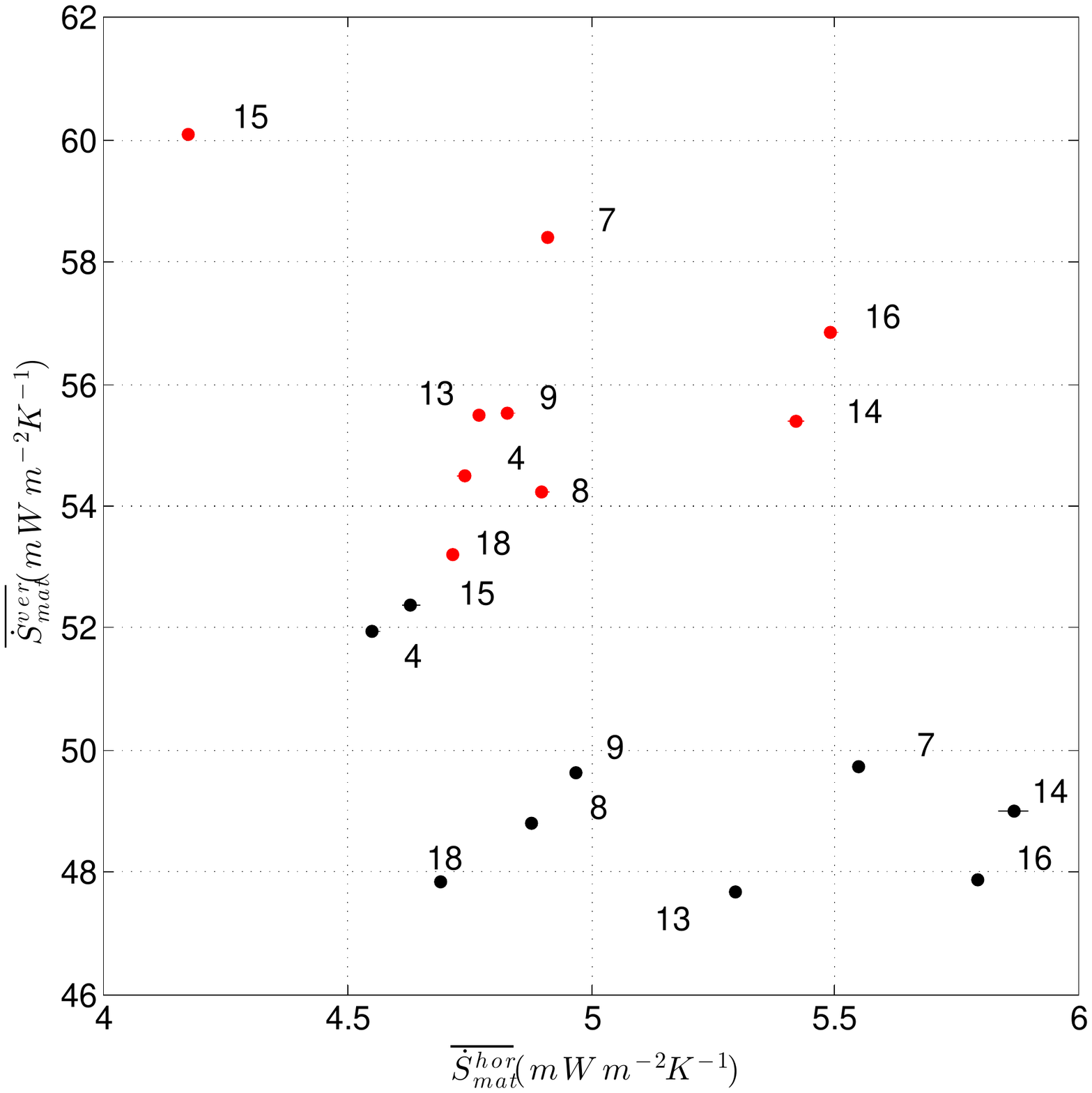} \label{fra3}
b)\includegraphics[width=18pc,trim= 2pc 10pc 2pc 10pc,clip]{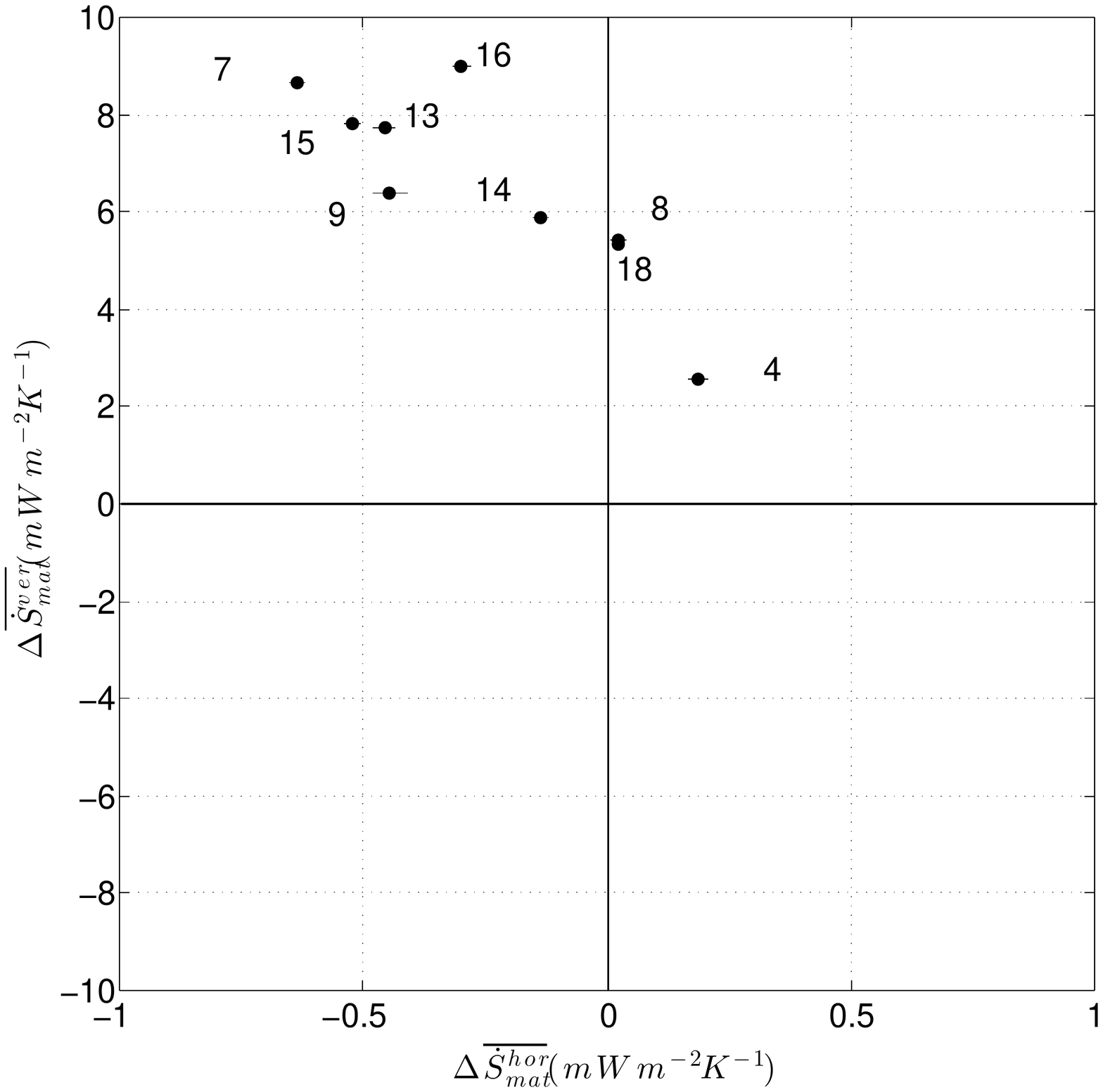} \label{fra4}
\caption{a) {Scatter plot of contributions to the rate of material entropy production due to horizontal (x-axis) and vertical (y-axis) processes. Each point corresponds to a GCM from the CMIP3 dataset in pre-industrial (black) and post-industrial (red) scenarios (updated from \cite{Lucarini11})}. b) Difference between the SRESA1B scenario run (average of the last 30 years of the XXIII century and the pre-industrial climatology.
\label{scatter_smat}
}
\end{figure}

Figure \ref{scatter_smat} shows a scatter-plot of the globally averaged annual mean values of the vertical and horizontal components of the material entropy production computed from the outputs of several GCMs from the CMIP3 dataset in pre-industrial and post-industrial conditions (updated from \cite{Lucarini11}, limiting to the models for which the data availability made possible the comparison). The post-industrial case corresponds to the first 100 years after the stabilization of the $CO_2$ in the A1B climate change scenario. Issues related to the effective non-stationarity of the system have been treated as in \cite{LucariniRagone}. 
 
Comparing with the direct computation \citep{Pascale11} of $\overline{\dot{S}_{mat}}$ for the case of Had-CM3 (model 13) in the pre-industrial case shows that the relative error on the estimate is less than 5\% \citep{Lucarini11}. The typical values of the annual material entropy production in pre-industrial conditions range for most models between 47 and 53 $mW m^{-2} K^{-1}$, matching well the approximate estimate by \cite{Ambaum}. The contribution due to vertical processes is dominant by about one order of magnitude with respect to the contribution due to horizontal processes. This suggests that from the point of view of the entropy production, the climate system approximately behaves as a collection of weakly coupled vertical columns where mixing takes place \citep{lucapasca13}.

In increased CO$_2$ concentration conditions, the rate of material entropy production increases for all the models between 10\% and 20\%. Such a change is dominated by the increase in the vertical component, while the horizontal component sees in most cases a reduction of up to 10\%, despite the fact that large scale horizontal enthalpy transports increase for all models \citep{LucariniRagone}. This implies - see Eq. (\ref{eq:shor}) - a projected strong reduction in the large scale gradients of emission temperature, thus suggesting that in warmer conditions the climate system becomes more homogeneous in terms of meridional and zonal temperature differences. This fits well with what reported in \citep{LucACP,Luchyst} in terms of climate response to global warming-like conditions, and hints to a dominant role of the latent heat release due to convective processes in the response to the climate change \citep{LucariniRagone}.

\begin{figure}
\centering   a)\includegraphics[width=18pc,trim= 12pc 1pc 10pc 1pc,clip]{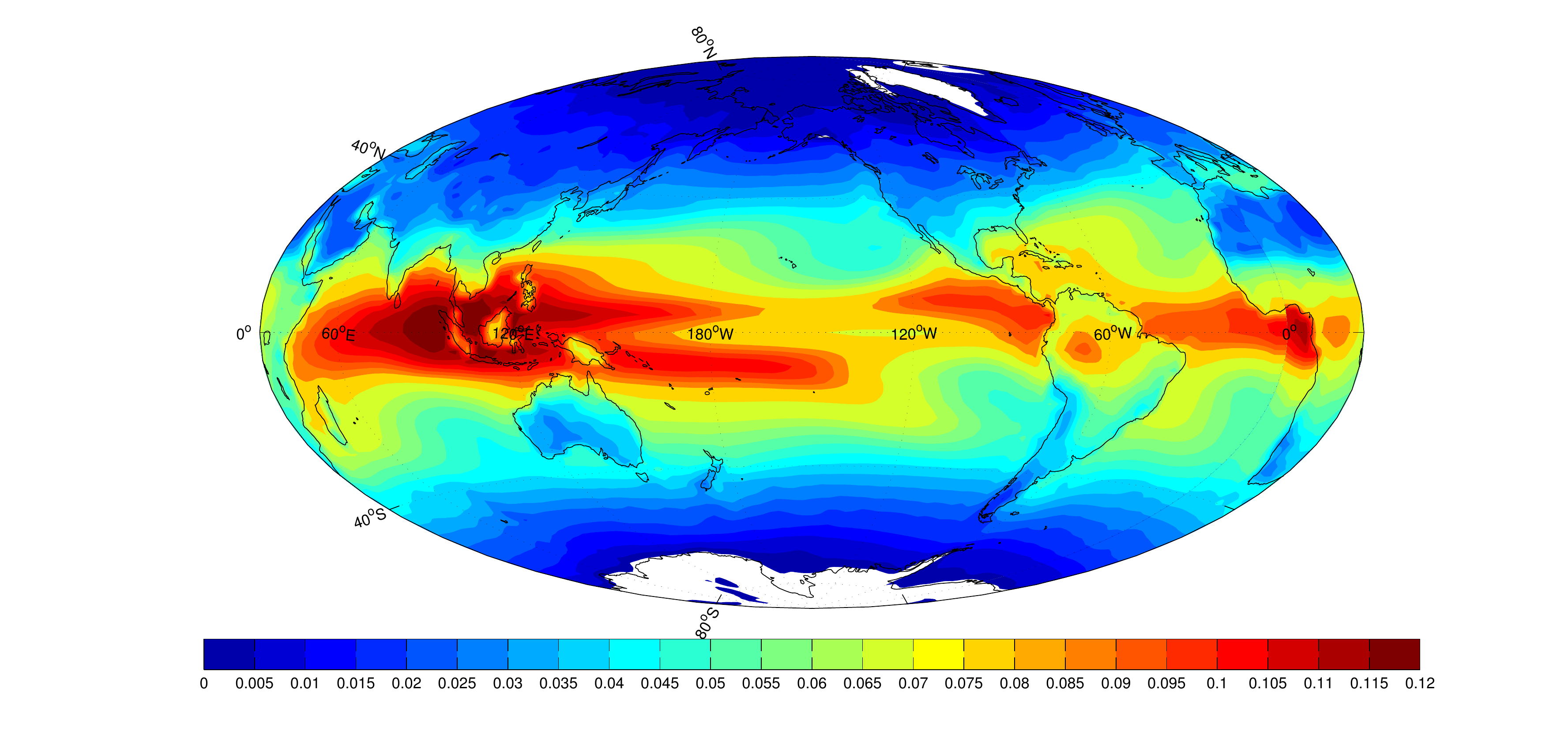}\\
  b) \includegraphics[width=18pc,trim= 12pc 1pc 10pc 1pc,clip]{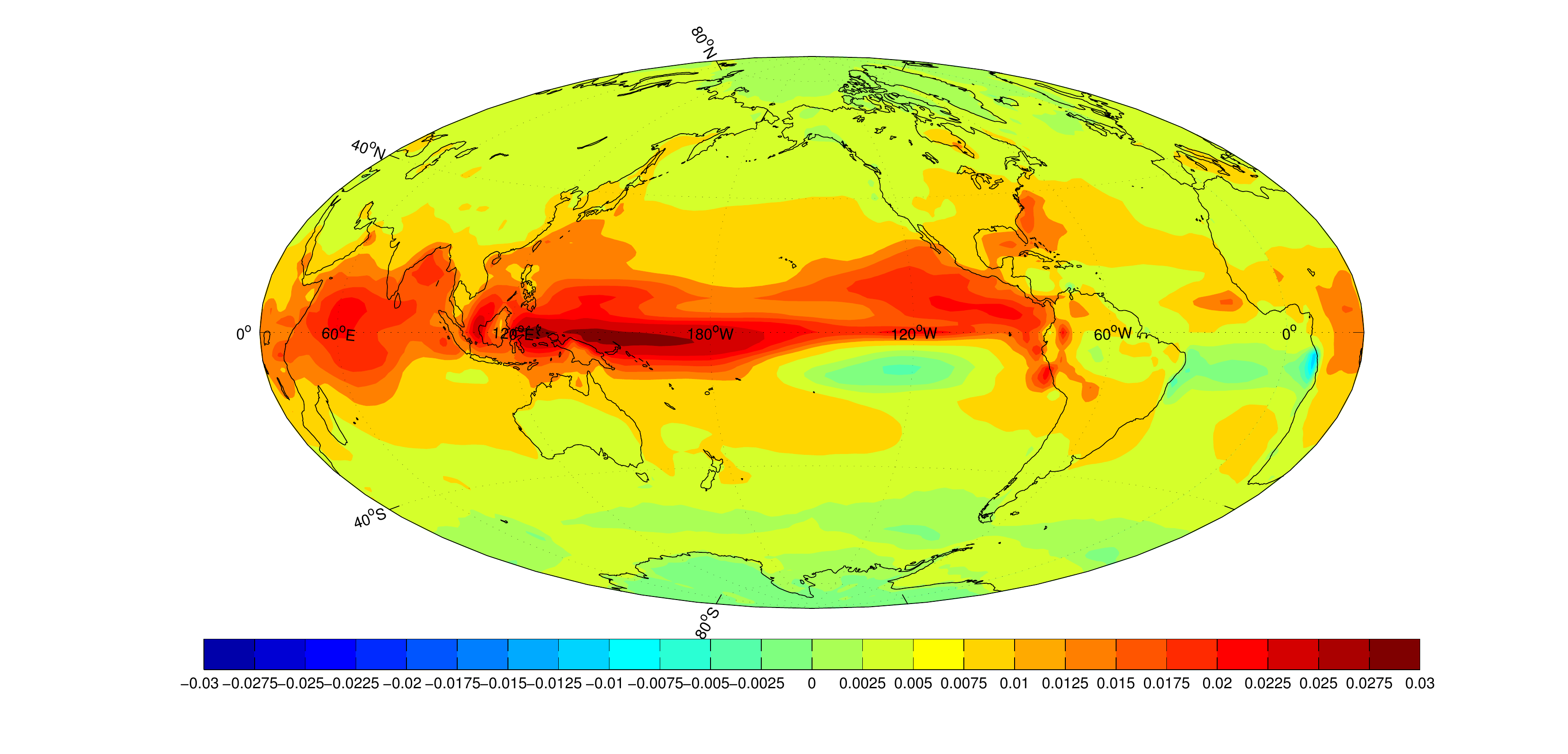}
\caption{(a) Spatial distribution of the contribution to the rate of material entropy production due to vertical processes in pre-industrial scenario for Had-CM3 (model 13 in figure \ref{scatter_smat}). (b) Anomalies in the post-industrial scenario with respect to the pre-industrial case for the same model (updated from \cite{Lucarini11}).}
\label{map_svert}
\end{figure}

\textcolor{black}{Figure \ref{map_svert}a shows the spatial distribution of the integrand of Eq. (\ref{eq:sver}) for the Had-CM3 model in pre-industrial conditions \citep{Lucarini11}, \textit{i.e.},the local contribution to the vertical component of material entropy production. Overall, the local material entropy production due to vertical processes seems to be a good indicator of the geographical distribution of convective activity: the highest values are observed in the warm pool of the western Pacific and Indian Ocean and in land areas characterized by warm and moist climates, while relatively low values are instead observed in the cold tongue of the eastern Pacific, near western boundary currents, and in the temperate and cold oceans, as well as on deserts and middle and high latitudes of terrestrial areas. Note that also in this case the role of latent heat releases is fundamental in determining the characteristics of the system, showing how the hydrological cycle is a crucial component of a thermodynamically consistent representation of the climate system.}

Figure \ref{map_svert}b shows the difference between the post-industrial and pre-industrial cases. The local vertical component of material entropy production increases almost everywhere, with negative anomalies confined  to polar regions and to limited areas of the Southern Hemisphere, with very small values. The positive anomalies are extremely high in the tropical regions, particularly in the eastern and western Pacific ocean. Note that the pattern of increase does not strictly follow the pattern of the absolute value in the pre-industrial case. \textcolor{black}{In particular the maximum of the increase is located eastward to the maximum of the entropy production in the pre-industrial case, a signature of a shifting of the warm pool and a modification of the Walker circulation \citep{Bayr,DiNezio,ipcc2013}. High values are also found in the Indian ocean, suggesting an increase of the convective activity connected with the Monsoon \citep{monsoonnat,ipcc2013}. }Significant local maxima are also observed in the Gulf of Mexico and along the Gulf Stream, and in the Mediterranean Sea. 

The local entropy production due to vertical processes behaves as a robust indicator of the impact of the climate change on large-scale features connected to convective activity. The pattern of increase is correlated to the pattern of variation of the surface temperature only to a minor extent. The reason is that this indicator contains in a synthetic way the information of the change in the surface temperature, in the vertical stability of the atmosphere, and in the intensity of the energy fluxes connected to the vertical processes. Therefore, it could be used in order to define robust indexes for large-scale processes for which strong convection is an important component. Moreover, the range of variation due to climate change of the local vertical entropy production is rather high if compared to the range of variation of standard fields like surface temperature or pressure. Therefore, one could expect a better signal-to-noise ratio and a more distinctive signature of climate change from indicators based on this quantity compared with what obtained with indicators based on simpler observables, similarly to what discussed by \citet{LucACP,Luchyst} and by \citet{Boschi} in the context of the identification of multi-stable regimes of the climate system.



\subsubsection{Bistabiliy and tipping points}
\label{bist}

Based on the evidence supported by  \citet{HoffmanSchrag} and from numerical models \citep{Budyko, Sellers, Ghil}, it is expected that the Earth is potentially capable of supporting multiple steady states for the same values of some parameters such as, for example,  the solar constant. Such states are the presently observed \textit{warm} state (W), and the entirely ice covered \textit{Snowball Earth} state (SB). This is due to  the presence of two   disjoint strange (chaotic) attractors.  The W$\rightarrow$SB and SB$\rightarrow$W transitions are due, mathematically,   to the catastrophic disappearance of one of the two strange attractors \citep{Arnold} and, physically, to  the positive ice-albedo feedback. 
The SB condition, which  might be a common feature also of Earth-like planets, hardly allows for the presence of life, so this issue is of extreme relevance  for defining habitability condition in extraterrestrial planets.

\begin{figure}[tbhp]
\centering
 \noindent\includegraphics[width=20pc,trim= 1pc 2pc 0pc 0pc,clip]{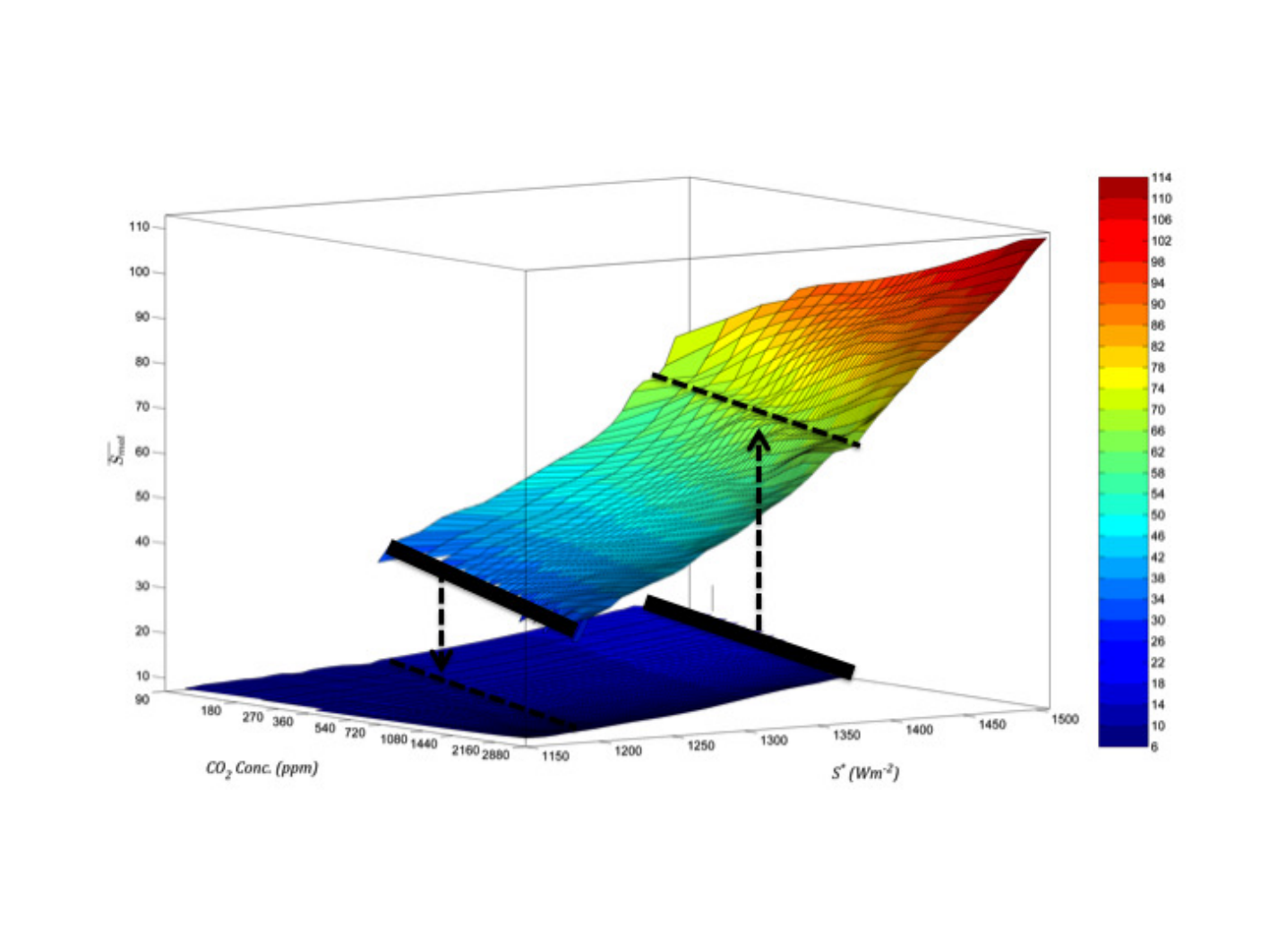}
 \caption{ Material entropy production  (mW\,m$^{-2}$\,K$^{-1}$) as a function of solar constant  $S^*$ and the CO$_2$ concentration. The transition SB$\rightarrow$W and W$\rightarrow$SB are marked with dashed arrows starting from the tipping point  regions (courtesy of Robert Boschi, Universit\"at Hamburg). 
 \label{leaves}
}
\end{figure}

PLASIM \citep{fraedrich2005},  a general circulation model of intermediate complexity, was used by \citet{Boschi} and by \citet{LucAstr} to reconstruct an extensive portion of the region of multistability in the plane described by the parameters $(S^*, [\textrm{CO}_2 ]$). The  surface temperature $T_s (S^*, [\textrm{CO}_2 ])$  is shown in Fig.~\ref{leaves}. The boundary of the domain in the parametric space where two states are admissible correspond to the tipping points of the system.

\begin{figure}[tbhp]
\centering
 a) \includegraphics[width=18pc]{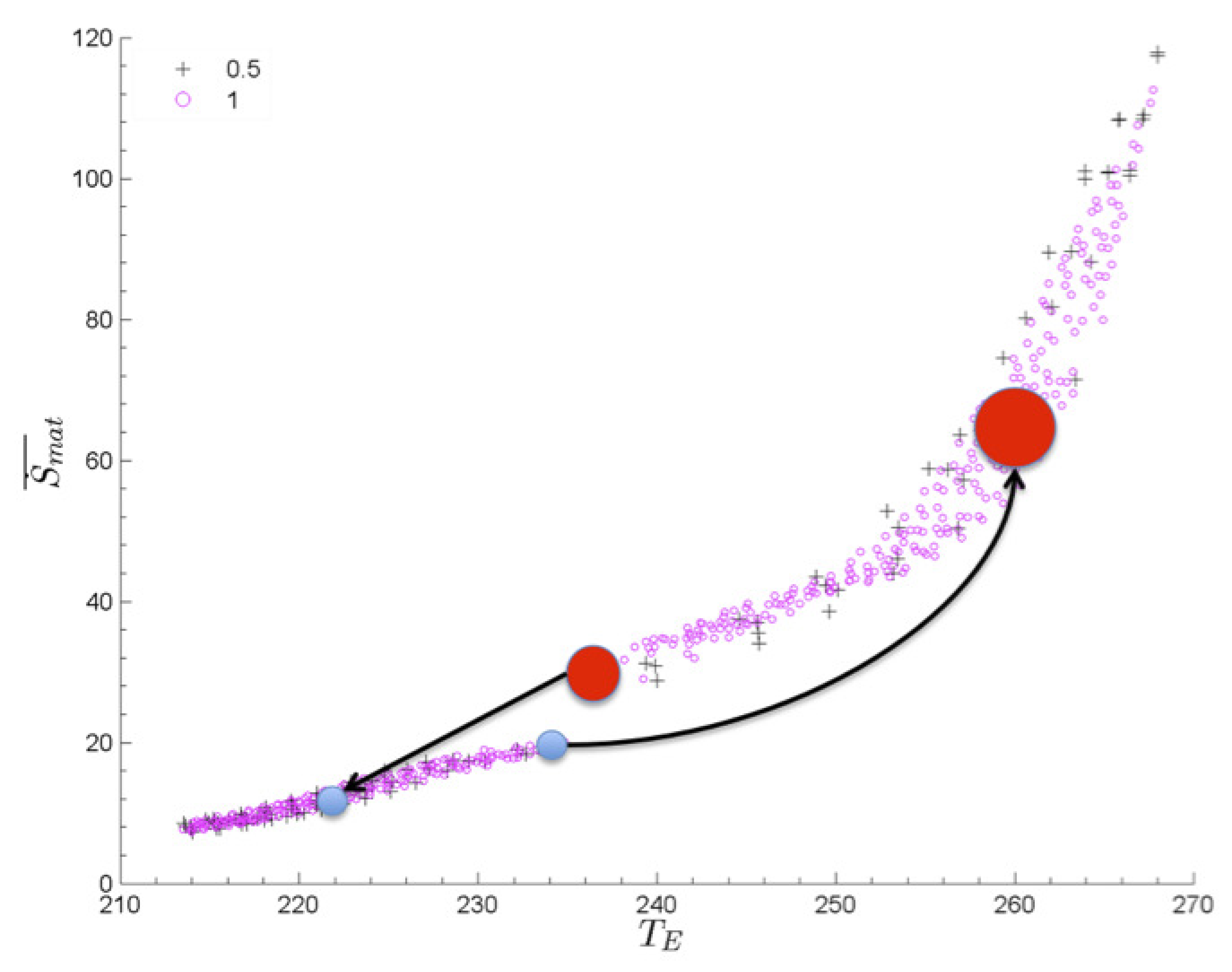} \label{ya_bobby1}\\
b) \includegraphics[width=18pc]{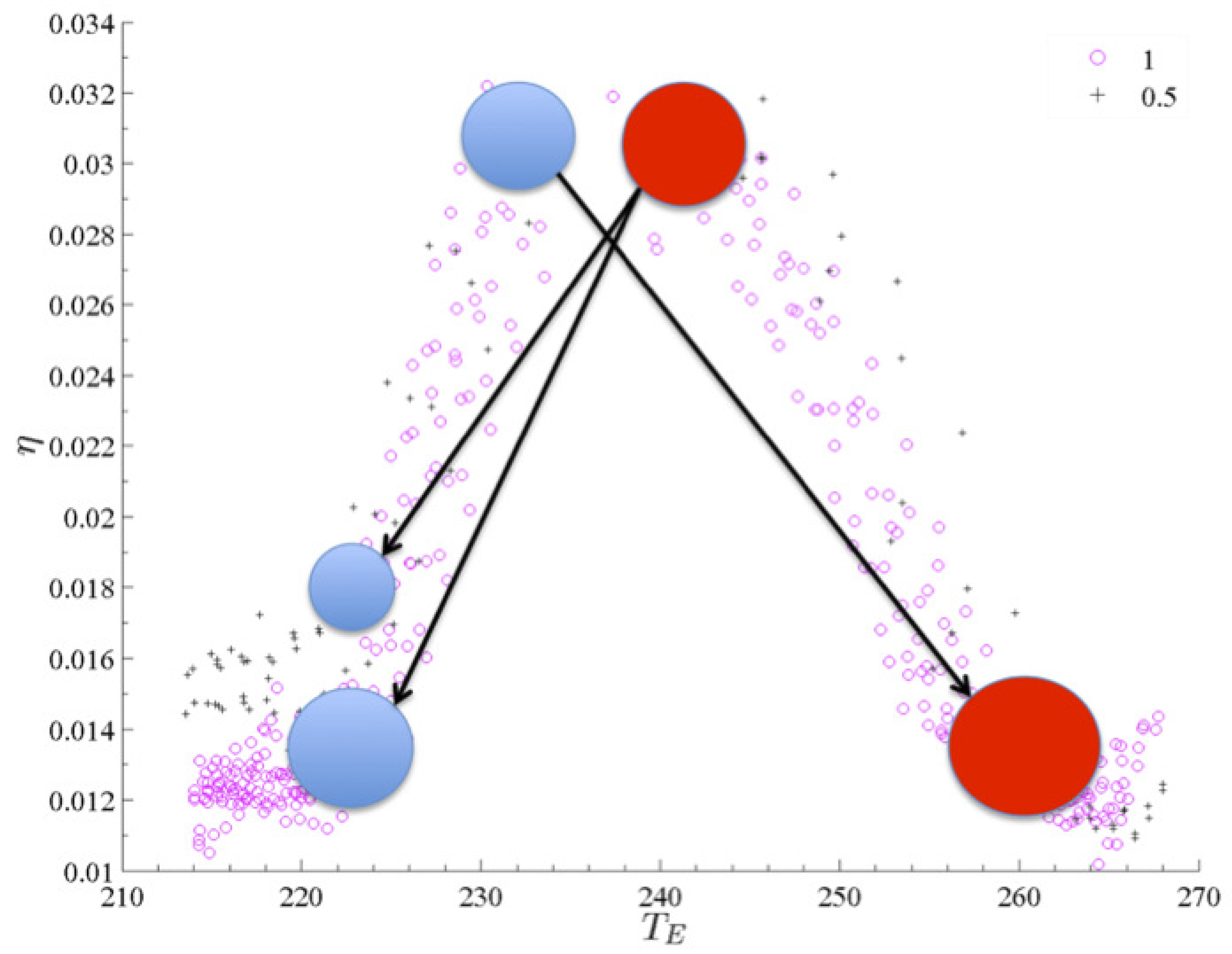} \label{ya_bobby2}
\caption{a): Rate of material entropy production ($10^{-3}$ W\,m$^{-2}$\, K$^{-1}$)  vs. emission temperature $T_E (K)$ for $ \Omega= \Omega_{earth}$ (magenta) and $\Omega = 0.5\Omega_{earth}$ (black). b):  as in a) figure but for efficiency (courtesy of Robert Boschi, Universit\"at Hamburg).
}
\end{figure}

The thermodynamical and dynamical properties of the W and SB states are largely different.  In the W states,  surface temperature are $40-60$ K higher than in the corresponding SB state and    the hydrological cycle dominates the dynamics. This  leads to  a  material entropy production (Fig.~\ref{ya_bobby1})  larger by a factor of 4 -- order of $(40-60) \times 10^{-3}$ W\,m$^{-2}$\, K$^{-1}$ vs. $(10-15) \times 10^{-3}$ W\, m$^{-2}$\, K$^{-1}$  --  with respect to the corresponding SB states \citep{Boschi}.  The SB state is eminently a dry climate, with entropy production mostly due to sensible heat fluxes and dissipation of kinetic energy. 

The response to increasing temperatures of the two states is rather different: the W states feature a decrease of the efficiency of the climate machine, as enhanced latent heat transports reduces energy availability by dampening temperature gradients, while in the SB states the efficiency is increased, because warmer states are associated to lower static stability, which favors large scale atmospheric motions (Fig.~\ref{ya_bobby2}). The entropy production increases for both states, but for different reasons: the system become more irreversible and less efficient in the case of W states, while stronger atmospheric motions lead to stronger dissipation and stronger energy transports in the case of SB states.
A general property which has been found is that, in both regimes, the efficiency $\eta$ increases for steady states getting closer to tipping points and dramatically drops at the transition to the new state belonging to the other attractor (Fig.~\ref{ya_bobby2}). In a rather general thermodynamical context, this can be framed as follows: the efficiency gives a measure of how far from equilibrium the system is. The negative feedbacks tend to counteract the differential heating due to the stellar insolation pattern, thus leading the system closer to equilibrium. At the bifurcation point, the negative feedbacks are overcome by the positive feedbacks, so that the system makes a global transition to a new state, where, in turn, the negative feedbacks are more efficient in stabilizing the system \citep{Boschi}.

Another interesting aspect  is the determination of empirical functional relations  between the main thermodynamical quantities and globally averaged  emission temperature $T_E=(LW_{toa}/\sigma)^{1/4}$, as shown in Fig.~\ref{ya_bobby1}. This  would permit to  express  non-equilibrium thermodynamical properties of the system in terms of parameters which are more directly accessible through measurements \citep{LucAstr}.

\subsubsection{Applications to planetary sciences}
\label{plan}

The discovery of hundreds of planets outside the solar system (exoplanets)  \citep{Seager} is  extending the scope of planetary sciences towards the study of the so-called \textit{exoclimates} \citep{Heng3}. 
A large number of the exoplanets discovered so far  are tidally locked to their parental star,  experiencing extreme stellar forcing on the dayside  where temperature up to $2000$ K can be reached.  Starlight energy,   deposited  within the atmosphere at the planet's dayside,  is then  transported  by atmospheric circulation  to the night side.  Such a system, similarly to the Earth's climate, works like a heat engine (Sect.~\ref{trasp}, Sect. ~\ref{secondlaw}). 

The strength of the day-to-night  enthalpy flux  controls the ratio  of outgoing longwave energy fluxes from the night and day side $\xi= LW_{night}/LW_{day}$, called efficiency of heat redistribution in the astrophysical literature. Observations through infrared light curves  show that  the hotter the planet, the more inefficient is the  atmospheres  at redistributing stellar  energy  leading to  larger day-night temperature differences.   Numerical simulations \citep{Perna} show  that $\xi$ varies between $0.2$ (low heat redistribution) and $1$ (full heat redistribution) and depends critically on the atmospheric optical properties  and  the intensity of the stellar irradiance. \textcolor{black}{Relating this definition of efficiency with the many different definitions used to characterize global circulations \citep{Johnson00, Schubert, Perna, Ambaum} and understanding their differences would be useful to provide a link between energy conversion and energy transport in planetary atmospheres.}

A thorough  understanding of dissipative processes  is fundamental for dealing with planetary atmospheres  \citep{Goodman, Pascale13}.  Dissipative processes  are poorly known on Solar System planets  and    on exoplanets. Let us make some examples. {In hot Jupiters temperatures may be very high ($\ge 1500$ K),  allowing for thermal ionization (governed by the \emph{Saha equation}) and thus   fast-moving (in hot Jupiters winds $\sim 1$ km\,s$^{-1}$)  electric  charges. This  induces an electric current towards the interior of the planet, where energy is then converted into heat by ohmic dissipation.}. Another dissipative mechanism  believed to be a common feature in planetary atmospheres is  shock wave breaking \citep{Batygin, Heng}. \textcolor{black}{Note that the indirect method (Eq. \ref{claus2}) could, in principle, be applied in order to infer information about the dissipative processes in the interior of exoplanets \citep{Schubert}, where  radiative fluxes are the only piece of information we can access.}

\section{Climate Response and Prediction}
\label{ruelle}
In the previous section, we have investigated the climate as a non-equilibrium physical system and have emphasized the intimate relation between forcing, dissipation, energy conversion, and irreversibility. The same approach can be brought to a more theoretical level by taking the point of view of non-equilibrium statistical mechanics. 

Non-equilibrium statistical mechanics provides the natural setting for investigating the mathematical properties of  forced and dissipative chaotic systems, which live in a non-equilibrium steady state (NESS). In this state, typically, the phase space  contracts, entropy is generated, and the predictability horizon is finite. Deviations from this behavior are possible, but extremely unlikely. Conceptually, non-equilibrium steady states are generated when a system is put in contact with reservoirs at different temperatures or chemical potentials, and one disregards the transient behaviors responsible for the relaxation processes \citep{gallavotti2006stationary}. This fits well the description of the non-equilibrium properties of the climate system given in section \ref{prigogine}.

The science behind non-equilibrium statistical mechanical systems is still in its infancy, so that, as opposed to the equilibrium case, we  are not able to predict the properties of a system given the parameters describing its internal dynamics and the boundary conditions, except in special cases where the dynamics is trivial. 

It is then important to choose a suitable mathematical setting for being able to state some useful general results and compare numerical experiments with theory. The mathematical paradigm we will consider is the one of so-called \textit{Axiom A systems} \citep{eckmann85,ruelle89}, which, according to the \textit{Chaotic Hypothesis}  \citep{gallavotti_chaotic_1996}, can be considered as good \textit{effective} models of chaotic systems with many degrees of freedom.

In general, we can say that an (time-continuous) Axiom A system \citep{eckmann85,ruelle89} obeys an  evolution equation of the  form $\dot{x} = F(x)$, $x\in R^n$, and  possesses an invariant measure $\rho(dx)$ supported on its attractor, which is, roughly speaking, the set of points where the system is asymptotically attracted to. 

If forcing and dissipation are present, the attractor is \textit{strange}, \textit{i.e.}, it does not look locally at all like a smooth manifold, so that we cannot write $\rho(dx)=\rho(x)dx$, where $\rho(x)$ is the density. Instead,  in the very intuitive language of Lorenz, it looks like the Cartesian product of a smooth manifold and a fractal set. The smooth manifold corresponds to the unstable directions of the flow, which make the system chaotic, while the Cantor set corresponds to the contracting directions, which result from dissipation. The invariant measure $\rho(dx)$ gives the weight to be used in phase space to compute the expectation of any observable $A$, which agrees, thanks to ergodicity, to the long-time average, so that $$\langle A \rangle =\rho(A)= \int  \rho(dx) A(x) = \lim_{T\rightarrow\infty} \frac{1}{T}\int_0^T dt A(x(t))$$ with probability 1 with respect to the choice of the initial conditions.The invariant measure of an Axiom A system is of Sinai-Ruelle-Bowen (SRB) type \citep{eckmann85,ruelle89,young_what_2002}. This has many consequences, including the fact that the measure is stable against weak stochastic forcing, see also the discussion in \citet{lucarini2012}.

\citet{ruelle_differentiation_1997,Ruelle:1998,ruelle_nonequilibrium_1998,ruelle_review_2009} recently proved that in the case of an Axiom A system, its SRB measure, despite the geometrical complexity of the attractor supporting it, has also an extremely fascinating degree of regularity. In fact, there is a smooth dependence of the SRB measure to small perturbations of the flow, and  it is possible to derive corresponding explicit formulas. 
This approach is
especially useful for studying the impact of changes in the internal parameters of a
system or of small modulations to the external forcing, and various studies have
highlighted the practical relevance of Ruelle theory for studying what we may call
the sensitivity of the system to small perturbations. We will here recapitulate some features of the Ruelle response theory and argue that it is a potentially useful tool for studying various classes of GFD problems, and, most notably for addressing rigorously and in an unified perspective  climate change prediction, climate response, and climate sensitivity.

\subsection{Response formulas and Fluctuation-Dissipation Theorem}
Let us consider an Axiom A dynamical system whose evolution equation can be written as $\dot{x} = F(x)$ and let's assume that it possesses an invariant SRB measure $\rho^{(0)}(dx)$.  \citet{ruelle_differentiation_1997,Ruelle:1998,ruelle_nonequilibrium_1998,ruelle_review_2009} has shown that if the system is weakly perturbed so that its evolution equation can be written as:
\begin{align}
\dot{x} = F(x)+\Psi(x)T(t)
\label{perturba}
\end{align}
where $\Psi(x)$ is a weak time-independent forcing and $T(t)$ is its time modulation, it is possible to write the modification to the expectation value of a general smooth observable $A$ as a perturbative series:
\begin{equation}
\rho(A)_t=\sum_{n=0}^\infty \rho^{(n)}(A)_t,
\end{equation}
where $\rho^{(0)}(A)_t=\rho^{(0)}(A)$ is the expectation value of $A$ according to the unpertubed invariant measure $\rho^0$ related to the dynamics $\dot{x} = F(x)$, while $\rho^{(n)}(A)_t$ with $n\geq1$ represents the contribution due to $n^{th}$ order  processes \cite{lucarini08}. %

Limiting our attention to the linear case we have:
\begin{equation}
\rho^{(1)}(A)_t = \int^{+\infty}_{-\infty}\textrm{d}\tau_1 G_{A}^{(1)}(\tau_1)T(t-\tau_1),
\label{risposta}
\end{equation}
where the first order Green function can be expressed as follows:
\begin{equation}
G^{(1)}_A(\tau_1)=\int \rho^0({d}x)\Theta(\tau_1)\Psi(x)\cdot \nabla A(x(\tau_1)),
\label{greenkubo}
\end{equation}
where $\Theta$ is the usual Heaviside distribution ($\Theta(x)=1$ if $x>0$, $\Theta(x)=0$ if $x<0$), whose derivative is the Dirac's delta. Equations \ref{risposta} and \ref{greenkubo} are key ingredients for studying climate response. Before continuing in this direction, 
we want to use these equations to discuss the celebrated Fluctuation-Dissipation Theorem (FDT) \citep{kubo57}. 

In systems possessing a smooth invariant measure (which, as discussed above, is \textit{not} typically the case for Axiom A systems), like when equilibrium conditions apply or stochastic forcing is imposed,  we can write $\rho^0(dx)=\rho^0(x)dx$, where $\rho^0(x)$ is the so-called \textit{density}. In this case, we can rewrite Eq. (\ref{greenkubo}) as follows:
\begin{align}
&\rho^{(1)}(A)_t=\int^{+\infty}_{-\infty}\textrm{d}\tau_1\Theta(\tau_1)\nonumber \times \\
&  \int dx \rho^0(x) B(x)A(x(\tau_1))T(t-\tau_1),
\label{FDTgen}
\end{align}
where $B(x)=-\nabla \cdot \left(\rho^0(x)\Psi(x)\right)/\rho^0(x)$. In other terms, one can predict the response at any time horizon $t$ from the knowledge of the lagged correlation between the chosen observable $A$ and the observable $B$, which depends on the invariant measure $\rho^0$ and on the perturbation vector field $\Psi$. See \citet{colaluca13} for a detailed discussion on the physical meaning of $B$. Equation (\ref{FDTgen}) provides a very general form of the FDT \citep{Ruelle:1998,vulpiani2007}, which extends the  results by \cite{kubo57}. Recently, the  FTD for system possessing a smooth invariant measure result has been extended to the nonlinear case \citep{lucarinicolangeli2012}.

The more common forms of the FDT can be obtained by taking one or more of the following assumptions:
:\begin{itemize}
\item the perturbation flow is the form $\Psi(x)=\epsilon \hat{x}_i$;
\item the observable is of the form $A(x)=x_j$.
\end{itemize}
where $x_k$ is the $k^{th}$ component of the $x$ vector and $\hat{x}_k$ is the corresponding unit vector. In this case,  Eq. (\ref{FDTgen}) takes the form:
\begin{align}
& \rho^{(1)}(x_j)_t =-\epsilon \int^{+\infty}_{-\infty}\textrm{d}\tau_1\Theta(\tau_1) \times \nonumber \\ &  \int dx \rho^0(x) \partial_i \log[\rho^0(x)] x_j(\tau_1)T(t-\tau_1),
\end{align}
If one takes the additional simplifying assumption that unperturbed invariant measure has a Gaussian form, so that $\rho^0(x)=1/Z \exp(-\tilde\beta\sum_{j=1}^N x_j^2/2)$, where $\tilde\beta>0$ and $Z$ is a normalizing factor, we obtain:
\begin{align}
&\rho^{(1)}(x_j)_t =\epsilon \tilde\beta \int^{+\infty}_{-\infty}\textrm{d}\tau_1\Theta(\tau_1) \int dx \rho^0(x) x_i  x_j(\tau_1)T(t-\tau_1) \nonumber \\ &=\epsilon \beta \int^{+\infty}_{-\infty}\textrm{d}\tau_1\Theta(\tau_1) C_{i,j}  (\tau_1)T(t-\tau_1),
\end{align}
where $C_{i,j}$ is the lagged  correlation  between $x_i$ and $x_j$ in the unperturbed state. 

Unfortunately, the link between linear response of the system to external perturbations and its
internal fluctuations seems more elusive when the unperturbed state has a singular invariant measure. \citet{ruelle_review_2009} shows that since the unperturbed invariant  $\rho^{(0)}(dx)$ is singular,
the response of the system contains two contributions, such that the first may
be expressed in terms of a correlation function evaluated with respect to the
unperturbed dynamics along the space tangent to the attractor (unstable manifold) and is formally identical to what given in Eq. (\ref{FDTgen}). This part of the response decays rapidly due to decay of correlations due to chaos. On the other hand, the second term, which has no equilibrium counterpart, depends on the dynamics along the stable manifold, and,
hence, it may not be determined from the unperturbed dynamics and is also quite
difficult to compute numerically. 
These properties suggest the basic fact, already suggested heuristically by  \citet{lorenz79}, that  in the case of non-equilibrium systems internal and forced fluctuations of the system are not equivalent,  the former being restricted to the unstable manifold only.

Despite such a serious mathematical difficulty, the application of FDT, even in extremely simplified, quasi-Gaussian, approximation, has enjoyed a good success in climate  \citep{langen_estimating_2005,gritsun_climate_2007} even if it is clear that the ability of FDT in predicting the response to perturbation depends critically on the choice of the observable of interest, on the length of the integrations needed for constructing the approximation of the invariant measure, and, of course, on the validity of the linear approximation \cite{cooper_climate_2011,Cooper2013}.

There are, in fact, various ways to circumvent the problem of the rigorous non-equivalence between forced and free fluctuations. Apart from the obvious smoothing  effect due to unavoidable physical or numerical noise,  when considering smooth, coarse-grained observables(like this of climatic interest), one expects to see little influence of the fine structure of the invariant measure of chaotic deterministic systems, as projections from high-dimensional spaces to lower dimensional ones are involved\citep{marconi2008} and coarse-graining effects can be invoked \citep{wouters_multi-level_2013}. One expects that the FDT will perform better  in predicting the response of the system if one considers as observable $A$ quantities like  the globally averaged surface temperature rather than, \textit{e.g.}, the surface temperature in an individual grid point. Further comments can be found at the end of  section \ref{morizwanzig}

\subsection{Computing the Response}
\subsubsection{Spectroscopic method}
\label{spectro}
\begin{figure}
\centering
 \noindent\includegraphics[width=20pc]{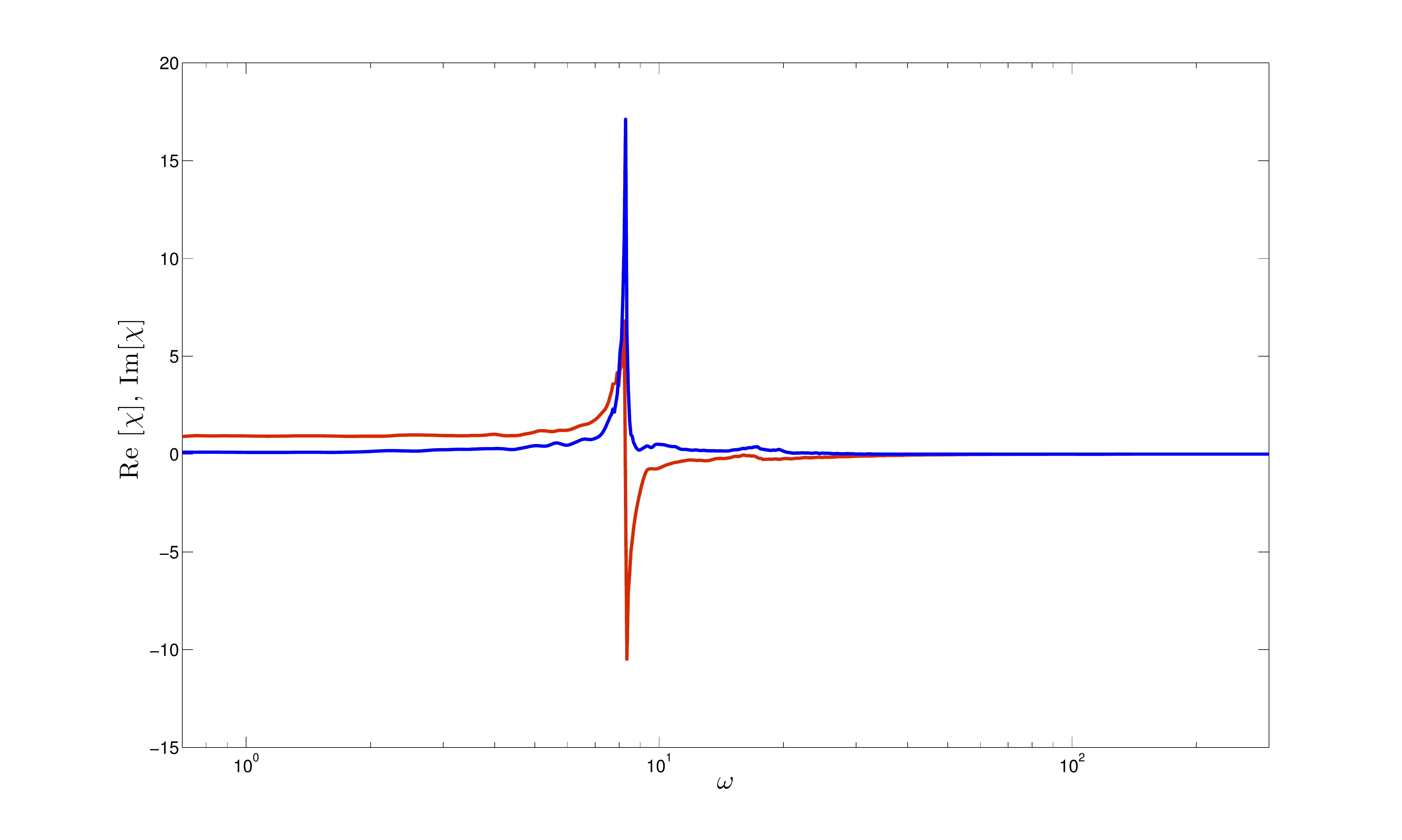}
 \caption{Measured real (blue line) and imaginary (red line) part of the susceptibility $z$ variable of the Lorenz 63 model. Data from \citet{lucarini09}}. 
 \label{ChiL63}
\end{figure}

\begin{figure}
\centering
 \noindent\includegraphics[width=20pc]{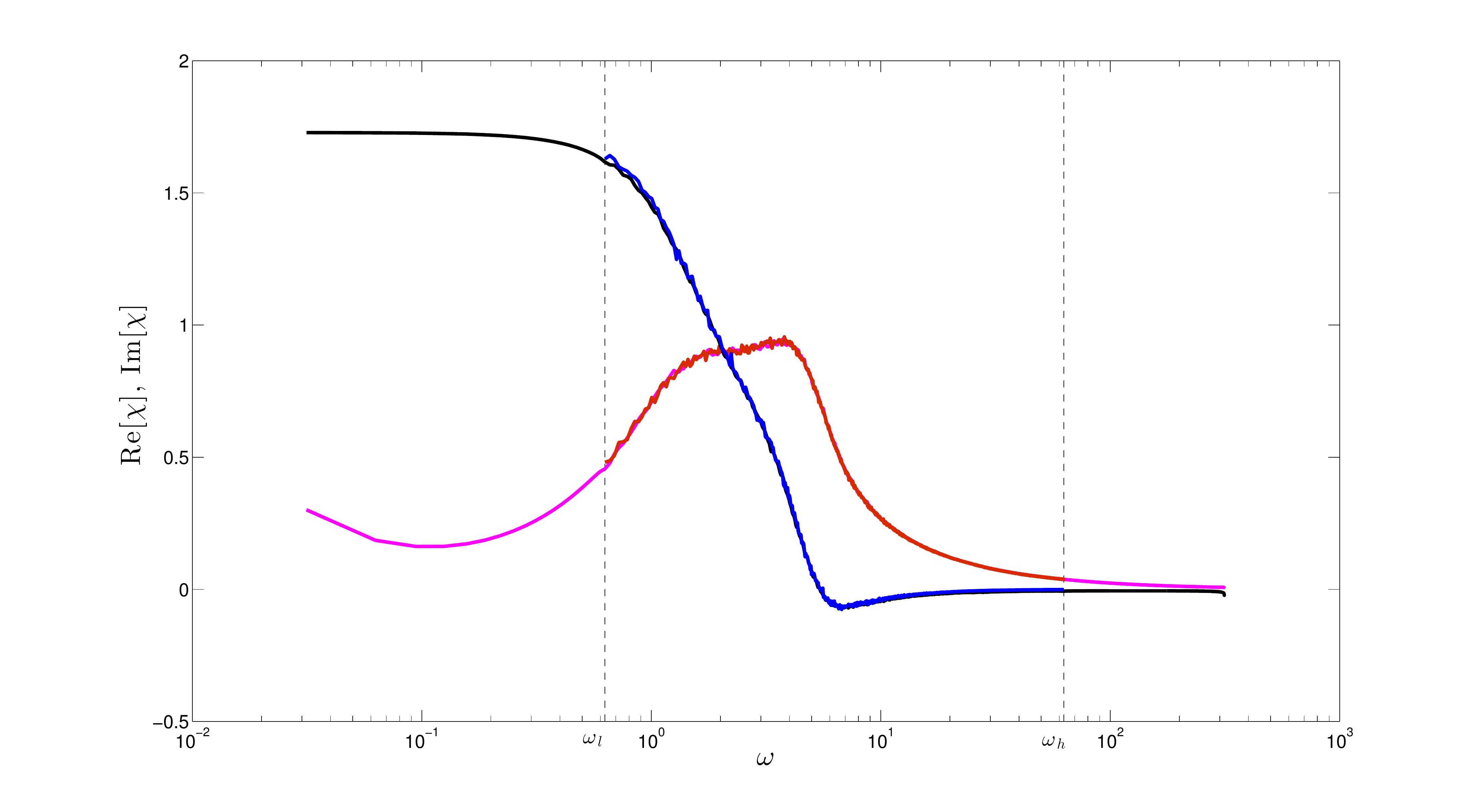}
 \caption{Measured real (blue line) and imaginary (red line) part of the susceptibility for the average energy of the Lorenz 96 model. The rigorous extrapolation of the susceptibility obtained via Kramers-Kronig analysis is reported (real part: black line; imaginary part: magenta line). Data from \citet{Lucarini:2011_NPG}}. 
 \label{ChiEn}
\end{figure}

If we select $T(t )=\epsilon\cos(\omega_0 t)=\epsilon/2(\exp(-i\omega_0 t+\exp(i\omega_0 t))$ as modulating factor of the perturbation field $\Psi(x)$, from equation Eq. (\ref{risposta}) we derive:
\begin{align}
\tilde\rho^{(1)}(A)_t &=\epsilon/2  \int^{+\infty}_{-\infty}\textrm{d}\tau_1 G_{A}^{(1)}(\tau_1)\exp(-i\omega_0(t-\tau_1))\nonumber\\&+\epsilon/2  \int^{+\infty}_{-\infty}\textrm{d}\tau_1 G_{A}^{(1)}(\tau_1)\exp(i\omega_0(t-\tau_1))\nonumber\\
&=\epsilon/2\exp(-i\omega_0 t)\chi_A^{(1)}(\omega_0)+c.c.
\label{risposta2b}
\end{align}
where $\chi_A^{(1)}(\omega_0)$ is the Fourier Transform of $G_{A}^{(1)}(t)$, usually referred to as linear susceptibility, evaluated at frequency $\omega=\omega_0$, and \textit{c.c.} indicates complex conjugate. Therefore, under the hypothesis of linearity, by performing an ensemble of experiments where the forcing is of the form $T(t )=\epsilon\cos(\omega_0 t)$, we can extract the linear susceptibility at frequency $\omega$ by selecting the $\omega_0$ component of the Fourier transform of the signal $\tilde\rho^{(1)}(A)_t $ obtained by taking the ensemble average of the difference between the time series of $A$ in the perturbed and unperturbed case.  By changing systematically the frequency $\omega$ of the forcing, one can reconstruct the susceptibility $\chi^{(1)}_A(\omega)$ on a chosen interval of frequencies. It is useful to recapitulate some useful features of the susceptibility:
\begin{itemize}
\item Resonances in the susceptibility function correspond to spectral ranges where the system is extremely sensitive to forcings. In Fig. \ref{ChiL63} we  show the real and imaginary part of the susceptibility for $z$ variable of the \cite{Lorenz:1963}  model  for the \textit{classical} values of the parameters ($m=1$, $\sigma=10$, $r=28$, $\beta=8/3$) and a given choice of the forcing ($\Psi(x)=[0; x; 0]^\top$, $T(t)=2\epsilon\cos(\omega t)$). We find that for $\omega\sim8.3$, a very peaked spectral feature is apparent. Such a resonance is due to the Unstable Periodic Orbits (UPO)  of the system with the corresponding period \citep{bruno1994}. UPOs populate densely the attractors of chaotic systems and constitute the so-called skeleton of the dynamics. In the case geophysical flows, UPOs have been associated to modes of low-frequency variability \citep{gritsun2008}. One can, more qualitatively, associate resonance to positive feedbacks acting on time scales corresponding to the resonant frequency.
\item While $|\chi^{(1)}_A(\omega)|$ measures the amplitude of the response of the system to perturbation at frequency $\omega$, $\arctan(\Im\{\chi^{(1)}_A(\omega)\}/\Re\{\chi^{(1)}_A(\omega)\})$ gives the phase delay between the forcing and the response, because  $\Re\{\chi^{(1)}_A(\omega)\}$ ($\Im\{\chi^{(1)}_A(\omega)\}$) gives the component of the response that is in phase (out of phase) with the forcing. Depending on the forcing, on the system, and on the observable, this angle can vary significantly even in a relatively small range of frequencies, as a result of resonances. 
\item The two components $\Im\{\chi^{(1)}_A(\omega)\}$ and $\Re\{\chi^{(1)}_A(\omega)\}$ are connected by integral equations, the so-called Kramers-Kronig relations \citep{arfken,lucarini2005,lucarini08,lucarini09,ruelle_review_2009}. Such relations have their foundation in the causality of the Green function (due to the presence of the Heaviside distribution in Eq. \ref{greenkubo}) and establish a fundamental connection between the response at different time scales: 
\begin{equation}
\Re\{\chi^{(1)}_A(\omega)\}=\frac{2}{\pi}\textrm{P}\int d\omega'\frac{\omega'\Im\{\chi^{(1)}_A(\omega')\}}{\omega'^2-\omega^2};
\end{equation}
\begin{equation}
\Im\{\chi^{(1)}_A(\omega)\}=-\frac{2\omega}{\pi}\textrm{P}\int d\omega'\frac{\omega'\Re\{\chi^{(1)}_A(\omega')\}}{\omega'^2-\omega^2}.
\end{equation}
where P indicates that the integral is taken in principal part \cite{arfken}. In particular one finds that:
\begin{equation}
\Re\{\chi^{(1)}_A(0)\}=\frac{2}{\pi}\int d\omega'\frac{\Im\{\chi^{(1)}_A(\omega')\}}{\omega'},
\label{sensitivity}
\end{equation}
which provides a  link between the static response - the sensitivity - and the out-of-phase response at all frequencies. A large literature exists in optics, acoustics, condensed matter physics, particle physics, signal processing on the theory and on the many applications of Kramers-Kronig relations and on the related \textit{sum rules}, which provide integral constraints related to the asymptotic behavior of the susceptibility \citep{lucarini2005}. 
\end{itemize}
In Fig. \ref{ChiEn} we present the real and imaginary part of the susceptibility of the mean energy $e$ of the celebrated \citet{lorenz_predictability:_1996} model:
\begin{equation}
\label{L96}
\frac{d x_i}{d t}=x_{i-1}(x_{i+1}-x_{i-2})-x_i+F
\end{equation}
where $i=1,2,.....,N$, and the index $i$ is cyclic so that $x_{i+N}=x_{i-N}=x_{i}$, and $e=1/N\sum_{j=1}^N x_j^2/2$. The quadratic term in the equations simulates advection, the linear one represents thermal or mechanical damping and the constant one is an external forcing. 
See details on the experiments in \citet{Lucarini:2011_NPG}, performed using $N=40$ and $F=8$. The system is perturbed by the vector field $\Psi(x)=[1; \ldots; 1]^\top$ modulated by $T(t)=2\epsilon\cos(\omega t)$. The resulting real and imaginary part of $\chi^{(1)}_e(\omega)$ are reported in Fig. \ref{ChiEn}, together with the output of the data inversion performed via Kramers-Kronig relations. Once we obtain the susceptibility, as discussed in \cite{Lucarini:2011_NPG}, it is possible to derive the corresponding Green function by applying the inverse Fourier Transform. This is the first application of the Kramers-Kronig theory in a geophysical context.

\subsubsection{Broadband forcing}
If, instead, we select $T(t)=\delta(t)$, we derive from Eq. (\ref{greenkubo}) that $\rho^{(1)}(A)_t =G^{(1)}_A(t)$, \textit{i.e.},the Green function corresponds to the relaxation of an ensemble of trajectories of the system after a finite displacement along $\Psi(x)$. Obviously, we have that $\tilde{\rho}^{(1)}(A)_\omega =\chi^{(1)}_A(\omega)$, where the $\tilde{}$ symbol, indicates, as customary, that a Fourier Transform has been applied, so that the Fourier Transform of the signal is the linear susceptibility. Therefore, using just one ensemble of experiments where the perturbation is described by an impulsive forcing, we can gather the same information on the response of the system which, in the previous case required an accurate sampling of different frequencies.  

Let us look at the problem from a slightly more general point of view. We apply the Fourier Transform to both sides of Eq. (\ref{risposta}) and obtain:
\begin{equation}
\tilde\rho^{(1)}(A)_\omega = \chi_{A}^{(1)}(\omega)\tilde{T}(\omega)
\label{risposta2}
\end{equation}
Choosing a sine or cosine function with argument $\omega_0 t$ for the function $T(t)$ amounts to selecting as  $\tilde{T}(\omega)$ the sum of two $\delta$'s  centered in $\omega=\pm\omega_0$. Therefore, the input (forcing) allows only a small portion of the information to derived on the system from  the output (response). Let us assume that we choose the modulation $T(t)$ such that $\tilde{T}(\omega)$ is not vanishing for any $\omega$, so that we have  a broadband modulation,  where \textit{e.g.} $|\tilde{T}(\omega)|$ for large values of  $\omega$  decreases  like a power law. 
 If we perform an ensemble of simulations of the forced system, measure  $\tilde\rho^{(1)}(A)_\omega$, we can invert Eq. \ref{risposta} and readily derive: 
\begin{equation}
\chi_{A}^{(1)}(\omega)=\frac{\tilde\rho^{(1)}(A)_\omega}{\tilde{T}(\omega)}
\label{risposta3}
\end{equation} 
Therefore, one single set of experiments is, in fact all we need to do to learn about the linear response properties of the system for the observable $A$. If we want to predict the response at finite and infinite time of the system to  forcing with the same spatial pattern $\Psi(x)$ but with different time modulation $R(t)$, we can derive $G_{A}^{(1)}(t)$  from $\chi_{A}^{(1)}(\omega)$ obtained via Eq. (\ref{risposta3}), and then plug it into Eq. (\ref{risposta}). Alternatively, one can write:
\begin{equation}
\tilde\rho^{(1)}(A)^R_\omega=\tilde\rho^{(1)}(A)^T_\omega \frac{\tilde{R}(\omega)}{\tilde{T}(\omega)}
\label{rispostanew}
\end{equation} 
where the upper indices $R$ and $T$ have been inserted for clarity, and then compute the inverse Fourier transform to derive the response at all times, or, if we  apply the inverse Fourier transform to Eq. \ref{risposta3}, we can compute the response to the $R$ perturbation as:
\begin{equation}
\rho^{(1)}(A)^R_t = \int^{+\infty}_{-\infty}\textrm{d}\tau_1 G_{A}^{(1)}(\tau_1)R(t-\tau_1).
\label{risposta4}
\end{equation}

\subsection{Prediction via Response theory}
The real test of the quality of an experimentally derived linear Green function $G^{(1)}_A$  is the assessment of its ability to support predictions about the system's response to any temporal pattern of forcing $R(t)$. The real benefit of the broadband approach described here relies on exploiting linearity, and so deriving $G^{(1)}_A$from just one ensemble of simulations, each performed with the same modulation $T(t)$. Computing the $G^{(1)}_A$ \textit{per se} might be, in fact of little relevance.

\begin{figure}
\centering
 \noindent\includegraphics[width=20pc]{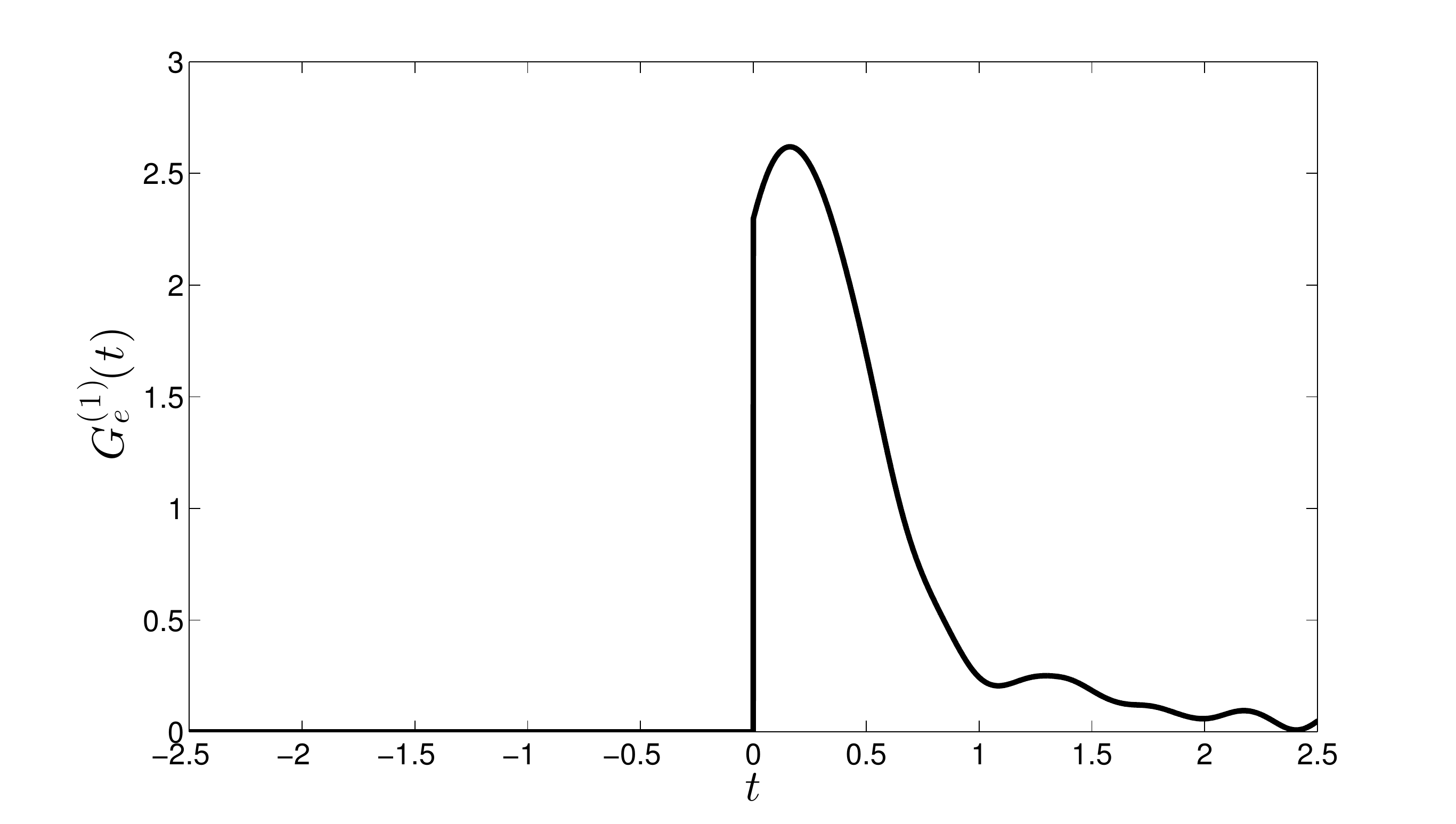}
 \caption{Linear Green function $G^{(1)}_e(t)$  for the average energy $e$ of the \citet{lorenz_predictability:_1996} model obtained by considering a step-like perturbation and using Eq. (\ref{theta}). Compare with Fig. 4 in \citet{Lucarini:2011_NPG}}. 
 \label{G1L96}
\end{figure}

\begin{figure}
\centering
 \noindent\includegraphics[width=20pc]{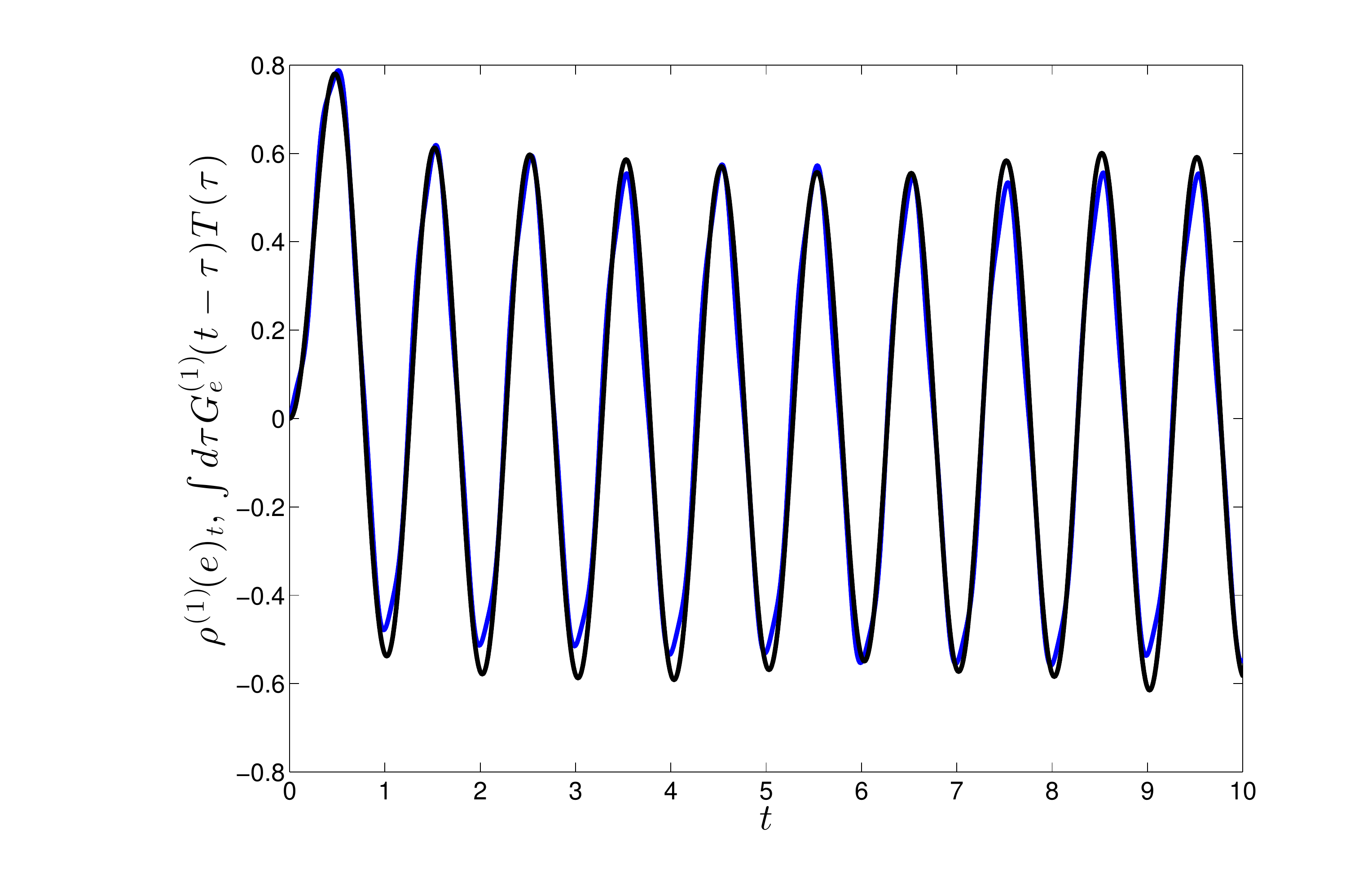}
 \caption{Prediction of finite-time response of the average energy $e$ of the \citet{lorenz_predictability:_1996} model to a forcing with modulation $T(t)=\epsilon \sin(2\pi t)$ ($\epsilon=0.25$). Observed response $\rho^{(1)}(e)_t$ (blue line) vs.  prediction obtained using the linear Green function $G^{(1)}_e(t)$ shown in Fig. \ref{G1L96}. }
 \label{PredeL96}
\end{figure}

At this regard, we have performed additional experiments on the \citet{lorenz_predictability:_1996} model mirroring what presented in  section \ref{spectro}. In this case, we have chosen as time modulation $T(t)=\epsilon \Theta(t)$, whose spectrum is indeed broadband ($\tilde{T}(\omega)/\epsilon=\pi\delta(\omega)+i\textrm{P}[1/\omega]$, where P indicates the \textit{principal part}) \cite{arfken}. In this case, we have:
\begin{equation}
G^{(1)}_e(t)=\frac{d}{dt} \rho^{(1)}(e)_t.
\label{theta}
\end{equation}
Using about 1/100 of the computing time needed in \citet{Lucarini:2011_NPG}, we have produced an estimate of the Green function of comparable quality; see Fig. \ref{G1L96}. Additionally, we decided to check the predictive power of the reconstructed Green function given in Fig. \ref{G1L96} by testing its performance in predicting, through Eq. (\ref{risposta}), the response of the system to a perturbation having temporal pattern given by $T(t)=\epsilon \sin(2\pi t)$ ($\epsilon=0.25$). The results are presented in Fig. \ref{PredeL96}. The agreement between the measured value of $\rho^{(1)}(e)_t$ and the value predicted using $\int d\tau G^{(1)}_e(\tau)T(t-\tau)$ is remarkable. One must emphasize that the agreement is comparable if one selects $\epsilon=1$, thus moving away from the linear regime.

\begin{figure}
\centering
a) \noindent\includegraphics[width=16pc]{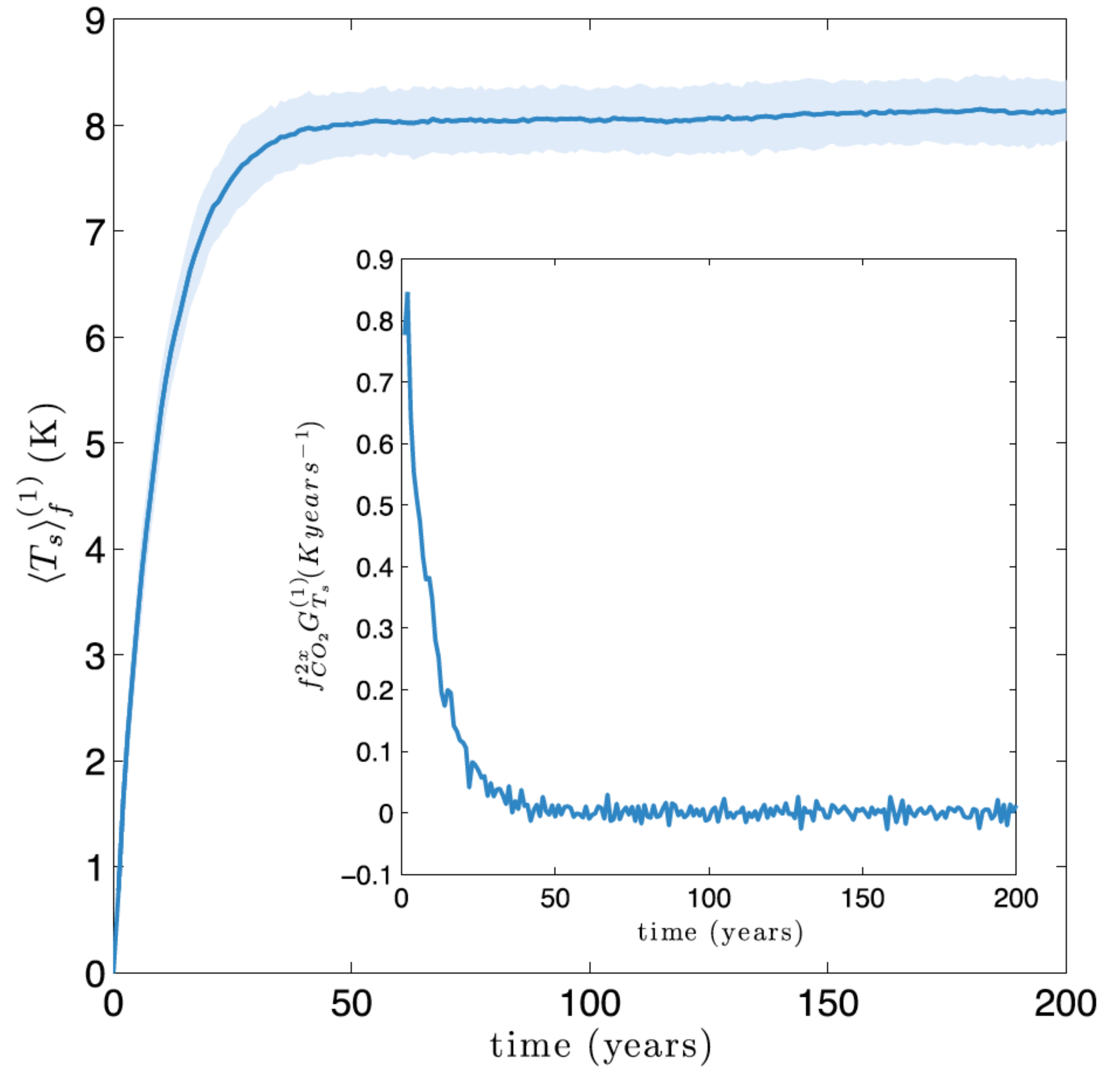}\\
b)  \noindent\includegraphics[width=16pc]{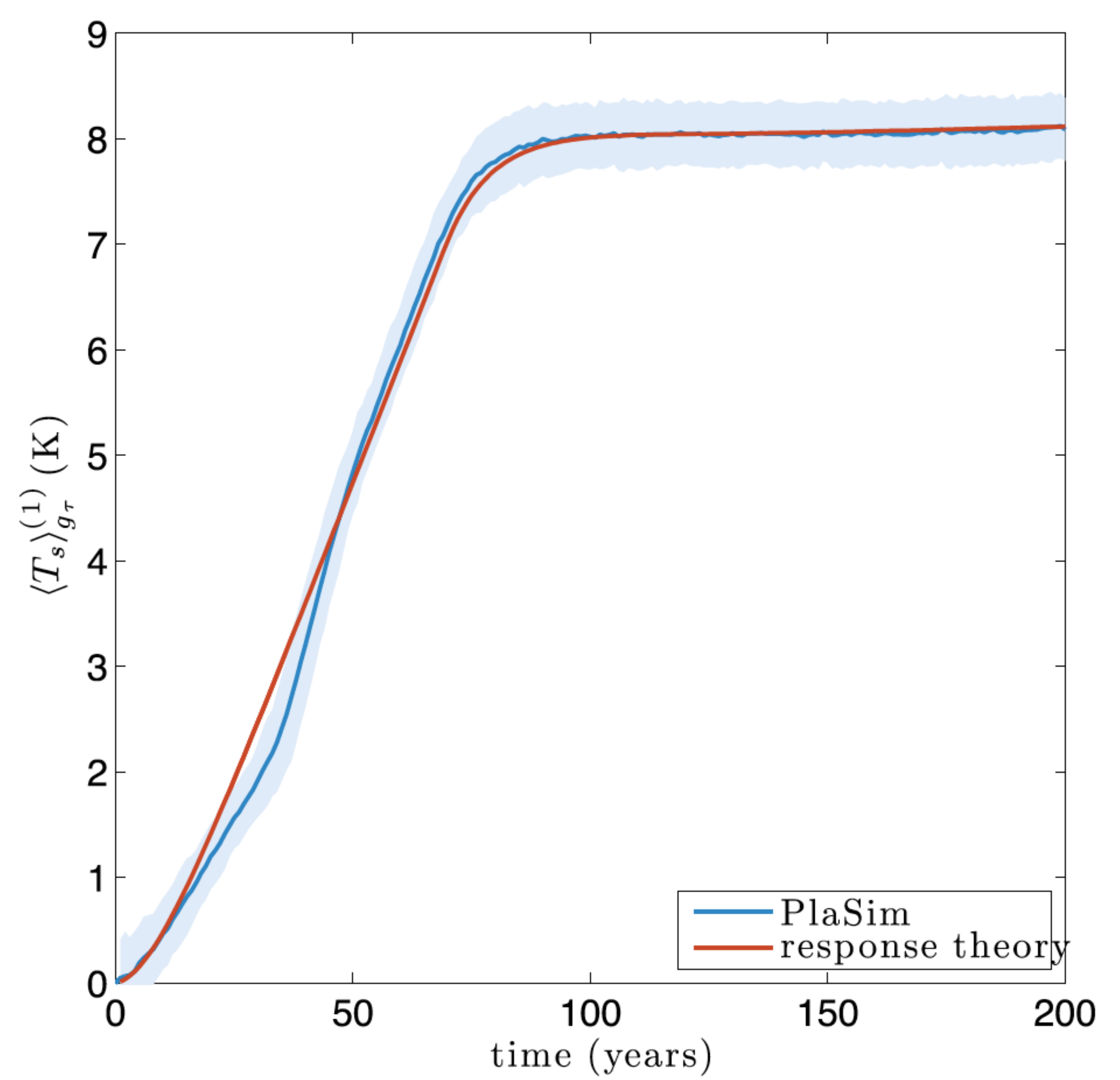}\\

 \caption{Studying climate change using  response theory. a) Change in  $T_S$  after an instantaneous doubling  of the $CO_2$ concentration. The lightly colored band indicates the  two-standard deviation range around the ensemble mean. Insert: Green function of the $T_S$. b) Comparison between GCM simulations (blue) and response theory prediction (red) for 1\% per year increase of the $CO_2$ concentration up doubling. Lightly colored band as in a).}
 \label{Tpred}
\end{figure}

 \subsection{Climate Response, Climate Change prediction}
Let us take inspiration from the previous example in order to get some results of stricter geophysical relevance: we want to perform predictions on the impact of increases in the $CO_2$ concentration on the globally averaged surface temperature as simulated by a climate model, the simplified yet Earth-like  PLASIM \citep{fraedrich2005}.  In what follows we present some new results (see also discussion in \citet{RLL2014}), with the goal of proving the feasibility of the proposed methodology.

\begin{itemize}
\item $\dot{x}=F(x)$ is the system of equations describing the discretized version of a given model of the continuum PDEs describing the evolution of the climate in a baseline scenario with set boundary conditions and values for, \textit{e.g.},  the $CO_2$ concentration and the value of the solar constant.  We assume, for simplicity, that system model does not feature daily or seasonal variations in the radiative input at the top of the atmosphere.  
\item Let us choose for the observable $A$ the globally averaged surface temperature of the planet $T_S$. 
\item We study the perturbed system $\dot{x}=F(x)+f(t)\Psi(x)$. Let us choose as  perturbation field $\Psi(x)$ the convergence of radiative fluxes due to change in the logarithm of the atmospheric $CO_2$ concentration. We want to be able to predict at finite and infinite time the response of the the system to one of the standard $CO_2$ forcing scenario given by the IPCC by performing an independent set of perturbed model integrations. 
\item The test perturbation is modulated by the function $f(t)=\epsilon\Theta(t)$, where $\epsilon$ is such that we double the amount of $CO_2$ concentration in the atmosphere. Our goal is to predict the climate response to the customary $1\%$ increase of $CO_2$ concentration from the baseline value to its double. We select as baseline concentration $[CO_2]=360$ $ppm$.
\item We perform 200 simulations, each lasting 200 years for both scenarios of $CO_2$ forcing.  Our experiments are performed using PLASIM \citep{fraedrich2005} with a T21 spatial resolution, 10 vertical layers in the atmosphere, and swamp ocean having depth of 50 m.    
\end{itemize}
From the time series of the ensemble mean of the change of $T_S$ - $\rho^{(1)}(T_S)_t$ - resulting from the sudden increase in the $CO_2$, we derive the Green function $G^{(1)}_{T_S}(t)$ using Eq. (\ref{theta}). See Fig \ref{Tpred} a). Climate sensitivity is, in fact, defined by Eq. (\ref{sensitivity}). Given the chosen pattern of forcing, we can rewrite is as follows:
 
\begin{equation}
\Delta_T=\Re\{\chi^{(1)}_{T_S}(0)\}=\frac{2}{\pi}\int d\omega'\Re\{\tilde\rho^{(1)}(T_S)_{\omega'}\},
\end{equation}

which relates climate response at all frequencies to its sensitivity. 

In order to test the predictive power of the response theory, we then convolute the Green function with the temporal pattern of forcing of the second set of experiments. 

We choose as test experiment the classical IPCC scenario of 1\% per year exponential increase of $CO_2$ concentration up to doubling of the initial concentration (realized in $\tau\sim 70$ years, and constant concentration afterwards. Since our relevant control parameter is the logarithm of the $CO_2$ concentration, the second pattern of forcing $g(t)$ is, in fact, a  ramp increasing linearly with time  from 0 to $\epsilon$ in $\tau\sim70$ years, with constant value equal to $\epsilon$ for larger times. The results are presented in Fig. \ref{Tpred} b), where we compare the predicted pattern of increase (blue line), obtained using Eq. \ref{risposta4}, with the measured one (black line). The agreement is remarkable, both on the short and on the long time scales, while a some discrepancy exists between 20 and 50 years lead time, where strong nonlinear effects due to ice-albedo feedback are dominant (not shown).
   
Apparently, despite all the nonlinear feedbacks of the climate model, the response to changes in the logarithm of $CO_2$ concentration can be accurately described by linear response theory at all time scales. Nonlinearity in the underlying equations and presence of strong positive and negative feedbacks do not rule out the possibility of constructing accurate methods for predicting the response. In fact, the methods described here could be extended to the nonlinear case by looking at the response in the frequency domain \citep{lucarini08,lucarini09}, even if the data quality requirement is obviously stricter.

The result presented here suggests that many of the scenarios of greenhouse gases concentration included in the IPCC reports \citep{ipcc2001,ipcc2007,ipcc2013} may in fact be partly redundant, as for certain variables  might be accurately described by linear response theory starting from just one scenario.  Equations \ref{risposta3} - \ref{rispostanew} constitute the basis for predicting climate response at all scales. 

Obviously, with a given set of forced experiments, it is possible to derive the sensitivity to the the given forcing for as many climatic observables as desired. It is important to note that, for a given finite intensity $\epsilon$ of the forcing, the accuracy of the linear theory in describing the full response depends also on the observable of interest. Moreover, the signal to noise ratio and, consequently, the time scales over which predictive skill is good may change a lot from variable to variable. The results presented in this section extend to a  more general setting and with stronger foundation the excellent intuition by \citet{hasselmann1993}  on the use of the linear response for addressing the problem of the so-called \textit{cold start} of coupled atmosphere-ocean models.

Here we have shown results from just one observable primary climatic interest. The analysis of other observables will shed light on the mechanisms determining the climate response to the forcing due to changes n the atmospheric composition. As an example, the analysis of the response of large-scale meridional gradients of temperature at surface and in the middle troposphere will provide information on changes in the midlatitude circulation. The existence of approximate functional relationship between the susceptibilities of different observables \citep{lucarini09} would provide the key for defining rigorously the so-called emergent constraints \citep{emergent}.

In practical terms, the applicability of response theory  corresponds to having smooth dependence of climate properties  with respect to some given parameters. Indeed, this is not the case in the vicinity of tipping points (see Fig. \ref{leaves}). Response theory, may, nonetheless, suggest rigorous ways for defining and detecting tipping points, because one expects that these are associated to a divergence of the linear response.

Finally, in order to talk about predictability, we need to specify what are the time scales over which we expect to have satisfactory predictive skills. In fact, linear response theory allows for deriving some scaling laws for addressing this matter. The main obstacle for achieving a good degree of predictability is the uncertainty on the estimate of   response signal given in Eq. (\ref{risposta2}) from the outcomes of the numerical experiments because of the finiteness of the ensemble and of the duration of each numerical simulation. See a detailed discussion of this issue in \citet{RLL2014}.

\section{Multiscale systems and \\parametrizations}
\label{morizwanzig}

The climate system features non-trivial behavior on a large range of temporal and spatial scales \citep{Peixoto:1992,vallis_atmospheric_2006,lucarini_modelling_2011,fraedrich_wavenumber-frequency_1978}
When representing such a complex system in a numerical simulation, the ratio of smallest to largest time scale determines the number of required time steps, and the number of interactions between scales that have to be calculated at each step can increase exponentially with the range of spatial variables. It is therefore clear that, no matter which are the available computing resources, we are able to simulate explicitly only the variables relevant for given ranges of spatial and temporal scales. Different choices of such ranges correspond to different approximate theories of geophysical fluid dynamics aimed at describing specific phenomenologies, a prominent case being that of quasi-geostrophic theory~\cite{klein2010}.

A manifestation of the inability to treat ultraslow variability can be found in the usual practice in climate modeling of choosing fixed or externally driven boundary conditions, such as done when assuming a fixed extent for the land-based glaciers, and, consequently, for the sea-level, or imposing a specific path of $CO_2$ concentration for the atmosphere. Instead, the impossibility of treating accurately fast processes requires the construction of  so-called parametrizations able to account, at least approximately, for the effect of the small scales on the large scales, as a function of the properties of the large scale variables, such in the case of several important physical processes, such as, \textit{e.g.}, deep and shallow atmospheric convection, gravity wave drag, clouds, mixing in the ocean.   

Parametrizing  small scale processes is important because such unresolved processes impact the dynamics of larger scales in terms of error growth, predictability, and climatic biases. Presently, most of the parametrizations used in climate models are deterministic, \textit{i.e.}, for given state of the resolved variables, the effect of the unresolved scale on the resolved scales is uniquely determined. We often refer to these as \textit{bulk parametrizations}. More recently, it has been emphasized that such a point of view should be modified for taking into account the fact that that many different states of the unresolved variables are compatible with a given state of the resolved variables. This leads to considering the possibility of using  stochastic parametrizations \citep{palmer_stochastic_2009}, which show promising abilities in reducing biases and reproducing more effectively the uncertainties associated to performing mode reduction.

\begin{figure}[th]
\centerline{
\hspace{10pt} \includegraphics[width=18pc]{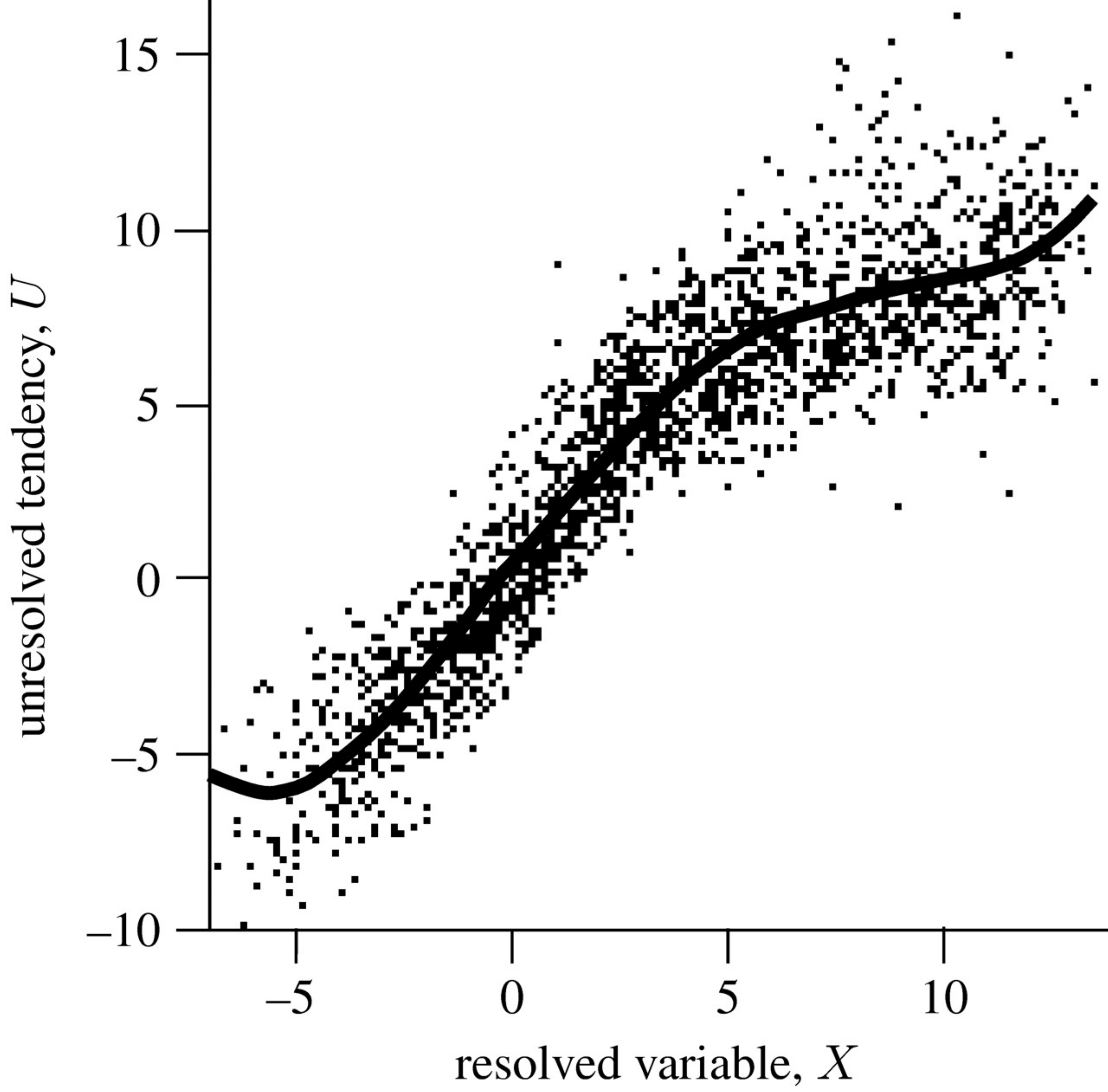}}
\caption{Diagram describing how to parametrize the effect of the fast variables on the tendency of the slow variables $X$. The solid line $U$ - see $y-$ axis - corresponds to $M_X$ in Eq. \ref{para}, while the variability associated to the cloud of points needs to be represented via a stochastic term like $\eta$ in  Eq. \ref{para}. From \citet{wilks_effects_2005}.}
\label{Wilks}
\end{figure}

When large time-scale separation exists between the resolved and non-resolved variables, the problem of parametrization can be cast as follows. We consider a system of the form $\dot{Z}=F(Z)$, $Z\in R^N$ and we divide the state vector $Z=(X,Y)$, where $X$ are the slow components we are interested into and $Y$ are the fast components we want to parametrize.  We rewrite the evolution equation as follows:
\begin{align}
\frac{dX}{dt} &= G_X(X,Y)=F_X(X) + \Phi_X(X,Y) \nonumber \\
\frac{dY}{dt} &= G_Y(X,Y)=F_Y(Y) +  \Phi_Y(X,Y) \label{eq:coupled_dyn_syst_a}
\end{align}
where we have split the dynamics of each set of variables into the autonomous part and into the coupling terms. The basic goal is to be able to write as an equation of the form 
\begin{equation}
\frac{dX}{dt} = F_X(X) + M_X(X)+\eta(X)\label{para}
\end{equation}
where $M_X$ and $\eta$ corresponds to the deterministic and stochastic components of the parametrization, respectively. A now classic example of empirical construction and of testing of stochastic parametrizations is given by \citet{wilks_effects_2005}, see Fig. \ref{Wilks}. 

It must be emphasized that many of the approaches used so far have been based on the existence of a time scale separation between microscopic and macroscopic processes, following, conceptually,  the pioneering point of view proposed by \citet{hasselmann_stochastic_1976}.  If one does assume such a vast time-scale separation between the slow variables $X$ and the fast variables $Y$, averaging and homogenization methods \citep{arnold_hasselmanns_2001, kifer_recent_2004, pavliotis_multiscale_2008} allow for deriving an effective autonomous dynamics for the $X$ variables, able to encompass the impact of the dynamics of the $Y$ variables. The motivation most often stated for the applicability of this theory to climate science is the setting considered by Hasselmann, where fast \textit{weather} systems influence slow \textit{climate} dynamics. 

Unfortunately, in many practical cases of interest in geophysical fluid dynamics, such a scale separation does not exist - see, \textit{e.g.}, the classical study by \citet{Mitchell1976} - so that there is no spectral gap able to support univocally the identification of the $X$ and $Y$ variable. In fact, when the resolution of a numerical model is changed,  all the parametrizations have to be re-tuned, because the set of resolved variables has changed. 

Here we will focus on analytical methods that allow one to derive reduced models from the dynamical equations of a full model. 
Projector operator techniques have been introduced in statistical mechanics with the goal of effectively removing the $Y$ variables. In particular, considerable interest has been raised by the Mori-Zwanzig approach, through which a formal - albeit practically inaccessible - solution for the evolution of the $X$ variables is derived \citep{zwanzig_memory_1961,mori_transport_1965,zwanzig_nonequilibrium_2001}. These equations in general contain both a correlated noise term and a memory term. Some attempts have been made to make approximation to the Mori-Zwanzig projected equations to obtain practically useful equations. In applications of stochastic mode reduction in climate science, the memory term is usually not taken into account. This term could however be very relevant in systems without a time-scale separation, as for example in the parametrization of cloud formation in an atmospheric circulation model. The presence of memory in such systems has been discussed by \citet{bengtsson_stochastic_2013,davies_simple_2009,piriou_approach_2007}. Note that, when we consider coupled systems where asynchronous is used, memory effects are implicitly present in the dynamics.

Besides considering the limit of infinite time scale separation, another point of view can be taken, namely considering the limit of weak coupling between the dynamical processes occurring at different scales. In this limit, the dynamics retains the correlated noise and memory dependence that appeared in the Mori-Zwanzig reduced equations. The advantage of looking at this limit is however that the noise autocorrelation function and memory kernel can now be written as simple correlation and response functions of the unresolved dynamics. 


\subsection{Averaging and homogenization}
\label{sec:averaging}

When applying averaging and homogenization techniques, one considers dynamical systems where a small parameter $\epsilon$ controls the time scale separation between a slow and fast evolution in the system. The prototypical set of equations for such a problem is
\begin{align*}
	\frac{dX}{dt}=& G_X(X,Y) \\
	\frac{dY}{dt} =& G_Y(X,Y)=\frac{1}{\epsilon}\tilde{G}_Y(X,Y)
\end{align*}
The parameter $\epsilon$ controls the time scale separation between the variables $X$ and $Y$, which becomes infinite as $\epsilon \rightarrow 0$.

%
As the time scale separation becomes large, on the typical time scale for the variable $Y$, $X$ will remain almost constant. The fast variable $Y$ will obey an evolution defined by $(X,Y)$ for the current fixed value of $X$. On the much longer time scale connected to the slow system, the evolution of $X$ integrates out the rapid fluctuation of $Y$. As in the law of large numbers, the overall effect of all these integrated fluctuations can be substituted by one single value. It can be shown that for finite time $T$, the following applies:
\begin{itemize}
\item the trajectory $X(t)$ converges to a solution of:
\begin{align*}
	\frac{d\bar{X}}{dt}=\bar{G}_{\bar{X}}(\bar{X})
\end{align*}

where $\bar{G}_X(\bar{X})=\rho_{\bar{X}}(G_X(\bar{X},Y))$ is the averaged value of the tendency;
\item  the average is taken over the invariant measure $\rho_{X}$ of the $Y$ variable  of the dynamical system
\begin{align*}
	\frac{d{Y}}{dt}=& G_Y(\bar{X},Y)
\end{align*}
resulting when $\bar{X}$ is considered as a fixed forcing parameter. 
\end{itemize}

Let us consider a simple example system. 
\begin{align*}
	\frac{d{X}}{dt}=& (1-Y^2) x \\
	\frac{d{Y}}{dt}=& -\frac{1}{\epsilon} Y + \sqrt{\frac{2}{\epsilon}}\frac{dW}{dt}
\end{align*}
The $Y$ system is here independent of $X$. The invariant measure of the fast $y$ system is a Gaussian distribution with zero mean and unit variance. Taking the average of $G_X(X,Y)= (1-Y^2) x$ under the invariant measure of $Y$, we see that the averaged equation in this case is the uninteresting equation $\dot{X}=0$.

This simple example immediately motivates the use of homogenization methods. Here one scales the equation to a longer time scale $\theta = \epsilon t$, the so called diffusive time scale and then performs the asymptotic expansion. Similarly to how correctly rescaling the sums of the law of large number leads to the more interesting central limit theorem, which describes the fluctuations around the average value, also in the setting of time scale separated systems, we get stochastic behavior on the diffusive time scale. For the example considered above, we get a \textit{weak convergence} to a reduced stochastic differential equation for the $X$ variable instead of the trivial dynamical system obtained before \cite{pavliotis_multiscale_2008}.

The theory for averaging and homogenization in time-scale separated stochastic differential equations is well understood, with results for both one way and two way couplings between the levels \cite{bakhtin_diffusion_2004}. As usual, the theory is more complicated for deterministic systems. Examples of dynamical systems can be constructed where for a large set of initial conditions of $Y$, the solution for $X$ does not converge to the averaged solution \citep{kifer_convergence_2008}. Furthermore, if the $Y$ system has long time correlations, such as in a system with regime behavior, the homogenized system may converge badly and an extension based on a truncation of the transfer operator has been proposed \citep{schutte_averaging_2004}.

\citet{abramov_suppression_2011} has recently presented a study of uncertainty and predictability of the slow dynamics for a system of geophysical relevance. A study of averaging and homogenization for idealized climate models, with a range of examples, can be found in \citet{monahan_stochastic_2011}. Another rather successful attempt in this direction is given in \citet{majda_mathematical_2001}. In \citet{strounine_reduced_2010} stochastic mode reduction is applied to a three-level quasi-geostrophic model whereas in \citet{arnold_reduction_2003} the authors perform mode reduction on a simple coupled atmosphere-ocean model. Another application of homogenization to a toy model for the large-scale dynamics of the atmosphere can be found in \cite{frank_stochastic_2013}. Averaging for the case where one deals with partial differential equations, as is relevant for climate modeling, is discussed by \citet{dymnikov_mathematics_2012}.


A study of homogenization for geophysical flows was performed in \citet{bouchet_kinetic_2013}. The slow system is considered to be the evolution of zonal jets of a barotropic flow, which is forced by noise. The fast degrees of freedom are those representing the fast non-zonal turbulence. Homogenization has also been applied in \cite{dolaptchiev_stochastic_2012} to the Burgers equation, where the slow variables are taken to be averages over large grid boxes and the fast variables are the subgrid variables.

When one wants to consider very large time scales (for examples times of the order of $\exp(1/\epsilon)$), one needs to look beyond the central limit type theorems of homogenization and consider so called large deviation results. These describe for example the transitions between disconnected attractors of the averaged equations \citep{kifer_large_2009} and are of great relevance for studying tipping points \citep{Lenton2008}, going beyond simple one-dimensional approximate theories (see, \textit{e.g.}, discussion in \citet{LFW2012}).

\subsection{Projection operator techniques}
\label{sec:mz}

Projection operator techniques do not constitute a mode reduction per se, but are a way to rewrite the dynamical equation of a multi-level equations to depend only on a subset of variables. A projection is carried out on the level of the observables to remove unwanted, irrelevant and usually fast degrees of freedom. The price one has to pay for this apparent reduction is the appearance of additional terms that are as difficult to compute as the original system. It can however be a useful starting point for further approximations. These of techniques are also known as the Mori-Zwanzig approach \citep{zwanzig_ensemble_1960,zwanzig_memory_1961,mori_new_1974}. 

If a dynamical system is defined on a manifold $\mathcal{M}$, one defines a projection $\mathcal{P}$ from the space of observable functions on the full phase space $\mathcal{M}$ to a space of observables which are considered to contain only the interesting dynamics. Many different choices are possible; if the manifold $\mathcal{M}$ consists for example of a product of submanifolds $\mathcal{K}$ of relevant and $\mathcal{L}$ of irrelevant variables, one can take a conditional expectation with respect to a measure on $\mathcal{M}$, given the value of the relevant variables $X \in \mathcal{K}$:
\begin{align*}
(\mathcal{P} A) (X) = \frac{\int_{\mathcal{N}} A(X,Y) \rho(X,Y) dY}{\int_{\mathcal{N}} \rho(X,Y) dY} \,.
\end{align*}
Another possible choice is a projection onto a set of functions on $\mathcal{M}$, such as linear functions of the coordinates in a Euclidean phase space. In general, one can think at various ways of performing \textit{coarse graining}.

Let us go back to our general formulation of a dynamical system of the form $\dot{Z}=F(Z)$, $Z\in R^N$. The evolution of an observable $A(Z)$ can be written as $\dot{A}(Z) = F(Z)\cdot \nabla_Z A(Z)$, which can be written as $\dot{A} = L A$, often referred to as Liouville equation. The evolution operator $L$ is split into its projection $\mathcal{P}L$ onto the relevant space of observables and the complement $\mathcal{Q}L := (1-\mathcal{P})L$. As described by \citet{zwanzig_nonequilibrium_2001}, a generalized Langevin equation can then be derived based on Dyson's formula for operator exponentials
\begin{equation}
e^{t L} = e^{t \mathcal{Q} L} + \int_0^t e^{(t-s) L} \mathcal{P} L e^{s \mathcal{Q} L} ds \label{eq:dyson2}
\end{equation}
We write the Liouville equation for an observable $A$ as
\begin{equation*}
\frac{d A(t)}{dt} = LA(t) = e^{tL} LA  = e^{tL} \mathcal{P} L A  +  e^{tL} \mathcal{Q} L A
\end{equation*}
The factor $\exp(tL)$ in the second term can be further expanded by making use of Eq. (\ref{eq:dyson2}). This gives the following equation
\begin{equation*}
\frac{d A(t)}{d t} = e^{tL} \mathcal{P}L A + ( e^{t\mathcal{Q}L} + \int_0^t ds \;  e^{(t-s)L} \mathcal{P}L e^{s \mathcal{Q}L}) \mathcal{Q}LA
\end{equation*}
\citet{zwanzig_nonequilibrium_2001} proposes the following interpretation of this equation. The first term on the right hand side corresponds to the regular, deterministic dynamics of the system. The second term can be seen as describing a contribution from correlated noise, dependent on the initial conditions of the irrelevant degrees of freedom. The third term (under integral) represents the memory of the system due to the presence of irrelevant variables that have interacted with the relevant ones in the past. In other term, the price we pay by separating somewhat arbitrarily relevant from irrelevant degrees of freedom s that the irrelevant degrees of freedom act as a stochastic component and, somewhat counter-intuitively, as proxies for the past state of the relevant degrees of freedom. Note that we have done nothing more than manipulating the original evolution equation $\dot{A}= LA$. Correspondingly, the Mori-Zwanzig equation in itself does not simplify the problem. In order to derive a set of equations that are useful for numerical simulations, assumptions need to be made about the dynamical system. 

Several approximations to the Mori-Zwanzig equations have been proposed in the literature. There are the short and long memory approximations made in the method of optimal prediction \citep{chorin_stochastic_2013,defrasne_gute_2004,hald_convergence_2001,chorin_prediction_2006,chorin_problem_2006,chorin_optimal_2000,park_stochastic_2007,chorin_optimal_1998,bernstein_optimal_2007,chorin_optimal_2002}.

In the limit of an infinite time-scale separation between the relevant and irrelevant variables, the stochastic component of the parametrization can be represented as a white noise term, while the memory (also known as \textit{non-Markovian}) term vanishes, as the irrelevant variables decorrelate quickly. Therefore, in such a limit the Mori-Zwanzig decomposition is equivalent to the  homogenization method of section \ref{sec:averaging}. For a comparison of the short memory approximation of Mori-Zwanzig to homogenization for climate-relevant models, see \cite{stinis_comparative_2006}. We also refer the reader to recent results of \citet{CLW2013a,CLW2013b}, where general  mathematical results for the procedure of mode reduction, with thorough geometrical and dynamical interpretations, are given. 

Applications of the Mori-Zwanzig approach to fluid dynamics can be found in \cite{stinis_higher_2007,chandy_t-model_2009,hald_optimal_2007,hou_organized_2007}. 
A simple approximation to Mori-Zwanzig has been applied to jet formation on a $\beta$ plane in \cite{tobias_direct_2013}.

\subsection{Weakly coupled systems}
\label{sec:weak}

We now consider dynamical systems consisting of two systems with a weak coupling. In this case an expansion of the dynamics can be made in orders of the coupling, giving insight into what properties of the coupled systems determines the memory kernel and correlated noise that appeared in the Mori-Zwanzig approach \citep{wouters_disentangling_2012,wouters_multi-level_2013}, because no assumptions are taken regarding time-scale separation.

A possible application of this theory in climate science can be found in the
interaction between cloud formation and large scale atmospheric flow, where there is no distinct time scale separation, but instead the coupling could be considered as weak. The weak coupling limit of a tropical ocean-atmosphere model has also been considered in the literature \cite{neelin_modes_1993}.

Let us go back to Eq. \ref{eq:coupled_dyn_syst_a}. In this setting, the background vector field $F$ consists of a Cartesian product $(F_{X},F_{Y})^\top$ of the vector fields $F_{X}$ and $F_{Y}$ defining the autonomous $X$ and $Y$ dynamics. The perturbing vector field $\delta F$ is a coupling $(\Psi_{X},\Psi_{Y})^\top$ between the two systems. We rewrite the full dynamical system as:
\begin{align}
\frac{dX}{dt} &= F_X(X) + \epsilon \Psi_X(X,Y) \nonumber \\
\frac{dY}{dt} &= F_Y(Y) + \epsilon \Psi_Y(X,Y) \label{eq:coupled_dyn_syst}
\end{align}
where $\epsilon$ is added in order to clarify what kind of perturbative expansion we consider. For simplicity of presentation, for now we consider the case where $\Psi_{X}(X,Y)=\Psi_{X}(Y)$ and $\Psi_{Y}(X,Y)=\Psi_{Y}(X)$. We will come back to the general case later.

Given that the coupling term $\epsilon \Psi$ can be seen as a small perturbation to the uncoupled system, one can make use of response theory to study the change of long time means under a change in the coupling parameter $\epsilon$. We can therefore use the response formalism  described in Section \ref{ruelle}. After lengthy calculations, one obtains the explicit expression for  
\begin{equation}
\rho(A)_t=\rho^0(A)_t+\rho^{(1)}(A)_t+\rho^{(2)}(A)_t+O(\Psi^3)
\end{equation}

\begin{figure}[th]
\centerline{
\hspace{10pt} \includegraphics[scale=0.6]{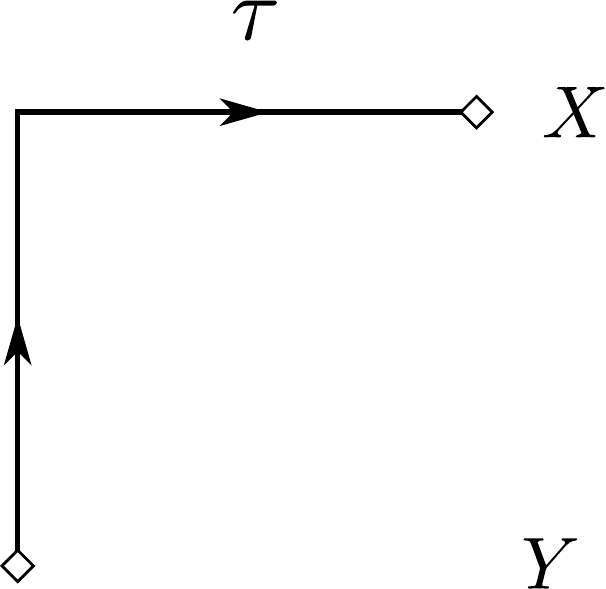}}
\caption{Diagram describing the mean field effect of the $Y$ variables on the $X$ variables. Term $M$ in Eq. (\ref{eq:second_modelb}).}
\label{fig:1st-order}
\end{figure}

\begin{figure}[th]
\centerline{
\hspace{10pt} \includegraphics[scale=0.6]{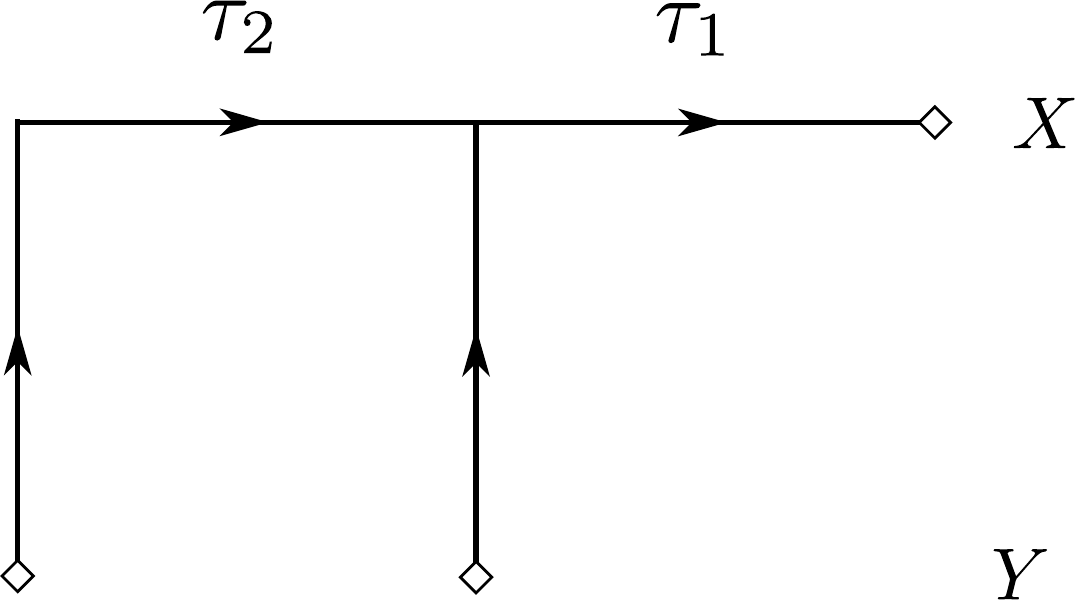}}
\caption{Diagram describing the impact of fluctuations of the $Y$ variables on the $X$ variables. Term $\sigma$ in Eq. (\ref{eq:second_modelb}).}
\label{fig:2nd-order1}
\end{figure}

\begin{figure}[th]
\centerline{
\hspace{10pt} \includegraphics[scale=0.6]{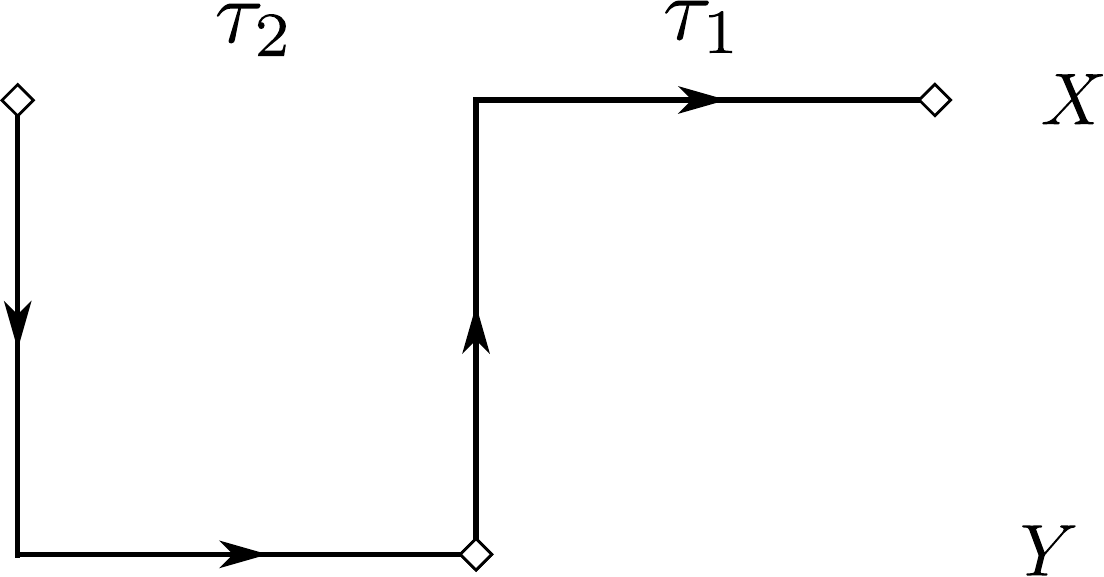}}
\caption{Diagram describing the non-Markovian effect of the  on the $X$ variables on themselves, mediated by the $Y$ variables. Term $h$ in Eq. (\ref{eq:second_modelb}).}
\label{fig:2nd-order2}
\end{figure}

As shown by \citet{wouters_disentangling_2012}, if one collects these first and second order responses to the coupling $\Psi$, an identical change in expectation values from the unperturbed $\rho_0$ up to third order in $\Psi$ can be obtained by adding a $Y$-independent forcing to the tendency of the $X$ variables as follows:
~ \begin{align}
~ \frac{dX(t)}{dt} =& F_X(X(t)) + M + \sigma (t) \nonumber \\ &  + \int_0^\infty d\tau h (\tau,X(t-\tau))\label{eq:second_modelb}
~ \end{align}
~ where $M=\rho_{0,Y}(\Psi_{X})$ is an averaged version of the $Y$ to $X$ coupling, $\sigma$ is a stochastic term, mimicking the two time correlation properties of the unresolved variables and $h$ is a memory kernel that introduces the non-Markovianity. A diagrammatic representation of processes responsible that these three additional terms are parametrizing is given in Figs. \ref{fig:1st-order}-\ref{fig:2nd-order2}. Figure  \ref{fig:1st-order} refers to the mean field effect, which is captured by the first order correction, and corresponds to the deterministic parametrization. Figure  \ref{fig:2nd-order1} describes the effect of the fluctuations of the unresolved variables, which results into an effective stochastic term in the parametrization. Finally, figure \ref{fig:2nd-order2}  clarifies how memory effects enter into the picture of the parametrization: the resolved variables at a given time impact the resolved variables at a later time through a transfer of information mediated by the unresolved variables. The memory effect is present due to the finite time scale difference between resolved and unresolved variables, which also ensures that the stochastic contribution shown in Fig. \ref{fig:2nd-order1} cannot be represented by a white noise process. In \citet{wouters_multi-level_2013} this reduced equation was shown to be related to an expansion in the coupling strength of a Mori-Zwanzig equation.

 If the coupling functions $\Psi_{X}$ and $\Psi_{Y}$ are allowed to be dependent on both $X$ and $Y$, the above analysis can still be carried out. In practical terms, this accounts for the possibility that the coupling terms are function of both the variables we want to parametrize and of those we want to keep explicitly represented in our model. For the case of separable couplings $\Psi_{X}(X,Y)= \Psi_{X,1}(X)\Psi_{X,2}(Y)$ and $\Psi_{Y}(X,Y)= \Psi_{Y,1}(X)\Psi_{Y,2}(Y)$ the average term becomes $X$ dependent and the noise term becomes multiplicative instead of additive. 
An expression for more general couplings can be derived be decomposing the coupling functions into a basis of separable functions, see  \cite{wouters_disentangling_2012}, and then the same procedure can be applied.

\section{Summary and Conclusions}
\label{conclusions}

The goal of this review paper is the provision of an overview of some ideas emerging at the interface between  climate science, physics,  and mathematics, with the objective of contributing to bridging the gap between different scientific communities. The topics have been selected by the authors with the goal of covering (at least partially) relevant aspects of the deep symmetries of geophysical flows, of the processes by which they convert and transport energy, and generate entropy, and of constructing relevant statistical mechanical models able to address fundamental issues like the response of the climate system to forcings, the representation of the interaction across scales, the definition of relevant physical quantities able to describe succinctly the dynamics of the system. This review also informs the development and testing of climate models of various degrees of complexity, by analyzing their physical and mathematical well-posedness and for constructing parametrizations of unresolved processes, and by putting the basis for constructing diagnostic tools able to capture  the most relevant climate processes.   

The Nambu formulation of geophysical fluid dynamics explored in  section \ref{nambu} emphasizes the existence, in the inviscid and unforced case, of non-trivial conserved quantities that are embedded in the equations of motion. Such quantities play a fundamental role, analogous to energy's, in the description of the state and of the dynamics of the system and can be regarded as observables of great relevance also in the case where dissipation and forcing are present. Moreover, the Nambu formalism suggests us ways for devising very accurate numerical schemes, which do not have spurious diffusive behavior.

The symmetry properties of the flow in the inviscid limit allow the construction of the ensembles describing the equilibrium statistical mechanical properties of the geophysical flows (section \ref{onsager}), where the vorticity - in the two dimensional case - plays the role of the most important physical quantity. Starting from the classical construction due to Onsager of the gas of interacting vortices, the theory leads us to construct a theory of barotropic and baroclinic QG turbulence. 

Taking the point of view of non-equilibrium systems, we have that thanks to the presence of gradients of physical quantities like temperature and chemical concentrations - in first instance due to the inhomogeneity of the incoming solar radiation, of the optical properties of the geophysical fluids, and of the boundary conditions - the climate system can transform available potential energy into kinetic energy via internal instabilities, resulting in organized fluid motions. In  section \ref{prigogine} the analysis of the energy and entropy budgets of the climate system is shown to provide a comprehensive picture of climate dynamics, new tools  for testing and auditing climate models and measuring climate change, for investigating of the climate tipping points, and for studying the properties of general planetary atmospheres.

Section \ref{ruelle} introduces some basic concepts of non-equilibrium statistical mechanics, connecting the macroscopic properties described in the previous section to the  features of the family of chaotic  dynamical systems which constitute the backbone of the mathematical description of non-equilibrium systems. For such systems, the relationship between internal fluctuations and response to forcings is studied with the goal of developing methods for predicting climate change. After clarifying the conditions under which the FDT is valid, we present some new results such as a successful climate prediction for decadal and longer time scales. In this sense, we show that the problem of climate change is mathematically well-posed.

Non-equilibrium statistical mechanics is also the subject of  section \ref{morizwanzig}, where we show how the Mori-Zwanzig formalism supports the provision of rigorous methods for constructing parametrizations of unresolved processes. It is possible to derive a surrogate dynamics for the coarse grained variable of interest for climatic purposes, incorporating, as result of the coupling with the small scale, fast variables, a deterministic, a stochastic, and a non-Markovian contribution, corresponding to memory effects, which add to the unperturbed dynamics. The same results can be obtained using the response theory described in  section \ref{ruelle}, thus showing that the construction of parametrizations for weather and for climate models should have common ground. 

Among the many topics and aspects left out of this review, we need to mention recent developments aimed at connecting the complementary, rather than opposing \cite{Lorenz:1963} and \cite{hasselmann_stochastic_1976} perspectives on complex dynamics dynamics, which focus on deterministic chaos and stochastic perturbations to dynamical systems, respectively. We refer in particular to the idea of constructing time-dependent measures for non autonomous dynamical systems \citep{Chekroun2011} through the introduction of the so-called \textit{pullback attractor}, which is the geometrical object the trajectories initialized in a distant past  tend to at time $t$ with probability 1 as a result of the contracting dynamics. Such an object is not invariant with time, as a result of the time-dependent forcing, but, under suitable conditions on the properties of the dynamical system, the supported measure has at each instant properties similar to those of the (invariant) SRB measure one can construct for, e.g. autonomous Axiom A dynamical \citep{ruelle89}. Such an approach allows for treating in a coherent way the presence of modulations in the dynamics of the system, without the need of applying response formulas or of assuming time-scale separations, and in particular allows for analyzing the case where the forcing is stochastic, leading to the concept of random attractor \citep{arnold1988}. On a different line of research, it is instead possible to use Ruelle response theory for computing the impact of adding stochastic noise on chaotic dynamical systems \citep{lucarini2012}. One finds the rate of convergence of the stochastically perturbed measure to the unperturbed one, and discovers the general result that adding noise enhances the power spectrum of any given observables at all frequencies. The difference between the power spectrum of the perturbed and unperturbed system can be used, mirroring an FDT, for computing the response of the system to deterministic perturbations.

The methods, the ideas, the perspectives presented in this paper are partially overlapping, partially complementary, partly in contrast. In particular, it is not obvious, as of today, whether it is more efficient to approach the problem of constructing a theory of climate dynamics starting from the framework of hamiltonian mechanics and quasi-equilibrium statistical mechanics or taking the point of view of dissipative chaotic dynamical systems, and of non-equilibrium statistical mechanics, and even the authors of this review disagree. The former approach can rely on much more powerful mathematical tools, while the latter is more realistic and epistemologically more correct, because, obviously, the climate is, indeed, a non-equilibrium system. Nonetheless, the experience accumulated in many other scientific branches (chemistry, acoustics, material science, optics, etc.) has shown that by suitably applying perturbation theory to equilibrium systems one can provide an extremely accurate description of non-equilibrium properties. Such a lack of unified perspective, of well-established paradigms, should be seen as sign of the vitality of many research perspectives in climate dynamics.

\begin{acknowledgments}
The authors acknowledge interactions with U. Achatz, M. Ambaum, F. Cooper, M. Ghil, J. Gregory, K. Heng, T. Kuna, P. N\'{e}vir, O. Pauluis. D. Ruelle, A. Seifert, and R. Tailleux. The authors wish to thank F. Lunkeit and F. Sielmann for helping in data analysis, and R. Boschi, A. M\"{u}ller and M. Sommer for providing useful figures. VL, RB, SP, and JW wish to acknowledge the financial support provided by the ERC-Starting Investigator Grant NAMASTE (Grant no. 257106).  The authors wish to acknowledge support by the Cluster of Excellence CliSAP. The National Center for Atmospheric Research is sponsored by the National Science Foundation.
 \end{acknowledgments}

\appendix

\section{Glossary}

For the benefit of the reader, we report here the  most relevant symbols used in this paper, indicating if the same symbol is used with different meaning.
\vspace{10 pt}

$\mathcal{H}$ Hamiltonian functional

$\{\bullet,\bullet\}_P$ Standard Poisson Brackets

$\bf u$ Velocity vector (two- or three-dimensional)

$\bf u_a$ Absolute velocity vector (including planetary rotation)

$\nabla \cdot$ Divergence operator (two- or three-dimensional)

$\nabla_h \cdot$ Horizontal divergence operator for three-dimensional vectors

$\nabla $ Gradient operator (two- or three-dimensional)

$\nabla_h$ Horizontal gradient operator for three-dimensional fields

$\psi$ streamfunction 

$\omega$ Vorticity function (two-dimensional dynamics)

$\mbox{\boldmath{$\omega$}}$ Vorticity vector (three-dimensional dynamics)

$\mbox{\boldmath{$\omega_a$}}$ Absolute vorticity vector (including planetary vorticity)

$S$ Symplectic matrix [0, -1; 1, 0];

$\mathcal{J}$ Jacobian operator

$\mathcal{X}$ Generic functional

$\delta \mathcal{X}/\delta a $ Functional derivative of $\mathcal{X}$ with respect to the function $a$.

$\{\bullet,\bullet,\bullet\}$ Nambu brackets

$\mu$ Horizontal divergence of the velocity field 

$h_T$ Total thickness of the fluid (shallow water equations)

$h$ Helicity

$h_a$ Absolute helicity (including planetary rotation)

$f=f_0+\beta y$ Planetary vorticity in $\beta-$plane approximation ($y$ indicates the South-North coordinate)

$\Phi$ Geopotential

$Q$ Quasi-geostrophic  potential vorticity

$q$ Potential vorticity for shallow water equations (Section \ref{nambu}); specific humidity (Section \ref{prigogine})

$N$ Brunt-V\"{a}is\"{a}l\"{a} frequency
 
 $\rho$ Density of the fluid (Sections \ref{nambu} and \ref{prigogine}); Invariant measure of the system (Sections \ref{onsager}, \ref{ruelle}, and \ref{morizwanzig}).

$\Omega$ Earth's angular velocity vector
 
$\Pi$ Ertel's potential vorticity

$\mathcal{E}$ Enstrophy functional

$\mathcal{E}_\rho$ Potential enstrophy functional

$\mathcal{S}$ Entropy functional 

$\mathcal{M}$ Mass functional 

$\partfun$ partition function

$R_D=2\pi/k_D$ Rossby deformation radius

$\meanfield{S}$ Mixing entropy

$e$ Specific energy  per unit mass

$i$ Specific internal energy  per unit mass

$p$ pressure

$H$ relative humidity

$g$ Gravity

$T$ Temperature

$C_v$, $C_W$ Specific heat at constant volume

$p$ pressure

$s$ ($\sigma=\rho s$) specific entropy per unit mass (per unit volume)

$\Omega(E)$ structure function

$L$ Latent heat of vaporization (Section \ref{prigogine}); Liouville operator (Section \ref{morizwanzig})

$\mathbf{F_R}$ Vector of  radiative   flux.

$\mathbf{F_S}$ Vector of  turbulent sensible heat flux.

$ \mathbf{F_L}$ Vector of  turbulent latent heat flux.

 $\mathbf{\tau}$ Surface stress tensor

$E$ Total energy

$P$ Total static potential energy

$K$ Total kinetic energy

$W$ Conversion rate between potential and kinetic energy

$D$ Rate of dissipation of the kinetic energy

$\dot{\Phi}^+$ Net positive heating rate taking place at average temperature $T^+$ 

$\dot{\Phi}^-$ Net negative heating rate  taking place at average temperature $T^-$

$\eta$ Climate efficiency

$\dot{S}_{mat}$ Rate of material entropy production of the climate system

$\dot{S}_{fric}$ Rate of material entropy production due to friction

$\dot{S}_{diff}$ Rate of material entropy production due to diffusion

$\dot{S}_{hyd}$ Rate of material entropy production due to the hydrological cycle

$\dot{S}^v_{mat}$ Rate of material entropy production of the climate system due to vertical processes

$\dot{S}^h_{mat}$ Rate of material entropy production of the climate system  due to horizontal processes

$\rho^{(1)}(A)$ First order correction to the expectation value of  the observable A 

$G^{(1)}_A(t)$ First order Green function for the observable A 

$\chi^{(1)}_A(\omega)$ First order susceptibility function for the observable A 

$\Delta_T$ Climate Sensitivity

$\mathcal{P} =1-\mathcal{Q} $ Projection operator performing \textit{coarse graining} on the dynamics and eliminates irrelevant degrees of freedom; $\mathcal{Q}$ is the complementary operator

\end{document}